\documentclass[useAMS,usenatbib]{mnras}
\bibpunct{(}{)}{;}{a}{}{,}
\usepackage{graphicx}
\usepackage{longtable}
\usepackage{color}

\newcommand{\magcir}{\raise
-2.truept\hbox{\rlap{\hbox{$\sim$}}\raise5.truept
\hbox{$>$}\ }}
\newcommand{\minmag}{\raise-2.truept\hbox{\rlap{\hbox{$<$}}\raise
6.truept\hbox
{$>$}\ }}

\newcommand{\lya}{Ly$\alpha$~}
\newcommand{\lyb}{Ly$\beta$~}

\newcommand{\kms}{km~s$^{-1}$}

\DeclareMathAlphabet{\mathsc}{OT1}{cmr}{m}{sc}
\def\testbx{bx}%
\DeclareRobustCommand{\ion}[2]{%
\relax\ifmmode
\ifx\testbx\f@series
{\mathbf{#1\,\mathsc{#2}}}\else
{\mathrm{#1\,\mathsc{#2}}}\fi
\else\textup{#1\,{\mdseries\textsc{#2}}}%
\fi}
\def\nhi{\mbox{$N_{\mathrm{\ion{H}{i}}}$}}

\def\civ{\mbox{\ion{C}{iv}}}

\def\ncivsys{\mbox{$N_{\mathrm{\ion{C}{iv}, \, sys}}$}}
\def\lognh{\mbox{$\log N_{\mathrm{\ion{H}{i}}}$}}
\def\lognc4{\mbox{$\log N_{\mathrm{\ion{C}{iv}}}$}}
\def\logno6{\mbox{$\log N_{\mathrm{\ion{C}{iv}}}$}}
\def\no6{\mbox{$N_{\mathrm{\ion{O}{vi}}}$}}

\def\nhisys{\mbox{$N_{\mathrm{\ion{H}{i}, \, sys}}$}}
\def\nhicl{\mbox{$N_{\mathrm{\ion{H}{i}, \, cl}}$}}
\def\ncivcl{\mbox{$N_{\mathrm{\ion{C}{iv}, \, cl}}$}}
\def\nhicomp{\mbox{$N_{\mathrm{\ion{H}{i}, \, comp}}$}}
\def\ncivcomp{\mbox{$N_{\mathrm{\ion{C}{iv}, \, comp}}$}}

\def\nciv{\mbox{$N_{\mathrm{\ion{C}{iv}}}$}}
\def\nhi{\mbox{$N_{\mathrm{\ion{H}{i}}}$}}

\title[The integrated \nhi\/--\nciv\/ relation at $z \sim 2.5$]
{Triple-ionised carbon associated with the low-density neutral hydrogen
gas at $1.7 < z < 3.3$: the integrated \nhi\/--\nciv\/
relation\thanks{Based 
on data obtained with UVES (Ultraviolet and Visual Echelle Spectrograph) at 
the VLT (Very Large Telescope), Paranal, Chile, from the ESO archive
and obtained with HIRES (HIgh Resolution Spectrometer) at Keck, Hawaii, USA, from 
the Keck archive.}}
\author[Kim et al.]{T.-S. Kim$^{1, 2, 3}$\thanks{E-mail: kim@oats.inaf.it}, 
R. F. Carswell$^{4}$,
C. Mongardi$^{5}$,
A. M. Partl$^{2}$, J. P. M\"ucket$^{2}$,
\newauthor
P. Barai$^{1}$, S. Cristiani$^{1}$ \\
$^1$ INAF, Osservatorio Astronomico di Trieste, Via G. B. Tiepolo, 11, 34143, Trieste, Italy\\
$^2$ Leibniz-Institut f\"ur Astrophysik Potsdam,
An der Sternwarte 16, D-14482 Potsdam, Germany\\
$^3$ Department of Astronomy, University of Wisconsin, 475 North Charter Street,
Madison, WI 53706, USA \\
$^4$ Institute of Astronomy, Madingley Road, Cambridge
CB3 0HA\\
$^{5}$ Dipartimento di Fisica, Sezione di Astronomia, Universit\`{a} di Trieste, Via G. B. 
Tiepolo 11, I-34143 Trieste, Italy
}

\begin{document}

\date{Accepted Recieved}

\maketitle

\begin{abstract}
From the Voigt profile fitting analysis of 183 intervening \ion{C}{iv} systems at
$1.7  < z  < 3.3$ in 23 high-quality UVES/VLT and HIRES/Keck
QSO spectra, 
we find that a majority of \ion{C}{iv} systems ($\sim$\,75\,\%) display
a well-characterised scaling relation between integrated column densities of 
\ion{H}{i} and \ion{C}{iv} with a negligible redshift evolution,
when column densities of all the \ion{H}{i} and \ion{C}{iv} components are 
integrated within a given
$\pm 150$\,\kms\/ range centred at the \ion{C}{iv} flux minimum. 
The integrated \ion{C}{iv} column density \ncivsys\/ increases with \nhisys\/
at $\log N_{\mathrm{\ion{H}{i}, \, sys}} \in [14, 16]$
and $\log N_{\mathrm{\ion{C}{iv}, \, sys}} \in [11.8, 14.0]$, then becomes almost 
independent of \nhisys\/ at $\log N_{\mathrm{\ion{H}{i}, \, sys}} \ge 16$,
with a large scatter:
at $\log N_{\mathrm{\ion{H}{i}, \, sys}} \in [14, 22]$,
$\log\,\ncivsys\/ = 
\left[\frac{C_{1}}{\log\,\nhisys\/ + C_{2}} \right] + C_{3}$, with 
$C_{1} = -1.90 \pm 0.55$, $C_{2} = -14.11 \pm 0.19$
and $C_{3} = 14.76 \pm 0.17$, respectively. 
The steep (flat) part is dominated by 
\ion{Si}{iv}-free (\ion{Si}{iv}-enriched)
\ion{C}{iv} systems. 
Extrapolating the \nhisys\/--\ncivsys\/ relation implies that most
absorbers with 
$\log N_{\mathrm{\ion{H}{i}}} \le 14$ are virtually \ion{C}{iv}-free.
The \nhisys\/--\ncivsys\/ relation does not hold for individual 
components, clumps or the integrated velocity range less than 
$\pm \,100$\,\kms\/.
This is expected if
the line-of-sight extent of 
\ion{C}{iv} is smaller than \ion{H}{i}
and $N_{\mathrm{\ion{C}{iv}, \, sys}}$ decreases more rapidly than 
$N_{\mathrm{\ion{H}{i}, \, sys}}$ at the larger impact parameter,  
regardless of the location of the \ion{H}{i}+\ion{C}{iv} gas in IGM filaments
or in intervening galactic halos.
\end{abstract}

\begin{keywords}
cosmology: observation -- intergalactic medium -- 
quasars: absorption lines
\end{keywords}

\section{Introduction}
\label{sec1}

The numerous, narrow absorption lines observed blueward of the Ly$\alpha$ emission 
line in spectra of background QSOs are mostly produced by the warm ($\sim 10^{4}$\,K),
photoionised, intergalactic neutral hydrogen (\ion{H}{i}) gas. These absorption
lines or {\it absorbers} 
are known as the Ly$\alpha$ forest or the intergalactic medium (IGM) 
and have a \ion{H}{i} column density 
($N_{\mathrm{\ion{H}{i}}}$) less than $10^{17}$\,cm$^{-2}$. 
Being a dominant reservoir of the baryons at all cosmic epochs and 
tracing the underlying dark matter in a simple manner, 
the Ly$\alpha$ forest has been used as a cosmological
tool to study the primordial power spectrum and the formation and evolution
of the large-scale matter distribution \citep{cen94, dave99, kim04, mcdonald06,
palanque13}.

Absorption lines redward of the Ly$\alpha$ emission are 
produced by metal species which also contribute a small fraction of the 
lines in the Ly$\alpha$ forest. The most common metal transition found 
in QSO spectra is the triply ionised carbon doublet, 
\ion{C}{iv} $\lambda\lambda$~1548.204, 1550.778. 
Roughly half of the Ly$\alpha$ forest with  
$N_{\ion{H}{i}} \! \ge \! 10^{14.5}$\,cm$^{-2}$ is
\ion{C}{iv}-enriched at $z \sim 3$ \citep{cowie95, tytler95, songaila98}.
The \ion{C}{iv} enrichment has even been suggested at lower \nhi\/
\citep{songaila98, ellison00, schaye03}. The triply ionised silicon doublet,
\ion{Si}{iv} $\lambda\lambda$~1393.760, 1402.772, is also common, but
is usually associated with a higher \nhi\/. 
An observational rule of thumb is that higher-\nhi\/ absorbers are 
associated with more metal species and stronger, multi-component metal lines.
Low-ionisation metal transitions, such as \ion{Mg}{ii} and \ion{C}{ii}, are
mostly found at $N_{\ion{H}{i}} \! \ge \! 10^{16}$\,cm$^{-2}$. 
Absorbers with $N_{\ion{H}{i}} \! \ge \! 10^{17.2}$\,cm$^{-2}$ 
(Lyman limit systems or LLSs) display a wide range of
metal species and ionisations \citep{steidel90, levshakov03b, prochaska06, lehner13}
and are thought to be associated with 
outflows/infall in outer halos
\citep{jenkins05, faucher11, kacprzak11, ribaudo11, bouche12, lehner13}, or with
extended disks/inner halos at $\le 10$--20\,kpc,
e.g. an analogue of intermediate/high velocity clouds of the Milky Way or merger
remnants \citep{thilker04, lehner09, stocke10}.

Heavy elements are produced in stars which occur in galaxies.
However, the \lya\/ 
forest does not have an {\it in situ} star formation due to its high temperature 
and low gas density at  $\sim \! 10^{-4}$~cm$^{-3}$, so any metals
associated with the Ly$\alpha$ forest must have been transferred there 
from galaxies in some way.
Therefore, since the discovery of metals associated with the Ly$\alpha$ forest,
studies on the IGM enrichment have been focused mainly on two topics:
{\it what} is the enrichment mechanism and {\it where} 
is the metal-enriched gas located. 

Among several proposed scenarios, such as the enrichment
by Population III stars at $10 \! < \! z \! < \! 20$ \citep{ostriker96, haiman97} 
and by dynamical removal of metals from galaxies through a merger or tidal
interaction \citep{gnedin97, gnedin98, aguirre01b}, the scenario with most
support from observations
is galactic-scale outflows or galactic winds
\citep{dave98, aguirre01a, schaye03, springel03, murray05, oppenheimer06}.

Although detailed outflow mechanisms are still far from being 
clear, in the simplified, qualitative picture, galactic outflows driven by 
supernovae or by young OB stars
disperse metal-enriched gas from disks to halos, from halos to the surrounding 
IGM. The extent of galactic winds are limited by the radiative and mechanical 
energy loss and the pressure of the infalling, surrounding medium. 
One of the predictions by the outflow models is the volume-averaged
overdensity--metallicity relation,
clearly shown in their Fig.~1 by \cite{aguirre01b} (but see also
\cite{springel03} and \cite{oppenheimer06}).
The IGM metallicity drops off sharply at overdensities
smaller than a drop-off overdensity. Above this drop-off ovderdensity, however,  
the IGM metallicity is independent of overdensities. The exact shape of this
overdensity--metallicity relation depends on the outflow velocity, its
onset time and the interaction with the surrounding IGM. 
With a higher outflow velocity, a longer traveling time, a low pressure of the local
IGM and a small potential well of outflow galaxies, 
outflows can enrich a lower-density IGM. With a smaller outflow velocity
and a strong galactic potential well, metals are likely to be located mostly inside
a virial radius of parents galaxies, never escaping in to the IGM, thus leaving
the typical IGM virtually metal-free

Galactic-scale outflows are 
a well-established phenomenon both
at lower and higher redshifts. In local starburst
galaxies, galactic outflows operate on scales of 10--100\,kpc, sometimes
even at $\sim \! 1000$\,\kms\/ scale, enough to escape the parents galaxies
\citep{strickland04, martin05, martin06, tremonti07}. 
The COS-Halos survey at $z \sim 0.2$ also shows that 
star-forming galaxies of $\sim$\,1\,L$_{\star}$ commonly have a large-scale
\ion{O}{vi} outflow up to 150\,kpc \citep{tumlinson11, werk13, werk14}.
At $z\!\sim\!3$, Lyman break galaxies often show outflows with velocities of
several hundred \kms\/ \citep{pettini02, shapley03, erb12}. 
High ions, such as \ion{O}{vi} and \ion{C}{iv}, in sub-damped
Ly$\alpha$ systems (sub-DLAs, $N_{\mathrm{\ion{H}{i}}} \sim 10^{19-20.3}$\,cm$^{-2}$) 
and
damped Ly$\alpha$ systems (DLAs, $N_{\mathrm{\ion{H}{i}}} \ge 10^{20.3}$\,cm$^{-2}$)
also reinforce the notion that the presence of inflows and outflows is common at $z \sim 3$
\citep{fox07a, fox07b, lehner14}.
In addition, \citet{steidel10} used
the close galaxy-galaxy pairs at $z \! \sim \! 2.2$ to study the gas surrounding
foreground galaxies in the spectra of background galaxies at the impact parameters
at 3--125 physical kpc. They found that foreground galaxies are surrounded by 
metal-enriched
gaseous envelopes and that the strength of metal ions, such as \ion{C}{iv} and
\ion{C}{ii}, decreases as a power law up to a certain
impact parameter,
then decreases rapidly beyond it. This point occurs at $\sim 250$\,kpc for
\ion{H}{i} Ly$\alpha$ at $N_{\mathrm{\ion{H}{i}}} \sim 10^{13.25}$\,cm$^{-2}$
and at $\sim 80$\,kpc for \ion{C}{iv} at 
$N_{\mathrm{\ion{C}{iv}}} \sim 10^{13.5}$\,cm$^{-2}$.

We present new results on the well-characterised 
\nhi\/--\nciv\/ relation of \ion{H}{i}
absorbers at $N_{\mathrm{\ion{H}{i}}} \sim 10^{12.5-22}$\,cm$^{-2}$
at $1.7 < z < 3.3$,
using 23 high-resolution ($\sim 6.7$\,\kms\/), high-signal-to-noise
($S/N \sim 50$ per pixel for \ion{H}{i} and $S/N \sim 100$
for \ion{C}{iv}) spectra obtained with UVES at the VLT and HIRES at Keck.
The observed \nhi\/--\nciv\/ relation is an equivalent to the theoretical
overdensity--metallicity relation, as
the overdensity is related to \nhi\/ and  
the metallicity can be estimated
from measured ion column densities,
assuming photoionisation equilibrium and the ambient UV background
\citep{cowie95, hui97, rauch97, dave99, schaye01}. Although our
study does not have a deep redshift survey to look for galaxies associated
with \ion{C}{iv} gas, our \ion{C}{iv} detection limit for a typical line of sight
is $N_{\mathrm{\ion{C}{iv}}} \sim 10^{12}$\,cm$^{-2}$. 
This limit is much lower than
$N_{\mathrm{\ion{C}{iv}}} \, \sim 10^{13.5}$\,cm$^{-2}$ of the study by
\citet{steidel10}. Our data explore a lower-\nciv\/ gas, i.e. far away from nearby
galaxies and provide new observational constraints  
on the outflow mechanisms and the metal abundance
at low-density \ion{H}{i} absorbers.

To obtain a robust \nhi\/ for \ion{C}{iv}-enriched \ion{H}{i} absorbers, 
we performed the Voigt profile fitting analysis including all the available
high-order Lyman series, cf. \citet{kim13}. To derive the physical conditions
of \ion{C}{iv}-producing gas, 
we performed the photoionisation modelling using the
code {\tt CLOUDY} version 13.03 \citep{ferland13}.

\begin{figure}
\vspace{0.2cm}
\hspace{-0.2cm}
\includegraphics[width=8.8cm]{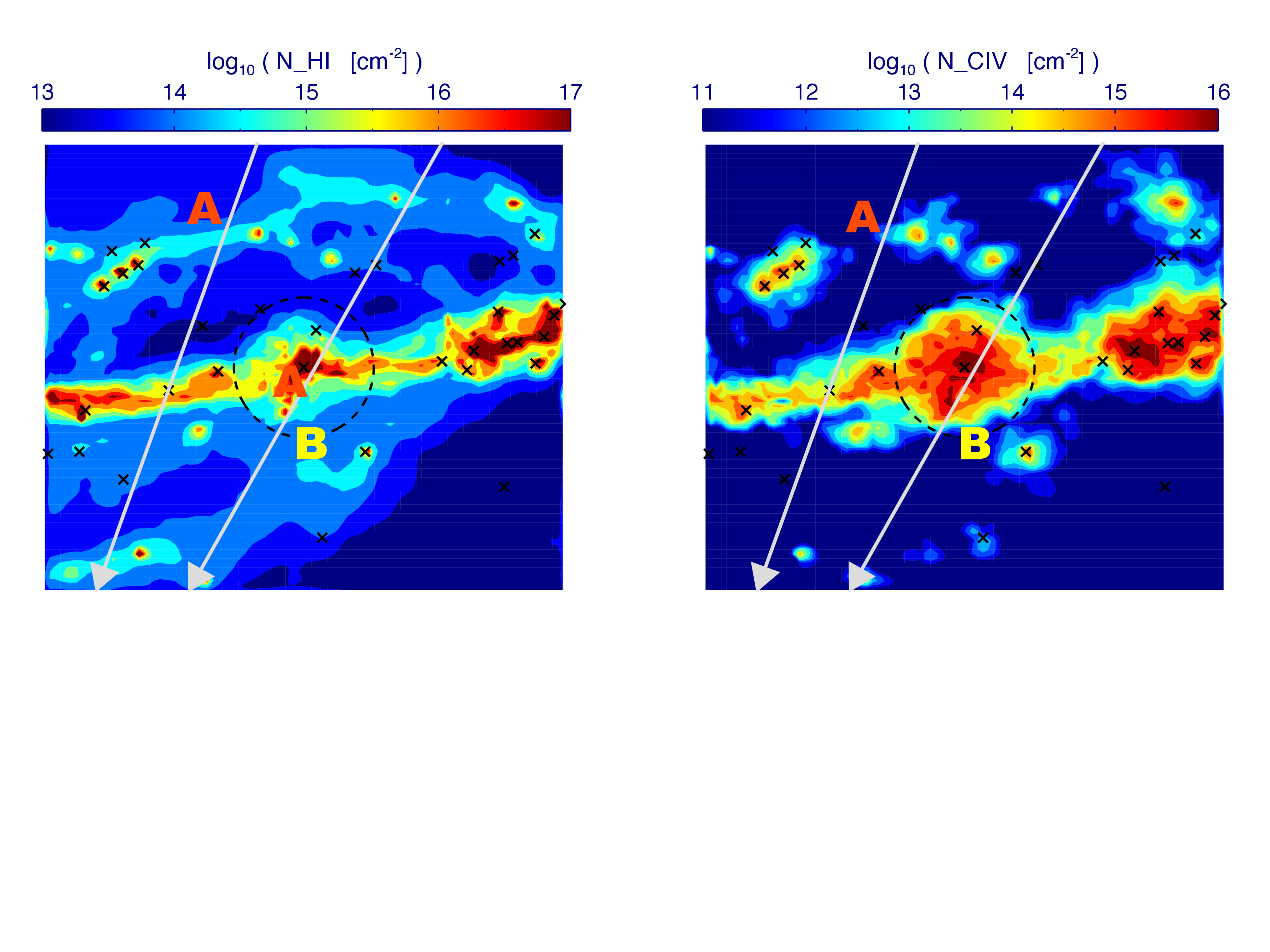}\\

\vspace{-2.2cm}
\caption{The simulated distribution of \ion{H}{i} (the left panel) and \ion{C}{iv}
(the right panel) at $z = 2$, centred around a galaxy with a halo mass
of $10^{11}$\,M$_{\odot}$ in the box size of 1 comoving Mpc. The snapshots
are taken from the run M25std of \citet{barai15}. The black
dashed circle denotes the galaxy virial radius of 150\,kpc. The black crosses
mark the centre of other galaxies or stellar subhalos identified within this
volume. The two arrows represent a typical line of sight in observations, 
which passes through the central galaxy as  well as halos of other galaxies,
metal-enriched filaments and metal-free filaments. The gases at the positions marked
as ``A" and ``B" have a similar $N_{\mathrm{\ion{H}{i}}}$ and  $N_{\mathrm{\ion{C}{iv}}}$,
but are located in a different place.
A coloured, online version provides a better view.
}
\label{fig1}
\end{figure}

Figure~\ref{fig1}
illustrates a snapshot from a 
typical cosmological hydrodynamic simulation taken from \citet{barai15}
at $z = 2$. The left and right panels display the distribution of \ion{H}{i} 
and \ion{C}{iv}, their column density range indicated as a color bar on top.
The two arrows in each panel sample a typical line of sight in observations,
which passes through massive galaxies, dwarf galaxies, IGM filaments
close and far from galaxies and voids. A gas having
a similar $N_{\mathrm{\ion{H}{i}}}$ and $N_{\mathrm{\ion{C}{iv}}}$ can
be located in a different environment.

In practice, the
observed lines of sight provide an ensemble of the gas at a variety of
\nhi\/ and \nciv\/ as well as many different locations, i.e. filaments vs halos within
a virial radius. Therefore, a scatter \nhi\/--\nciv\/ relation might be expected, as seen
by \citet{simcoe04}. Indeed, our \ion{H}{i}+\ion{C}{iv} components
aligned within 5\,\kms\/ from each other
display a scatter plot on the \nhi\/--\nciv\/ plane, but the scatter can be
used to constrain the physical condition of absorbing gas. We also find that
a well-characterised, integrated \nhi\/--\nciv\/ relation exists for
the low-density \ion{Si}{iv}-free \ion{C}{iv} {\it systems} defined as \ion{H}{i}
and \ion{C}{iv} profiles within a fixed velocity ranged centred at the \ion{C}{iv} 
flux minimum. 

This paper is organised as follows. Section \ref{sec2} describes
the analysed data. In Section \ref{sec3}, we present the
Voigt profile fitting analysis in detail. Section \ref{sec4} introduces  
a new working definition on a system and a clump to 
describe our \ion{H}{i}+\ion{C}{iv} sample. 
As Sections \ref{sec3} and \ref{sec4} are rather technical, casual readers might skip 
directly to Section \ref{sec5} in which  
our main results on the integrated \nhi\/ and \nciv\/ relation of \ion{C}{iv}
systems and clumps are presented. Results on the aligned components
are presented in Section~\ref{sec6}. 
Implications of our results are explored in
Section~\ref{sec7}. Section~\ref{sec8} presents the summary.

The logarithmic column density of an {\it ion} M, $\log N_{\mathrm{M}}$,
is expressed as $\log (N_{\mathrm{M}} / {\mathrm{cm}^{-2}})$. 
Throughout the paper, the cosmological parameters are assumed to be
the matter density $\Omega_{m}=0.24$,
the cosmological constant $\Omega_{\Lambda}=0.73$ and
the present-day Hubble constant  $H_{\mathrm{0}} = 100\,h$
km s$^{-1}$ Mpc$^{-1}$ with $h=0.7$ in accord with WMAP measurements
\citep{jarosik11}. 
We also ask readers to look at the online, coloured
versions of any figures to resolve any ambiguities.

\begin{table*}
\caption{Analysed QSOs}
\label{tab1}
{\scriptsize{
\begin{tabular}{lclccll}
\hline
\noalign{\smallskip}
QSO & $z_{\mathrm{em}}^{\mathrm{a}}$ & $z_{\mathrm{\ion{H}{i}, \ion{C}{iv}}}$  & 
   S/N$^{\mathrm{b}}$ & Inst. & ref.$^{\mathrm{c}}$ & Notes \\

\noalign{\smallskip}
\hline
\noalign{\smallskip}

Q0055--269       & 3.656 & 2.663--3.054, 3.077--3.390 & [29, 62] & UVES & 1, 2, 3 & \\
PKS2126--158   & 3.280 & 2.413--2.608, 2.669--3.055, 3.075--3.208 & [100, 125] & UVES & 1, 2, 3 &\\
HS1425+6039    & 3.180 & 2.667--3.110                      & [83, 71] & HIRES & 4 & 
               DLA at $z = 2.827$, sub-DLA at $z = 2.770$ \\
Q0636+6801     & 3.175 & 2.863--3.105                       & [50, 67] & HIRES & 4 & \\
Q0420--388      & 3.115 &  2.670--2.615, 2.670--3.045 & [125, 111] & UVES  & 1, 2, 3 &
               sub-DLA at $z = 3.087$ \\
HE0940--1050  & 3.082 & 2.498--2.716, 2.776--3.014 & [83, 111] & UVES & 1, 2, 3 & \\
HE2347--4342  & 2.873 & 2.080--2.710, 2.770--2.809 & [100, 100] & UVES & 1, 2, 3 & \\
HE0151--4326  & 2.781$^{\mathrm{d}}$ & 2.072--2.710  & [100, 125] & UVES & 1 &
                          mini-BAL \\
Q0002--422    & 2.768 & 2.016--2.705                         & [87, 137] & UVES & 1, 2, 3 & \\
PKS0329--255  & 2.704 & 2.091--2.643                       & [38, 63] & UVES & 1, 2, 3 & \\
Q0453--423    & 2.657 & 1.978--2.595                         & [67, 100] & UVES & 1, 2, 3 &
               sub-DLA at $z = 2.305$ \\
HE1347--2457  & 2.612$^{\mathrm{d}}$ &  1.987--2.552 &  [63, 67] & UVES & 1, 2, 3 & \\
Q0329--385      & 2.435 & 2.001--2.378                         & [45, 83/45] & UVES & 1, 2, 3 & \\
HE2217--2818  & 2.413 & 1.979--2.355                         & [67, 100] & UVES & 1, 2, 3 & mini-BAL \\
Q0109--3518    & 2.405 & 1.976--2.348                         & [67, 91/140] & UVES & 1, 2, 3 & \\
HE1122--1648  & 2.404 & 1.975--2.346                         & [111, 200/71] & UVES & 1, 2, 3 & \\
HE0001--2340  & 2.264 & 1.993--2.211                         & [67, 80] & UVES & 1 &
               sub-DLA at $z = 2.187$ \\               
J2233--606$^{\mathrm{e}}$ & 2.251 & 1.588--2.197      & [33, 45] & UVES & 1, 2, 3, 5 & mini-BAL \\
PKS0237--23    & 2.222 & 1.975--2.167                      & [99, 136] & UVES & 1, 2, 3 &
                sub-DLA at $z = 1.673$  \\
PKS1448--232  & 2.219     &  1.986--2.168                  & [57, 122/70] & UVES & 1, 2, 3 & \\
Q0122--380      & 2.191$^{\mathrm{d}}$ & 1.977--2.140  & [48, 77] & UVES & 1, 2, 3 & \\
Q1101--264   & 2.141 & 1.800--2.090$^{\mathrm{f}}$    & [67, 77] & UVES & 1, 2, 3 &
                              sub-DLA at $z = 1.839$ \\
HE1341--1020  & 2.138$^{\mathrm{d}}$ & 1.972--2.086   & [50, 63] & UVES & 1 & 
                               mini-BAL \\
\noalign{\smallskip}
\hline
\end{tabular}
\begin{list}{}{}
\item[$^{\mathrm{a}}$]
The redshift is measured from the observed Ly$\alpha$ emission line of the QSOs.
\item[$^{\mathrm{b}}$]
The first and the second numbers separated by a comma in the brackets are 
the S/N per pixel estimated from the central parts of the \ion{H}{i} and \ion{C}{iv} regions, 
respectively. Due
to an instrument setup which determines the wavelength ranges to be overlapped, 
some QSOs have a much higher S/N for part of the \ion{C}{iv} region. In this case, 
two numbers are listed separated by ``/".
\item[$^{\mathrm{c}}$]
1: Kim et al. (2004); 2: Kim et al. (2007); 3. Kim et al. (2013); 4: Boksenberg \& Sargent (2015);
5: Savaglio et al. (1999)
\item[$^{\mathrm{d}}$]
The redshift estimated from the Ly$\alpha$ emission profile is uncertain as there are many
absorption lines in the peak of the Ly$\alpha$ emission line or as the emission profile
is non-Gaussian.
\item[$^{\mathrm{e}}$]
The {\it HST}/STIS E230M spectrum (Savaglio et al. 1999) is included in the analysis 3
since its observed
wavelength at 2550--3057 \AA\/ covers high-order Lyman lines of saturated \ion{H}{i}
lines in the analysed redshift range. 
\item[$^{\mathrm{f}}$]
Although \ion{H}{i} Lyman lines higher than Ly$\beta$ are not covered at
$z < 1.972$, a robust \ion{H}{i} column density of a sub-DLA at $z = 1.839$ 
can be obtained due to the damping wing. In order to increase the high-$N_{\mathrm{\ion{H}{i}}}$
absorbers in our sample, the redshift range for this QSO was extended to $z = 1.800$.
\end{list}
}}
\end{table*}


\section{Data}
\label{sec2}

Table~\ref{tab1} lists the 23 QSOs and their spectra analysed
in this study. The 21 raw spectra of the 23 QSOs were taken from the
ESO VLT/UVES archive, while the remaining 2 QSOs were taken from the
Keck/HIRES archive. As this sample was designed to study
the low-density IGM at $2 < z < 3.5$, the first selection criterion was 
QSOs without any strong
DLAs and only few LLSs 
in any one sightline. In addition, 
in order to cover high-order Lyman
lines to obtain a reliable column density of saturated \ion{H}{i}, the second
selection criterion was only QSOs with high S/N and a long,
continuous wavelength coverage.
18 of the 21 UVES spectra
were analysed in \citet{kim07, kim13} and the other three were described by
\citet{kim04}. Most UVES spectra cover from 3050\,\AA\/ 
(a natural cutoff due to the Earth's atmosphere)
to 10000\,\AA\/ (due to the limitation of the optical instrument), with
some gaps due to the CCD detector configuration. 
The 2 HIRES spectra are described by \citet{boksenberg14}. 
A {\it HST}/STIS echelle spectrum is available for 
J2233--606{\footnote{\tt http://www.stsci.edu/ftp/observing/hdf/hdfsouth/hdfs.html}}
\citep{savaglio99}, so it was used here to extend its wavelength coverage 
down to 2300~\AA\/. 

The wavelengths for all spectra are heliocentric corrected, and the
spectral resolution is 
$R \sim 45\,000$ (or $\sim$\,6.7\,km s$^{-1}$). 
The UVES/STIS spectra and the HIRES
spectra are sampled at 0.05\,\AA\/ and 0.04\,\AA\/, respectively. 

To avoid the proximity effect, the region of 5,000\,\kms\/ blueward of
the QSO's Ly$\alpha$ emission was excluded from the Ly$\alpha$ \ion{H}{i} study. 
Note that this velocity cut also eliminates any absorption lines due to
the ejecta of mini-BAL (Broad Absorption Line) QSOs in the sample. This
sets the highest redshift searched for \ion{C}{iv} in each QSO.
The lowest search redshift of \ion{C}{iv} doublets is set by the QSO's Ly$\alpha$
emission line itself. Below the Ly$\alpha$ emission, \ion{C}{iv} becomes blended with
the Ly$\alpha$ forest and its detection is likely to be incomplete regardless
of its detection limit. 
In addition, obtaining a reliable \nhi\/ of saturated \ion{H}{i} lines
requires high-order Lyman lines, which further limits the useful redshift range. 
The signal-to-noise (S/N) ratio
also affects a profile fitting process, with a higher S/N providing a more reliable 
deblending of saturated lines. 
These requirements
set the analysed redshift range for \ion{H}{i} and \ion{C}{iv} of each spectra,
which is listed in the third column
in Table~\ref{tab1}. Section~\ref{sec3} discusses the coverage of high-order lines
and the S/N for obtaining a robust column density in more detail. 

In addition to the CCD detector gaps, the UVES spectra are 
contaminated by numerous telluric lines above 6200\,\AA\/,
particularly at 6276$\sim$6319\,\AA\/ ($3.055 < z_{\mathrm{\ion{C}{iv}}} < 3.080$)
in the redshift range of interest. These regions are also excluded. Where
there are such gaps in the coverage, multiple search ranges are listed in the third
column in Table~\ref{tab1}. 
We did not exclude any other regions, such as the ones
around a DLA or a sub-DLA, and included all the detected 
high-$N_{\mathrm{\ion{H}{i}}}$ absorbers in our analysis.

The fourth column of Table~\ref{tab1}
lists the S/N per pixel in the central parts of the \ion{H}{i} and \ion{C}{iv} regions. 
Since the S/N varies from spectrum to spectrum and even along the same 
spectrum, the listed S/N is only a rough indicator of the data quality. 
The typical S/N is 30\,$\sim$\,50 in the Ly$\alpha$ forest region,
while it is 10\,$\sim$\,15 in the Ly$\beta$ forest region and even lower in the higher-order
forest regions. 
On the other hand, the S/N ratio in the \ion{C}{iv} region is
in general higher. 
The S/N of the J2233--606 STIS spectrum is $\sim$\,7 
at $\sim$\,2750\,\AA\/. 
As the S/N varies along the spectrum, the detection limit
of \ion{H}{i} Ly$\alpha$ and \ion{C}{iv} also varies locally. 

The 5th and 6th columns note the spectrograph used to obtain the spectrum and
the references in which the same spectrum was analysed for other scientific
objectives. Any strong intervening absorbers and/or
mini-BAL QSOs are noted in the 7th
column.
 
\section{Voigt profile fitting analysis}
\label{sec3}

\subsection{Brief description of the Voigt profile fitting procedure}
\label{sec3:1}

We have fitted Voigt profiles to the absorption lines 
to obtain the absorption line 
parameters: the redshift $z$, the column density $N$ in cm$^{-2}$
and the Doppler parameter $b$ ($=\sqrt{2}\sigma$, where $\sigma$ is 
the standard deviation for a Gaussian distribution) in \kms\/. 

Voigt profiles were fitted to the absorption lines using
{\tt VPFIT} \citep{carswell14},
using the rest-frame wavelengths and the oscillator strengths
provided with the program. 
The three versions of {\tt VPFIT} were used to produce the final
line parameters, versions 8.2, 9.5 and 10.2, depending on when the fitting analysis was
performed. For the application here, the final results are very similar whichever
version is used.
Note that the line lists used in this work are similar to, but not necessarily
exactly the same as, those used previously \citep{kim07, kim13}, since
small changes in the estimates for the local continuum or the removal/addition
of weak column density components affect the detailed results. 

Details of the methods used and caveats can be
found in the documentation accompanying the program, and also in 
Carswell, Schaye \& Kim (2002) and
\cite{kim07}. Here, we give a short description of the fitting procedure, following
\cite{kim13}. 

First, each spectrum was divided into several chunks and normalised locally by connecting
seemingly unabsorbed regions
using the {\tt CONTINUUM/ECHELLE} command in {\tt IRAF}. Second, possible metal
lines were searched for from the longest wavelength toward the 
shorter wavelength. All the identified 
metal lines were fitted first. Then, using them as presets, the rest of the
absorption features were fitted as \ion{H}{i}.
When metal lines were blended with \ion{H}{i} lines and/or other metal lines,
all the blended lines were included in the fit simultaneously.
To obtain
reliable line parameters of saturated \ion{H}{i} lines, all the available higher-order
Lyman series such as Ly$\beta$ and Ly$\gamma$ were also included in the fit. 
Since there is no unique solution to the profile fitting, we imposed only one
condition: a minimum number of necessary
components to reach a reduced $\chi^{2}$ value to be $\le 1.2$,
cf. Boksenberg \& Sargent (2014). Note that the STIS E230M line spread function
was used to fit the J2233$-$606 STIS spectrum.

The redshifts and $b$ parameters for \ion{H}{i} and \ion{C}{iv}
were allowed to vary {\it freely} to obtain a minimum $\chi^{2}$  in {\tt VPFIT}.
The \ion{C}{iv} absorbers found in QSO spectra are usually associated with 
a saturated Ly$\alpha$ \ion{H}{i} and have 
a multi-component unsaturated \ion{C}{iv}. Even the higher order Lyman
line profiles do not reveal velocity structures as detailed as those of \ion{C}{iv},
at least in part because the thermal line broadening is greater for hydrogen
than carbon due to the atomic mass difference. Consequently it is not
generally possible to use \ion{H}{i} to determine the \ion{C}{iv} redshift. Indeed
there is no physical reason to expect the velocity structure of \ion{C}{iv} to
closely follow that of \ion{H}{i}, given that the peculiar velocity and bulk motions
play a role in the kinematics of the absorbers.
Moreover, studies of close QSO
pairs show that \ion{H}{i} and \ion{C}{iv} display a different small-scale
structure \citep{rauch01a}. It has also been found that there is
a velocity difference at 5--18\,\kms\/ between the centroids of \ion{H}{i}
and \ion{C}{iv} \citep{ellison00}.

During the simultaneous fit of different transitions of the same ions, 
we have often adjusted a small amount of the continuum placement to
achieve a overall better fit result. Initially, the {\tt VPFIT} continuum adjustment
option was turned on. Then, comparing the unnormalised
spectrum with the adjusted continuum, we decided whether to take
the {\tt VPFIT} adjustment as it was or to apply a slightly different continuum. 
With this re-adjusted new 
continuum, the entire spectrum was fitted again. We iterated  
the continuum re-adjustment and fitting several times until
the satisfactory fits were obtained, finding
some more previously unidentified metal lines and fixing any wrong
line parameters previously obtained. The final iteration was done without
the {\tt VPFIT} continuum adjustment option. 

Note that all the quoted errors of
the fitted line parameters are directly from {\tt VPFIT}. These do not 
include continuum uncertainties usually adopted in the Milky Way
ISM studies, i.e. changing the continuum by $\pm 0.5$ times the
standard deviation of a local continuum \citep{sembach91}. 
We also note that {\tt VPFIT} does not work like the apparent optical
depth analysis \citep{savage91}, in that {\tt VPFIT} looks for the best-fit solution
by minimising $\chi^{2}$. This is not necessarily a lower limit column density
for saturated lines. 
If the {\tt VPFIT} column density errors are greater than 0.2--0.3\,dex,
the error estimates can be far from true
for single saturated lines and one should treat the errors with caution
(see the documents
provided with {\tt VPFIT} for more details, in particular Chapter~17). While
the decomposition of absorption features depends on the local S/N and
resolution, especially $b$ parameters, the total column density integrated
over a given velocity range is in general more securely determined.

\subsection{A robust \ion{H}{i} column density measurement}
\label{sec3:2}

In general,
\ion{C}{iv} associated with the Ly$\alpha$ forest is {\it not}
saturated, which provides a well-measured \nciv\/ as long as
blends by other lines are accounted for. 
On the other hand, 
\ion{H}{i} Ly$\alpha$ associated with \ion{C}{iv} is usually saturated.
Therefore, without other information it is not possible to
determine the \ion{H}{i} component structure or, unless there are measurable 
damping wings, to obtain a reliable 
\ion{H}{i} column density. For this reason we have chosen analyzed 
redshift ranges 
which ensure that at least Ly$\beta$ is always available. Where possible, 
other Lyman lines with lower oscillator strength are also included in the
Voigt profile fit to determine the reliable \ion{H}{i} parameters. 
The atmospheric cutoff at 3050\,\AA\/ 
sets a natural lowest redshift limit to cover the corresponding \lyb\/ 
to be $z > 1.98$. With an intervening
Lyman limit system, the lowest redshift bound increases, as listed in the
3rd column of Table~\ref{tab1}.
Note that if  \nhi\/ is high enough to show a damping wing,
Ly$\alpha$ line alone is enough to estimate a robust \nhi\/. 
We also note that well-resolved unsaturated \ion{H}{i} Ly$\alpha$
components do not require a high-order Lyman line to determine
the line parameter, especially for high-S/N spectra we have used in this study.
For relatively clean, unsaturated \ion{H}{i} Ly$\alpha$, the column density
estimates are reliable at $\log N_{\mathrm{\ion{H}{i}}} \le 14.3$.

\begin{figure}
\hspace{-0.3cm}
\includegraphics[width=8.9cm]{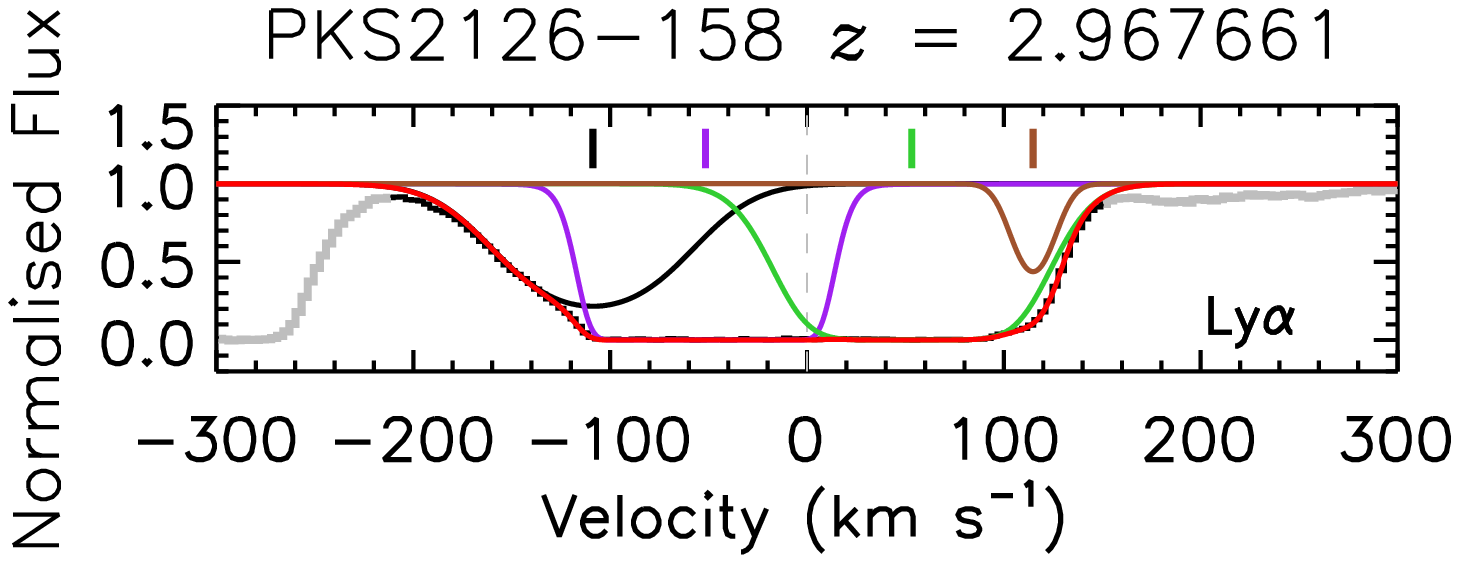}\\

\hspace{-0.3cm}
\includegraphics[width=8.9cm]{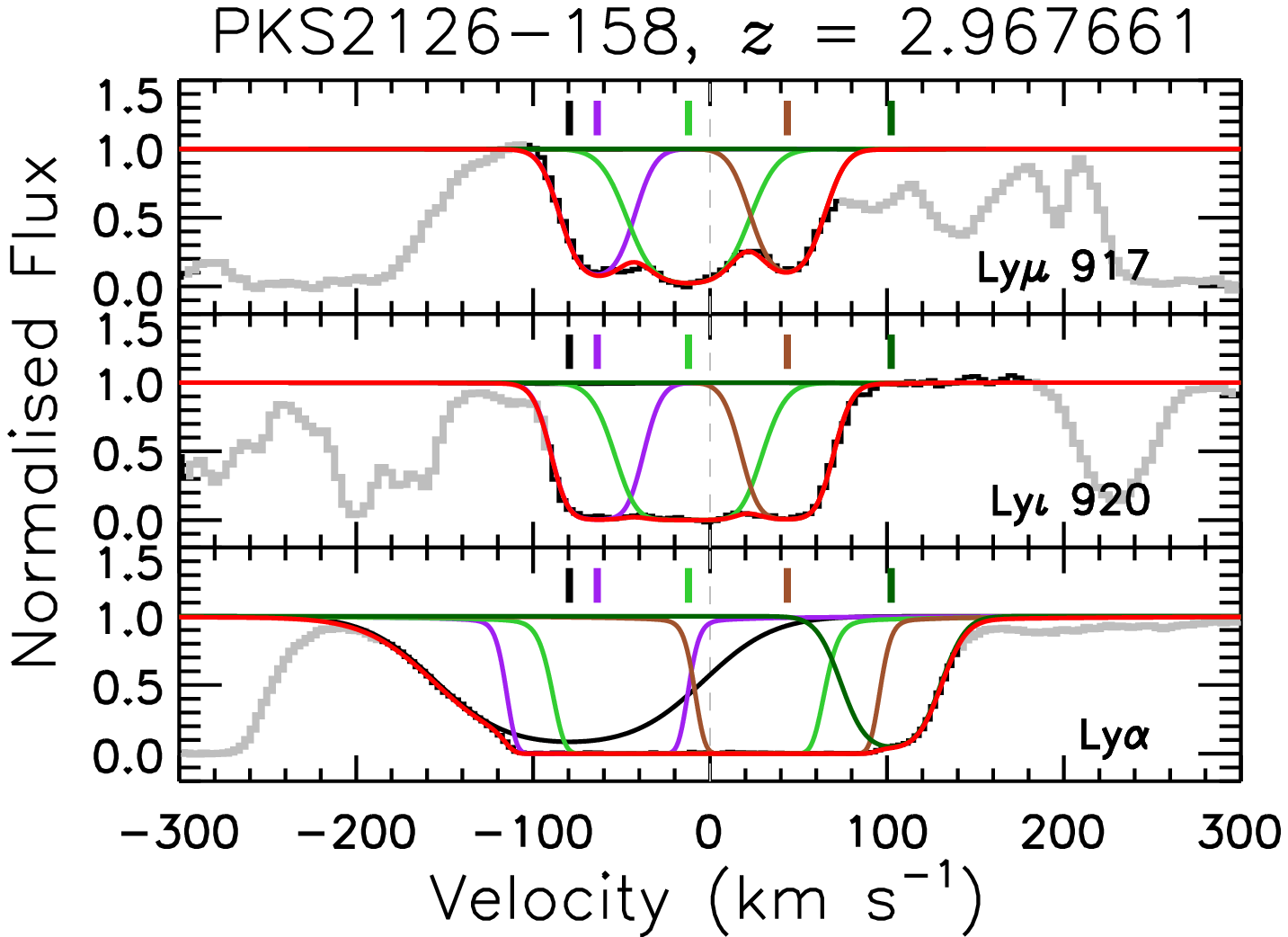}\\
\vspace{-0.3cm}
\caption{{\protect\footnotesize{The normalised \ion{H}{i} flux 
(black histogram) vs 
the relative velocity $v$ of a \ion{C}{iv} absorber at $z=2.967661$ 
toward PKS2126--158. The zero velocity is set to be
at the redshift of the \ion{C}{iv} flux minimum. 
The fitted \ion{H}{i} 
components are shown as coloured profiles with ticks
marking the component velocity centre in the same colour,
while the red
profile represents the entire profile constructed from all the fitted components
at $v \in [-150, +150]$\,\kms\/. The data are shown in gray where
the absorption is from blended metal lines or \ion{H}{i} at velocities
outside $\pm150$\,\kms\/. 
Upper panel: Only the saturated \ion{H}{i} 
Ly$\alpha$ is used in the Voigt profile fitting. The total \nhi\/ within 
$\pm 150$\,km s$^{-1}$ is $16.11\pm0.20$. Lower panel: All the 
available high-order Lyman
lines were fitted simultaneously. The \ion{H}{i} absorption centred at
$v \sim 0$\,km s$^{-1}$ becomes 
unsaturated at Ly$\mu$, revealing its 3-component nature. The total
\nhi\/ within $\pm 150$\,km s$^{-1}$ is $17.28\pm0.01$, which is 1.17\,dex
larger than the Ly$\alpha$-only fit. 
}}}
\label{fig2}
\end{figure}

However, the determination of the \ion{H}{i} column density is not always 
straightforward. Even if several of the Lyman series lines are accessible,
they are not always useful because of blending with lower redshift Ly$\alpha$. 
At $z \gg 3$, line blending is so severe that most high-order lines 
become blended with the lower-$z$ Ly$\alpha$ forest. Also, because of 
the higher absorption line density, the continuum level is uncertain. 
At $z < 2.5$, line blending becomes less problematic, but at lower redshifts 
the number of available high-order Lyman lines in optical spectra also decreases. 
A further complication may occur in high redshift QSOs, when there are systems 
with $\log N_{\ion{H}{i}} \ge 17.2$ in the observed spectrum. 
Since these have significant Lyman continuum 
absorption at the rest-frame wavelength $< 912$\,\AA\/ and
decrease the flux significantly, the higher order Lyman lines 
in the system of interest may not be measurable in that region.

Another difficulty which strongly affects the analysis of saturated 
lines is uncertainties in the zero level in the data \citep{kim07}. In particular, 
at short wavelengths in UVES data, the true zero level offset could  
be a few percent of the local continuum. If the true zero is above the adopted 
one, then the fitting program attempts to put in many unsaturated components.
If the true zero is below the adopted one, then the criterion for a satisfactory fit 
may never be satisfied. The zero level for each spectral region can be treated 
as a free parameter in {\tt VPFIT}, so this option was used if appropriate.

For $\log N_{\mathrm{\ion{H}{i}}} \sim 16.5$ and $b=30$\,\kms\/ 
(a typical line width of the Ly$\alpha$ forest), the Lyman lines start to
become unsaturated at higher order than Ly$\eta$ (926.23\,\AA\/). 
The residual central intensity of Ly$\theta$ (923.15\,\AA\/) is then $\sim$\,4\% 
of the continuum, and for Ly$\iota\ (920.96{\rm \AA})$,  $\sim$\,10\%.
We have found empirically that a
fairly robust \nhi\/ can be obtained within 0.1\,dex
if both Ly$\beta$ and Ly$\gamma$ are included in the fit and if
$\log N_{\mathrm{\ion{H}{i}}} \le 17.0$. Even if only \lya\/ and Ly$\beta$ 
are available, 
the \nhi\/ obtained assuming a single component is usually within 0.1--0.2\,dex.
However,  the difference could become larger if 
a saturated line reveals several components at higher orders than Ly$\beta$.
About 31\% of the 
saturated Ly$\alpha$ in our  \ion{H}{i}+\ion{C}{iv} sample breaks into 
several components in Ly$\beta$.
About 39\% of the saturated Ly$\alpha$  
breaks into several weaker components at higher orders than Ly$\beta$. 

Figure~\ref{fig2} illustrates the importance of incorporating high-order  
lines in the profile fitting procedure and the difficulty in obtaining reliable fit
parameters. As the S/N in the Ly$\alpha$
region is about 160 per pixel, it is easy to recognise even by eye that
a single-component fit for the saturated core
does not match the observed left wing profile.  For such a fit the normalised $\chi^{2} \sim 6.1$.
There are two saturated \ion{H}{i} absorbers, 
the $z=2.727849$ absorber toward PKS2126--158
and the $z=2.328908$ absorber toward HE1347--2457, for which a low-S/N
in the available higher-order regions makes it difficult to estimate the saturated
line parameters reliably. For the former case,
depending on a one- or two-component structure at 
$v = 11$\,\kms\/, the resultant column density
can be differ by 0.38\,dex. For the $z=2.328908$ absorber toward HE1347--2457, 
the saturated component at $v = 14$\,\kms\/ could have a \nhi\/ difference by 0.72\,dex,
depending on the $b$ value.  With no significant improvement in the normalised $\chi^2$ value,
we took the one-component fit and the smaller-$b$ fit, respectively, 
but increased their error to include the error range by the alternative fit. 
We note that lines in the absorption wings tend to have a larger error in $z$, $b$
and $N$.

In our \ion{H}{i}+\ion{C}{iv} sample, 
we included only those with a well-measured \nhi\/ obtained by including 
high-order lines or from unsaturated Ly$\alpha$. 
When Ly$\alpha$ is saturated and no constraint could be obtained from Ly$\beta$ or
other higher-order lines because of blending, that
\ion{H}{i}+\ion{C}{iv} pair was excluded. However, this resulted in the exclusion of only
$\sim$2\% of all the \ion{H}{i}+\ion{C}{iv} pairs.
Saturated \ion{C}{iv} components associated with the forest \ion{H}{i} are much less 
common. There are only two saturated \ion{C}{iv} components in our sample, 
and both of these are in systems with $\log N_{\ion{H}{i}} \ge 17$.
These are included only as lower limits for illustration purposes, but excluded 
from the actual analysis. 

\subsection{Detection of weak \ion{C}{iv} lines}
\label{sec3:3}

The detection of weak lines depends critically on the local S/N and the
absorption line width $b$. A weak, narrow line is more easily recognised than 
a strong, broad line. To construct a robust \ion{H}{i}+\ion{C}{iv} sample, 
we selected a \ion{C}{iv} component only when the stronger doublet \ion{C}{iv} 
$\lambda 1548$ is detected at the $\ge 3\sigma$ level. Following the
procedure described in \citet{sembach91}, a standard deviation, 
1 r.m.s. (1$\sigma$), was measured in the nearby, unabsorbed continuum
region. When a weak \ion{C}{iv} line was detected, the
equivalent width (EW) of \ion{C}{iv} $\lambda$1548 
was estimated over the region which the line profile
falls, including the $\pm0.5\sigma$ continuum fitting errors.
This EW with the continuum errors is then compared to the 1$\sigma$ 
continuum EW integrated over the same wavelength range to
obtain the detection significance.

For $S/N \sim 100$, 
the \ion{C}{iv} detection limit at 3$\sigma$ is $\log N_{\mathrm{\ion{C}{iv}}} \sim 12.0$.
In a similar way, a rough detection limit of \ion{H}{i} at 3$\sigma$ is 
$\log N_{\ion{H}{i}} \sim 12.5$ for $S/N \sim 60$.
All of our 21 UVES spectra were included in the 
high-$z$ \ion{C}{iv} study by \citet{dodorico10}. In our analysed redshift 
range, about 5\% of the \ion{C}{iv} absorbers
are not reported in \citet{dodorico10}. Most
of these are weak \ion{C}{iv} at $\sim$\,3$\sigma$. The discrepancies
occur mainly when one line of the doublet is blended with other lines or weak telluric
features, though in some cases misidentification of \ion{C}{iv}
could be responsible. We included the border-line detections 
in our \ion{H}{i}$+$\ion{C}{iv} sample, as there are so few
of them that the scientific results and conclusions
are unaffected.

\begin{figure*}

\hspace{-0.2cm}
\includegraphics[width=5.9cm]{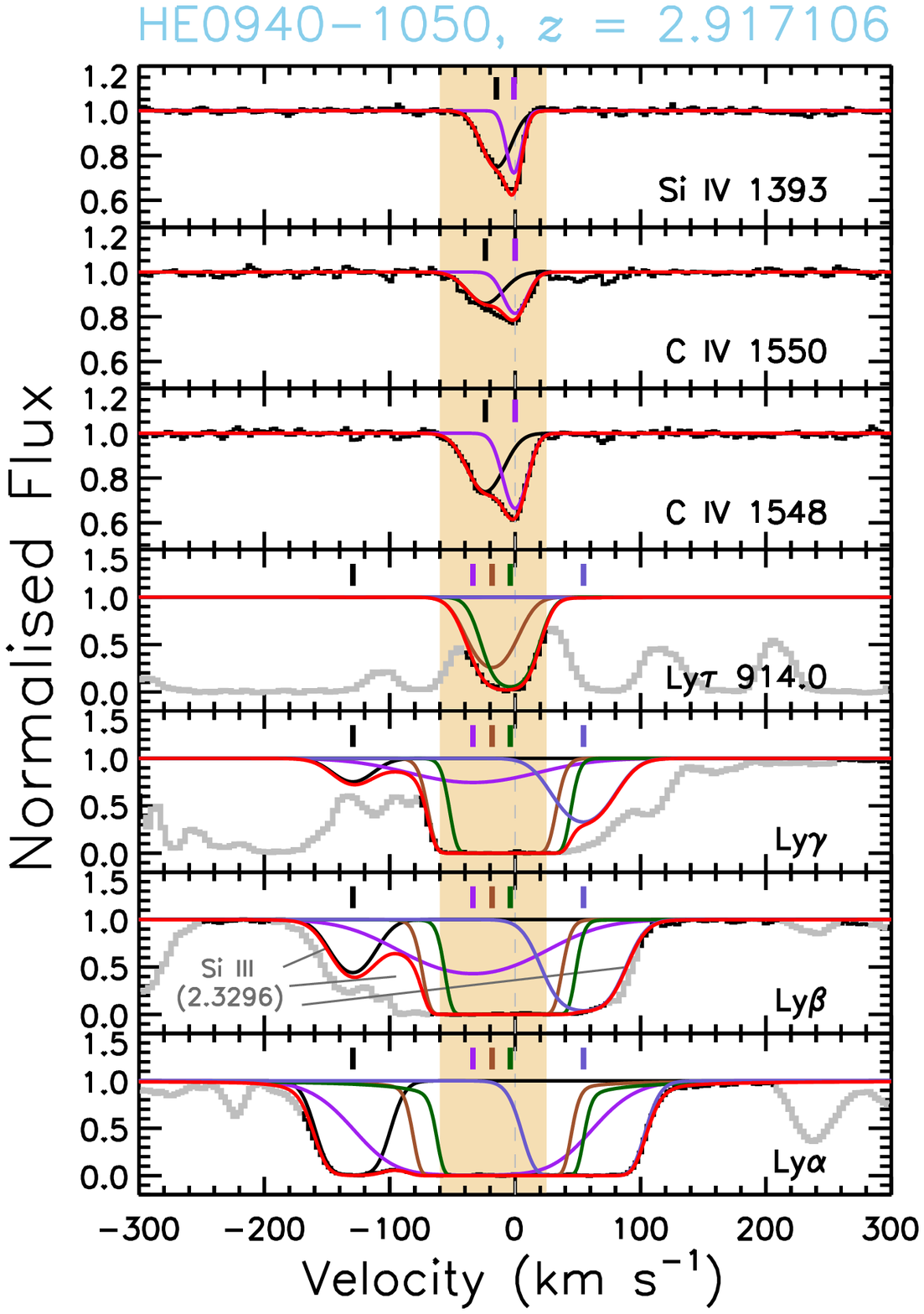}
\hspace{-0.3cm}
\includegraphics[width=5.9cm]{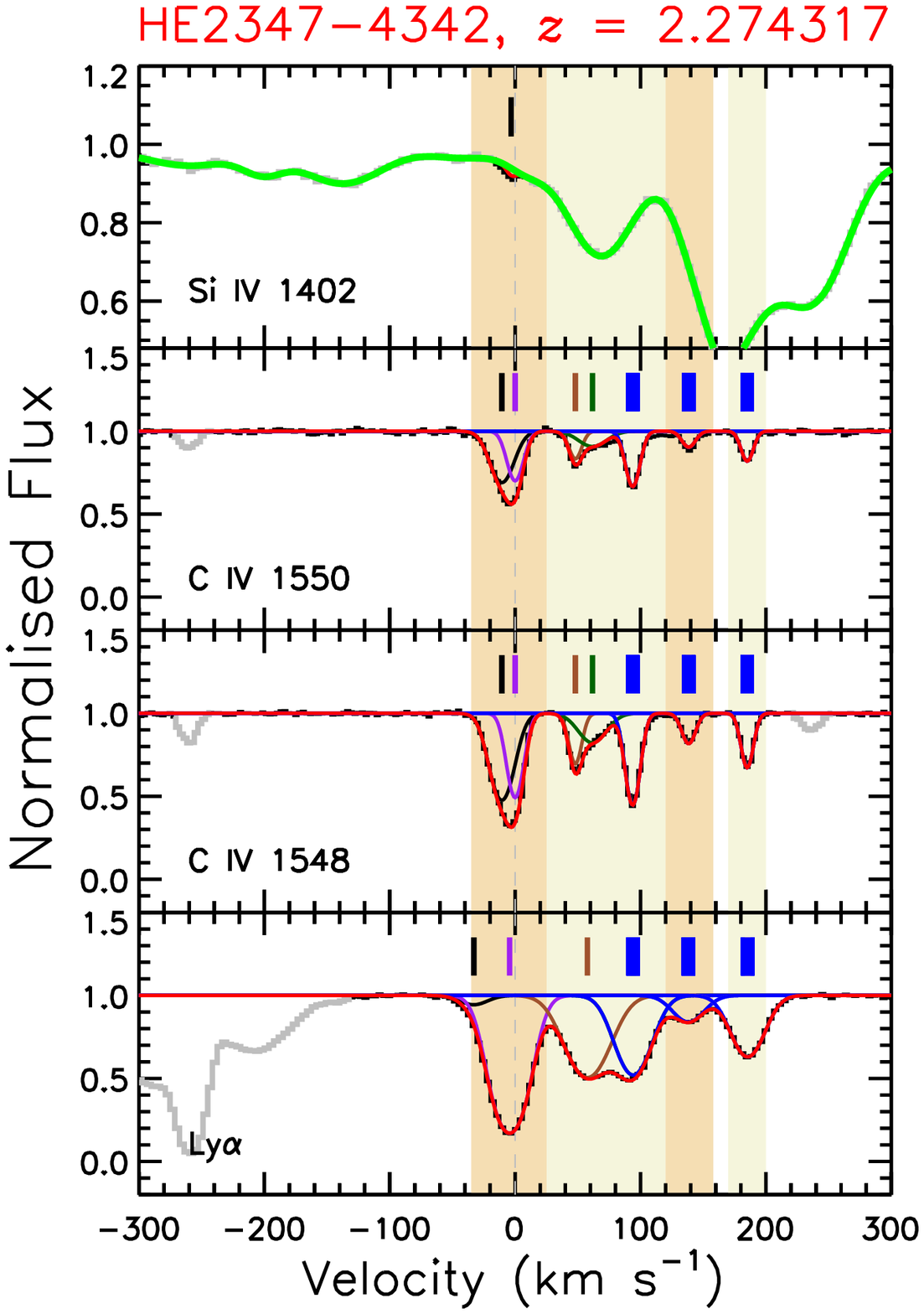}
\hspace{-0.3cm}
\includegraphics[width=5.9cm]{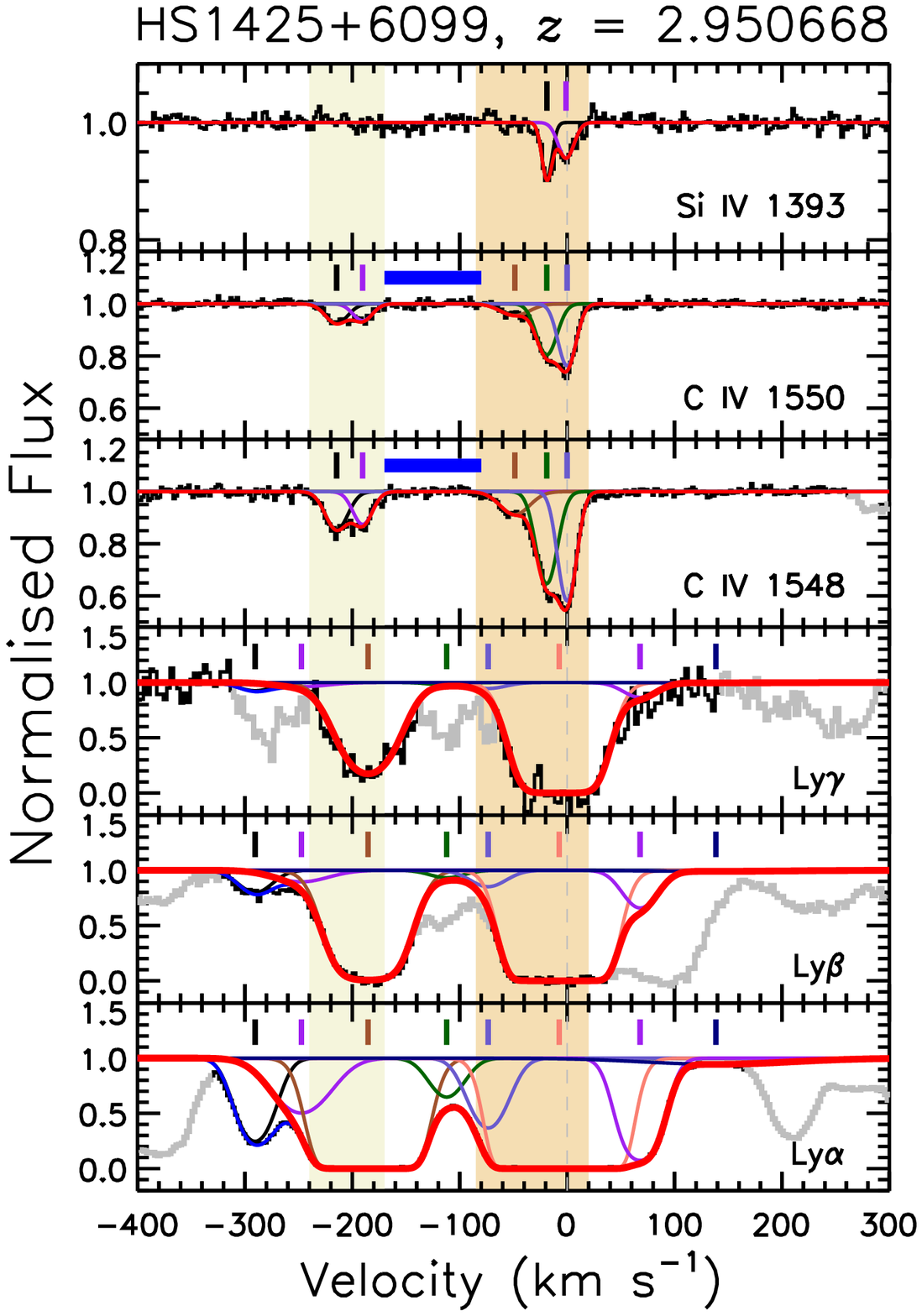}\\

\vspace{0.4cm}
\hspace{-0.2cm}
\includegraphics[width=5.9cm]{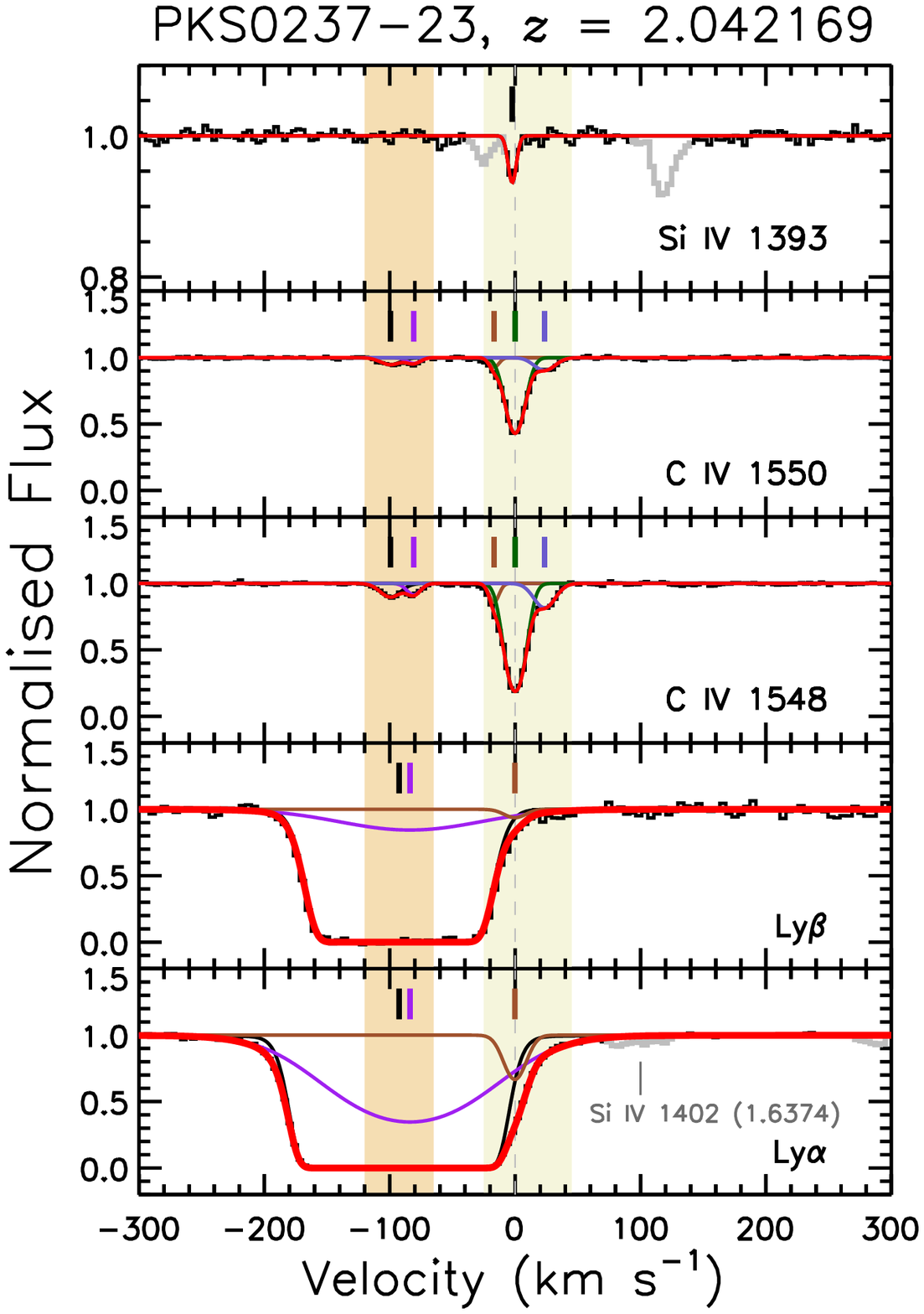}
\hspace{-0.3cm}
\includegraphics[width=5.9cm]{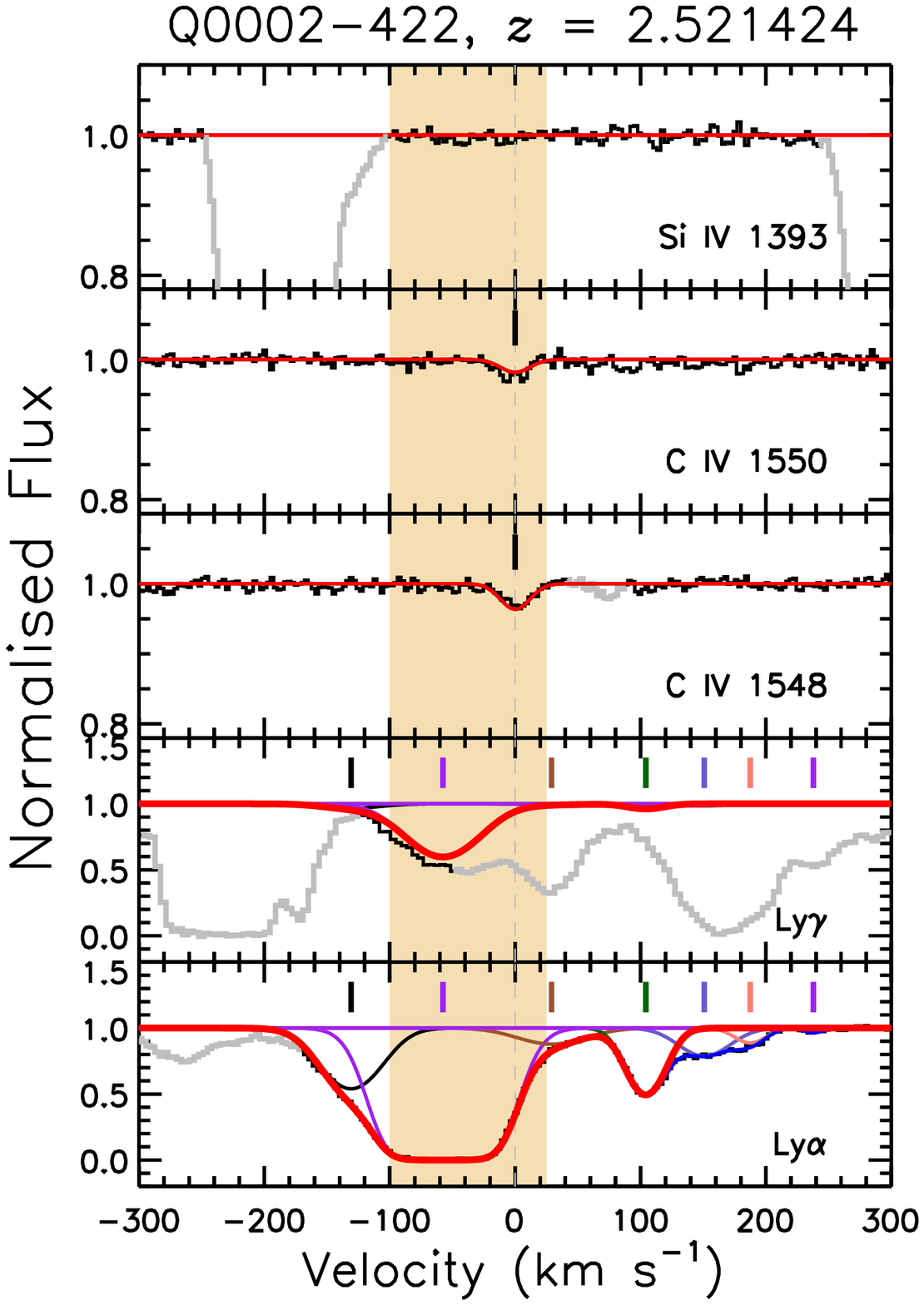}
\hspace{-0.3cm}
\includegraphics[width=5.9cm]{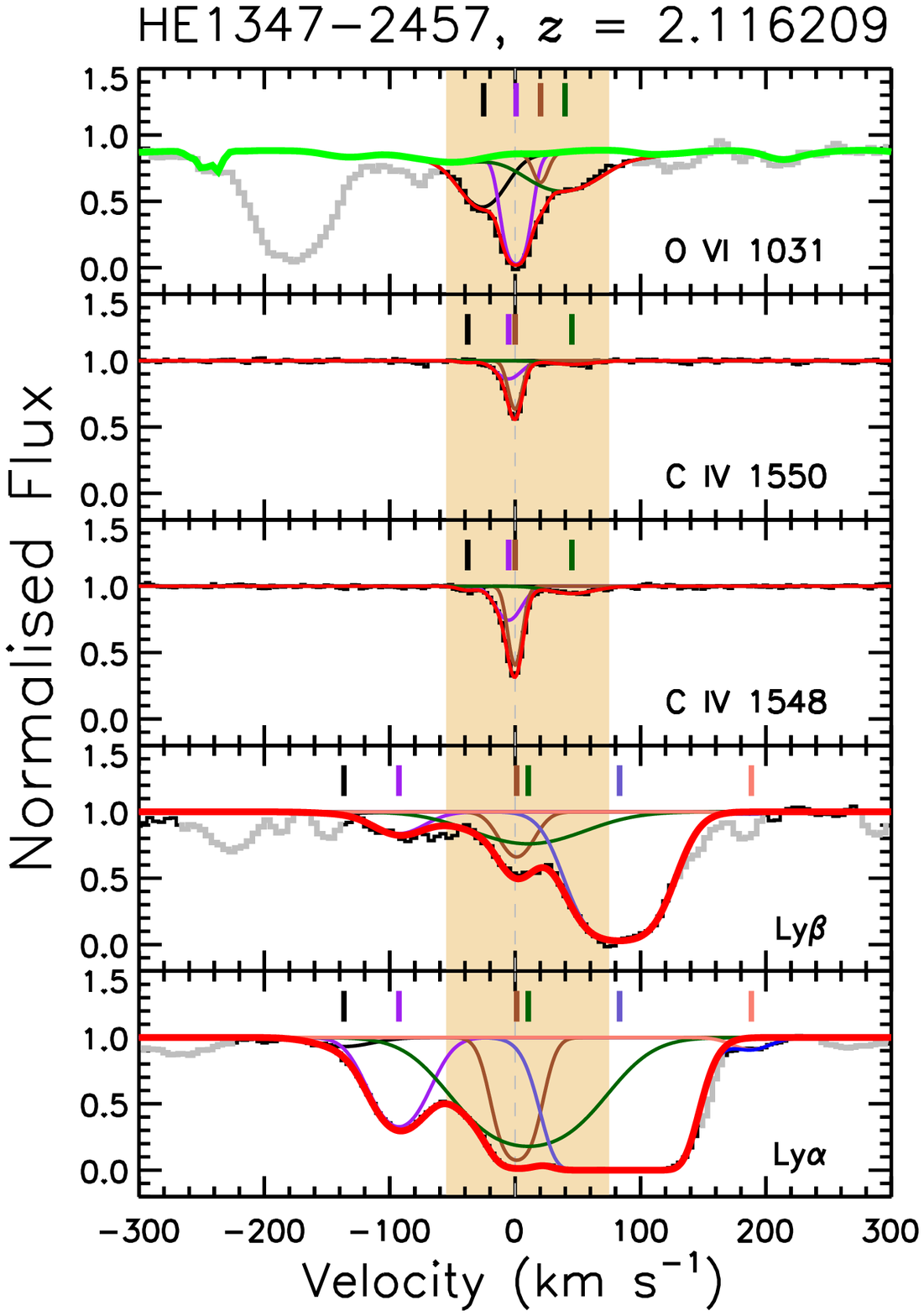}\\

\caption{
The velocity plots of 6 \ion{C}{iv} systems, which
is defined by a given velocity range centred at the \ion{C}{iv} flux minimum,
with a default range of $v \in [-150, +150]$\,\kms\/.
The plot format is the same as in Fig.~\ref{fig2}. For clarity, 
only the components within the integrated velocity range 
are shown, except when the \ion{H}{i} profile continues over the velocity
range. The thick red profile is 
generated by only the components within the velocity range,
while the thin blue \ion{H}{i} profile is by all the shown \ion{H}{i} components.
The overlaid green profile
in some panels illustrates the actual continuum due to the metal
lines, higher-order Lyman lines or the Lyman limit depression. The identified metal
line is indicated by its identification and redshift in parenthesis.
Shaded regions with a light orange colour
or a light lime colour delineate a \ion{C}{iv} clump in our working definition.
Upper-left panel: the \ion{C}{iv}
system contains only one clump defined at 
$v \in [-60, +25]$\,\kms\/ (light orange shade). 
Upper-middle panel: an example of higher-$N_{\mathrm{\ion{C}{iv}}}$
systems. Well-aligned \ion{H}{i} and \ion{C}{iv} components are 
indicated with thick blue ticks.
Upper-right panel: a \ion{C}{iv} absorption is at $\sim -200$\,\kms\/, outside
the default velocity range $\pm 150$\,\kms\/. Its right-wing is 90\,\kms\/ away
from the left wing of the \ion{C}{iv} at $v = 0$\,\kms\/,
indicated by the thick blue horizontal line in both \ion{C}{iv} panels. With the extended
velocity range $v \in [-250, +150]$\,\kms\/, the system contains two clumps. 
Lower-left panel: two distinct \ion{C}{iv} absorptions span at
$v \in [-120, -65]$ and $[+25, +45]$\,\kms\/, with two and one \ion{H}{i} components
in each \ion{C}{iv} clump velocity range.
Due to the saturated Ly$\alpha$ profile, this system is defined to have 
only one Ly$\alpha\beta$ clump at $v \in [-200, +45]$\,\kms\/.
Lower-middle panel: an example of ambiguous \ion{C}{iv} clumps, with no
\ion{H}{i} component within the \ion{C}{iv} clump velocity range at $v \in [-25, +25]$\,\kms\/.
Both \ion{C}{iv} clump and \ion{C}{iv} Ly$\alpha\beta$ clump are 
defined at $v \in [-100, +25]$\,\kms\/.
Lower-right panel: another ambiguous \ion{C}{iv} clump. 
The strongest \ion{H}{i} component at $v = +83$\,\kms\/ is not included
in the clump velocity range at $v \in [-55, +75]$\,\kms\/.
The coloured version is available online.
}
\label{fig3}
\end{figure*}

\section{Working definition of a system and a clump describing \ion{C}{iv} absorbers}
\label{sec4}

Various terms such as clouds, groups,
clumps and systems have been used
to describe QSO absorption features. These terms are often
used interchangeably.
Physically, it becomes meaningful to 
associate \ion{H}{i} with \ion{C}{iv} only if an effect on \ion{H}{i} triggers a
consequence on \ion{C}{iv} and vice versa. However, there is no independent way to
recognise this physical connection from the spectroscopic data alone.
Even though the \ion{H}{i} gas and the \ion{C}{iv} gas are very far in the
real space, their lines could be found to 
be very close in the redshift space in QSO spectra due to the
bulk motions and the peculiar velocity \citep{rauch97}. 

To describe our \ion{H}{i}+\ion{C}{iv} sample 
more clearly for this study, 
we re-define two terms commonly used in the literature, 
{\it systems} and {\it clumps}. Our definitions are solely based on the
{\it profile shapes}, 
assuming that a similar velocity structure between \ion{H}{i} and \ion{C}{iv}
is produced by similar underlying physical processes.
Other terms such as an absorber, a component or a line are
used loosely having a similar meaning as in the literature. 
We illustrate these working definitions in Figs.~\ref{fig3} and \ref{fig4},
which shows the normalised flux vs relative velocity 
of \ion{C}{iv} systems. The top panel
displays other associated metal species such \ion{Si}{iv}, \ion{N}{v} or \ion{O}{vi}.
When no clear metal species other than \ion{C}{iv} are detected in the 
observed wavelength range, \ion{Si}{iv} $\lambda$1393 is shown. The name of the
QSO and the redshift of the \ion{C}{iv} system listed on top
are colour-coded according to
$N_{\mathrm{\ion{C}{iv}}}$ compared to $N_{\mathrm{\ion{H}{i}}}$
as explained in Section~\ref{sec5:1}.
Note that even though we showed only a portion of fitted spectra 
in Figs.~\ref{fig3} and \ref{fig4} for clarity, we have fitted almost entire spectra
to obtain a reliable $N_{\mathrm{\ion{H}{i}}}$ and $N_{\mathrm{\ion{C}{iv}}}$,
as explained in Section~\ref{sec3}.

\subsection{Components}
\label{sec4:1}

A {\it component} is the most basic and simplest unit to 
describe absorption profiles. However, the component structure determined
by fitting Voigt profiles 
is not unique and is strongly dependent on the spectral 
resolution and S/N \citep{kim13, boksenberg14}.
So, for all but the simplest line profiles, the component
structure is not well-determined.
In this work, the components are strictly referred to the {\tt VPFIT} profile 
fitting results. 

Since {\tt VPFIT} profile fitting depends on the 
continuum and S/N, a slightly different velocity centroid can be 
obtained for \ion{H}{i} and \ion{C}{iv}, even if both are produced in the same absorbing gas.
Here, depending on the velocity difference, 
the presence of close components and 
the detection significance, \ion{H}{i}+\ion{C}{iv} component pairs 
were classified into well-aligned and reasonably-aligned components.

{\it Well-aligned} components refer to
the \ion{H}{i} and \ion{C}{iv} pairs when their velocity centroid
differs by $\le 5$\,\kms\/ for a unsaturated, single-component \ion{C}{iv} and 
both are relatively clean and well-measured. 
The \ion{H}{i}+\ion{C}{iv} component pairs marked with
a thick blue tick in the upper-middle panel of Fig.~\ref{fig3} are a good example
of well-aligned components.

{\it Reasonably-aligned} components refer to the ones which
also have a velocity difference at $\le 5$\,\kms\/, but if
nearby \ion{H}{i} components make 
the line parameter of the aligned \ion{H}{i} less reliable or if  a \ion{C}{iv} is 
located in a low-S/N region.

This classification is only applied for a \ion{H}{i}+\ion{C}{iv}
component pair which is isolated or separable from other nearby
\ion{H}{i} and \ion{C}{iv}. When several \ion{C}{iv} components are associated 
with a \ion{H}{i} absorption with a smaller number of \ion{H}{i} components, 
e.g. the \ion{C}{iv} components at $v \sim 0$ and $\sim 60$\,\kms\/ in the
upper-middle panel of Fig.~\ref{fig3}, there is no unambiguous
way to assign each \ion{C}{iv} component to \ion{H}{i}. In this case,
regardless of the small velocity difference between \ion{H}{i}+\ion{C}{iv} pairs,
they are not classified to be aligned.

Considering that 
the wavelength calibration of UVES and HIRES spectra is usually 
better than $\sim$\,1~\kms\/, this arbitrary choice of 5~\kms\/ is likely to be too generous
for strong and narrow components, and is reasonable for weak and broad
components. Our sample consists of 762 \ion{H}{i} components and 
628 \ion{C}{iv}
components. Only 6\% of the \ion{C}{iv} components (38/628) are
well-aligned. Reasonably-aligned \ion{C}{iv} components
are similar at $\sim$\,6\% (39/628). 
 
\subsection{Systems}

\begin{figure*}
\includegraphics[width=18.5cm]{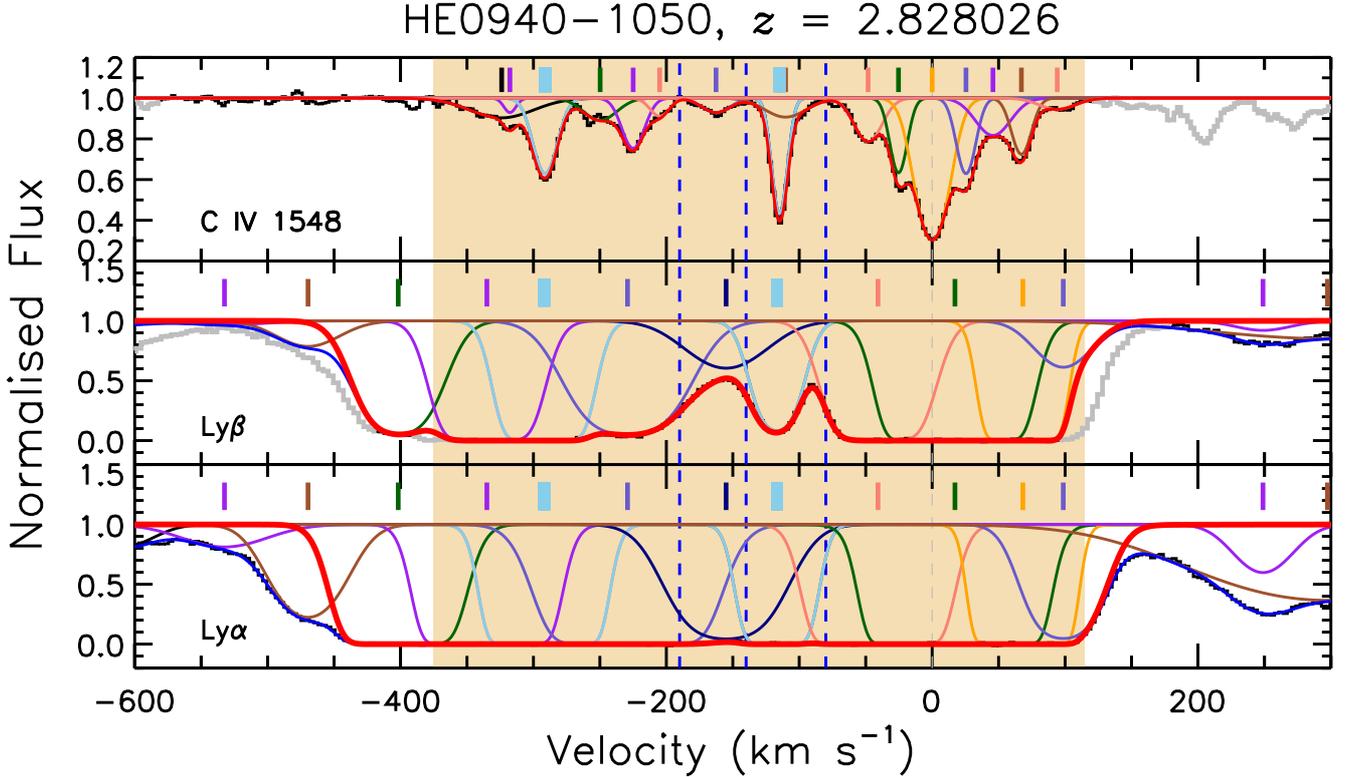}
\vspace{-0.6cm}
\caption{The velocity plot of the $z = 2.828026$ \ion{C}{iv} system toward 
HE0940--1050, defined at $v \in [-450, +250] $\,\kms\/. All the 
symbols are the
same as in Figs.~\ref{fig2} and \ref{fig3}. 
The three blue dashed, vertical lines delineate the
relative velocity at
$-190, -140$ and $+80$\,\kms\/,
where the normalised \ion{C}{iv} flux $F_{\mathrm{\ion{C}{iv}}}$ 
becomes $\sim 1$. Thick sky-blue ticks indicate the reasonably-aligned 
\ion{H}{i}+\ion{C}{iv}
component pairs. 
See the text for details.}
\label{fig4}
\end{figure*}

The \ion{C}{iv} {\it system} is defined by all the
\ion{C}{iv} components at $v \in [-150, +150]$\,\kms\/ centred at the 
\ion{C}{iv} flux minimum of a single or a group of several, closely-located
\ion{C}{iv} components. The redshift of a \ion{C}{iv} system defined in this way is not
necessarily the redshift of the strongest \ion{C}{iv} component, since a strongest 
component has sometimes a broader line width.
In some cases, a continuous
\ion{C}{iv} absorption profile extends beyond $\pm 150$\,\kms\/ or 
a \ion{C}{iv} absorption is outside the velocity range, but with a separation less
than 100\,\kms\/. In such cases,  
the integrated velocity range is extended by a 100\,\kms\/ step to include these \ion{C}{iv}.
If necessary, a new zero velocity
is defined to be at a \ion{C}{iv} flux minimum including the newly-added \ion{C}{iv}. 
The exactly same velocity range is used to assign {\it associated}
\ion{H}{i} or other metal components.

The upper-left 
panel in Fig.~\ref{fig3} shows a 
common \ion{C}{iv} system. For about 80\% of our sample, the
\ion{C}{iv} profile is rather simple and associated with saturated \ion{H}{i}. 
The upper-middle panel presents 
a rare \ion{C}{iv} system, in which the
\ion{C}{iv} absorption is stronger than the one of 
typical \ion{H}{i} absorbers having a similar $N_{\mathrm{\ion{H}{i}}}$.
These are the high-metallicity absorbers described by
\citet*{schaye07}. As the \ion{C}{iv} absorption continues
at $v > +150$\,\kms\/, the integrated velocity range is extended to 
$v \in [-150, +250]$\,\kms\/.
In the upper-right panel of Fig.~\ref{fig3},
an additional \ion{C}{iv} absorption occurs beyond
the $\pm150$\,\kms\/ range.
The separation between the wavelengths of 
the distinct \ion{C}{iv} profile wings to recover to the normalised  \ion{C}{iv} flux of 
$F_{\mathrm{\ion{C}{iv}}} \sim 1$ is
less than 100\,\kms\/. Therefore, these \ion{C}{iv}
absorptions are combined as a single \ion{C}{iv} system defined at
$v \in [-250, +150]$\,\kms\/.

This working definition of the \civ\/ system does 
not require that a \civ\/ centroid
should coincide with a \ion{H}{i} centroid. Moreover,
the \ion{C}{iv} system defined this way
often includes nearby \ion{H}{i} components clearly associated with no \ion{C}{iv}.
This definition is in fact more closely related to the
{\it volume-averaged} quantities commonly used in numerical simulations. 
Note that a conventional definition of a \ion{C}{iv} system 
in the literature would refer to a
group of \ion{H}{i} and \ion{C}{iv} absorption features seemingly associated 
in the velocity space, but
excluding any \ion{C}{iv}-free \ion{H}{i} components. 
We label \ion{H}{i} and
\ion{C}{iv} column densities of a system as the {\it integrated} \nhi\/ and \nciv\/,
or $N_{\mathrm{\ion{H}{i}, sys}}$ and $N_{\mathrm{\ion{C}{iv}, sys}}$.
When we want to emphasize the given integrated 
velocity range, we use   
$N_{\mathrm{\ion{H}{i}}} (\pm 150)$ and $N_{\mathrm{\ion{C}{iv}}} (\pm 150)$. 

On the face value, the fixed velocity range of $\pm 150$\,\kms\/ to define a system
could be considered rather arbitrary. This range was chosen on the basis 
of several observational findings.
Studies of close QSO pairs have
found a strong \civ\/ clustering signal within $\sim$\,200\,\kms\/, which might
indicate the outflow velocity could be less than $\sim$\,200\,\kms\/ 
\citep{rauch05}.
A significant clustering signal of \ion{H}{i} is also found at a transverse velocity separation
at $\sim$\,500\,km s$^{-1}$ \citep{dodorico06}.
The average velocity dispersions of high-$z$ galaxies
are $<\sigma> \,\, \sim 120$\,\kms\/ \citep{erb06}. As shown in Section~\ref{sec5:1},
the integrated \nhi\/--\nciv\/ relation depends on the integrated velocity range when
it is small.
However, as the integrated velocity range becomes $\ge \pm 100$\,\kms\/,
the integrated \nhi\/--\nciv\/ relation converges.

\subsection{Clumps}
\label{sec4:3}

Closely-located \ion{C}{iv} components often show 
visibly distinct, separable absorption features. When the absorption wing of
visibly separable \ion{C}{iv} profiles recovers to have a normalised flux  
$F_{\mathrm{\ion{C}{iv}}} = 1$ and a closest
\ion{C}{iv} absorption wing starts
at $\ge 5$\,\kms\/ away,
this distinct absorption feature is termed as a {\it clump}.
As with the \ion{C}{iv} system, all the \ion{H}{i} components 
within the \ion{C}{iv} clump velocity range are assigned to that clump.
Even when a \ion{H}{i}
component exists just outside the clump velocity range,
it is not extended to include this nearby \ion{H}{i} component.
Fortunately, such
\ion{H}{i} components are usually weak so that the clump \ion{H}{i}
column density, $N_{\mathrm{\ion{H}{i}, \, cl}}$,
does not increase significantly even if they are included.
If no \ion{H}{i} exists within the clump velocity range, the velocity range
is extended
to include a nearby \ion{H}{i} component based on both
\ion{H}{i} and \ion{C}{iv} absorption profile shapes.
A clump can consist of a single component or multiple components.

In the upper-left panel of Fig.~\ref{fig3}, the single \ion{C}{iv} 
absorption profile wings recover to the normalised 
\ion{C}{iv} flux of 1 at $v \in [-60, +25]$\,\kms\/.
As 3 \ion{H}{i} components exist in the \ion{C}{iv} clump velocity range,
this \ion{C}{iv} system consists of one \ion{C}{iv} clump. The clump column density of
\ion{H}{i} and \ion{C}{iv} is integrated over the clump velocity range. 
In the upper-right panel, the two distinct \ion{C}{iv} absorptions span at
$v \in [-240, -170]$ and $[-80, +20]$\,\kms\/. As only \ion{H}{i} components within
each clump velocity range are included to define $N_{\mathrm{\ion{H}{i}, \, cl}}$,
the \ion{H}{i} component at $v \sim -109$\,\kms\/
is not included in either two clump \ion{H}{i} column densities.

The lower panels of Fig.~\ref{fig3} present three
\ion{C}{iv} systems for which defining a clump is not
straightforward. In the lower-left panel, 
there exists a \ion{H}{i} component in each cleanly-defined
\ion{C}{iv} clump velocity range, even though both \ion{H}{i} and \ion{C}{iv} 
are fitted independently. Therefore, the system is classified to consist 
of 2 clumps. On the other hand, in the lower-middle
panel, no \ion{H}{i} component exists within the \ion{C}{iv} clump velocity 
range at $v \in [-25, +25]$\,\kms\/.
The closest \ion{H}{i} component is at 4\,\kms\/
away at $v = +29$\,\kms\/. However, the \ion{C}{iv} absorption
is more likely to be associated with the \ion{H}{i} absorption 
at $v = -58$\,\kms\/. The \ion{H}{i} profile
clearly reveals a separation of two distinct \ion{H}{i} 
absorptions at $v \sim 20$\,\kms\/ and the \ion{C}{iv} absorption occurs 
at $v < 20$\,\kms\/. Therefore, the clump velocity range is defined as
$v \in [-100, +25]$\,\kms\/ to include only \ion{H}{i} at $v = -58$\,\kms\/.

If the same reasoning is applied, in the lower-right panel, 
the broad \ion{C}{iv} 
absorption at $v = +45$\,\kms\/ should be assigned to the \ion{H}{i}
at $v = +83$\,\kms\/, the strongest \ion{H}{i} component. However, since
2 \ion{H}{i} components exist in the clump velocity range at $v \in [-55, +75]$\,\kms\/,
we strictly applied for the clump criterion without including the 
\ion{H}{i} component at $v =+83$\,\kms\/. 
Fortunately, only about 5\,\% of the clumps are ambiguous
for reasons similar to the last two examples.

Figure~\ref{fig4} illustrates another example of a clump.
It shows a velocity plot of one of the most complicated 
\ion{C}{iv} systems at $z = 2.828026$ toward HE0940--1050, defined at 
$v \in [-450, +250]$\,\kms\/. Considering the \ion{C}{iv} profile only at
$v \in [-375, +115]$\,\kms\/, the
\ion{C}{iv} flux is recovered to the normalised flux of 
$F_{\mathrm{\ion{C}{iv}}}= 1$ at
$v = -190$\,\kms\/, but the left wing of the \ion{C}{iv} profile due to the 
$v = -163$\,\kms\/ component does not allow the $F_{\mathrm{\ion{C}{iv}}} \sim 1$ 
region more than 5\,\kms\/.
The normalised flux does not reach to $F_{\mathrm{\ion{C}{iv}}} = 1$ at 
$v = -140$ and $-80$\,\kms\/.
Therefore, this system consists of only one clump. 
Most of high-$N_{\mathrm{\ion{H}{i}}}$
absorbers have a continuous \ion{C}{iv} absorption profile 
spanning over several hundred \kms\/ similar to the shown example, 
consisting of only 1 clump.

Unfortunately, the \ion{H}{i} component structure is not as 
well-determined as the \ion{C}{iv} due to the larger \ion{H}{i} thermal width and the 
non-uniqueness of the Voigt profile fitting. Therefore, assigning \ion{H}{i} components 
to a \ion{C}{iv} clump in a smaller velocity range than the system is not
necessarily robust. In addition, finding a velocity at which the \ion{C}{iv} flux 
recovers to a normalized flux $F_{\mathrm{\ion{C}{iv}}} =1$ 
is not always reliable, depending on the local S/N and the goodness of
the continuum placement. 

Due to these uncertainties, 
we used one supplementary definition to study \ion{C}{iv} clumps, 
a Ly$\alpha\beta$ clump. 
If a normalised \ion{C}{iv}
flux becomes at $F_{\mathrm{\ion{C}{iv}}} \ge 0.98$ 
at a relative velocity $v$ and both \ion{H}{i} Ly$\alpha$
and Ly$\beta$ absorption profiles are also clearly breakable at the similar velocity,
this distinct absorption feature is termed as a Ly$\alpha\beta$ clump.
The Ly$\alpha\beta$ clump is defined only in terms of the profile shape and is
closest to the 
{\it conventional} \ion{C}{iv} system commonly used in the literature. 
 
In the upper-left panel of Fig.~\ref{fig3}, 
the Ly$\alpha$ and Ly$\beta$ profiles show a smooth \ion{H}{i} absorption
at $v \in [-80, +200]$\,\kms\/, without displaying any distinct component structure.  
The \ion{H}{i} component at $v =  +55$\,\kms\/ becomes
distinguishable only at Ly$\gamma$. Therefore, even though the \ion{H}{i} at
$v = +55$\,\kms\/ does not associated with the \ion{C}{iv} directly, all the \ion{H}{i}
at $v \in [-80, +200]$\,\kms\/ are included for the \ion{H}{i} column density of
this Ly$\alpha\beta$ clump, $N_{\mathrm{\ion{H}{i}, \, \alpha\beta}}$. Similarly,
both Ly$\alpha$ and Ly$\beta$ profiles in the lower-left panel display a smooth
absorption at $v \in [-200, +40]$\,\kms\/, while two distinct \ion{C}{iv} absorptions 
are separated at $v \sim 50$\,\kms\/. Therefore, this system contains only one
Ly$\alpha\beta$ clump at $v \in [-200, +40]$\,\kms\/. In Fig.~\ref{fig4}, within the
system velocity range $v \in [-450, +150]$\,\kms\/, the \ion{C}{iv}
absorption flux becomes at $F_{\mathrm{\ion{C}{iv}}} \ge 0.98$ 
at $v \sim -375, -190, -140, -80$ and $+115$\,\kms\/,
while the \ion{H}{i} profile from Ly$\alpha$ and Ly$\beta$ breaks at 
$v \sim -450, -375, -250, -155, -150, -85$ and $+140$\,\kms\/. Therefore, this system
consists of 3 Ly$\alpha\beta$ clumps at $v \in [-375, -140]$, $[-140, -80]$ and
$[-80, +115]$\,\kms\/.

Ly$\alpha\beta$ clumps are
analysed only in Section~\ref{sec5:4}. 
In this work, clumps refer 
only to a clump defined by a velocity range of a distinct \ion{C}{iv}
absorption.


\begin{table*}
\caption[]{The integrated column densities of the \civ\/ systems at the $v = \pm$150\,\kms\/ range.
Only the beginning of the entire table is shown.
The full version of this table is available electronically on the MNRAS website.}
\label{tab2}
\vspace{-0.2cm}
\begin{tabular}{ccccccrl}
\hline

QSO  & $z_{\mathrm{abs}}$ & class\,$^{\mathrm{a}}$ &  $[v_{1}, v_{2}]^{\mathrm{b}}$
 & $\log N_{\mathrm{\ion{H}{i}, \, sys}}$\,$^{\mathrm{c}}$ & $\log N_{\mathrm{\ion{C}{iv}, \, sys}}$\,$^{\mathrm{c}}$
 & $\Delta_{\mathrm{\ion{C}{iv}, \, sys}}$ & Other ions\,$^{\mathrm{d}}$ \\[0.1cm]
   &    &    &  (\kms\/) &   &   &  (\kms\/)  &  \\
\hline
\noalign{\smallskip}

    Q0055$-$269 &   3.256200 &  2 &              & $15.33\pm0.02$ & $13.58\pm0.05$ &   36.0 &                     \ion{Si}{iv}, \ion{O}{vi} \\
    Q0055$-$269 &   3.248119 &  2 & [$-150, +250$] & $15.44\pm0.01$ & $13.04\pm0.03$ &   77.8 &                                               \\
    Q0055$-$269 &   3.190942 &  1 & [$-150, +350$] & $15.67\pm0.03$ & $14.58\pm0.03$ &  102.3 &                                  \ion{Si}{iv} \\
    Q0055$-$269 &   3.095658 &  2 &              & $15.06\pm0.06$ & $12.85\pm0.05$ &    6.9 &                                               \\
    Q0055$-$269 &   3.085889 &  2 &              & $15.37\pm0.02$ & $13.16\pm0.08$ &   11.2 &                                  \ion{Si}{iv} \\
    Q0055$-$269 &   3.038793 &  2 & [$-250, +150$] & $15.12\pm0.03$ & $13.00\pm0.02$ &   78.4 &                                               \\
    Q0055$-$269 &   3.004992 &  2 &              & $15.23\pm0.03$ & $12.81\pm0.03$ &   13.8 &                                        blends \\
    Q0055$-$269 &   2.950571 &  2 & [$-250, +150$] & $15.71\pm0.03$ & $13.87\pm0.03$ &   65.9&                                  \ion{Si}{iv} \\
    Q0055$-$269 &   2.945250 &  3 &              & $16.74\pm0.02$ & $12.91\pm0.02$ &   14.1 &                                  \ion{Si}{iv} \\
    Q0055$-$269 &   2.913867 &  2 &              & $15.30\pm0.02$ & $12.82\pm0.03$ &   19.0 &                                               \\
    Q0055$-$269 &   2.895563 &  2 &              & $15.48\pm0.06$ & $12.98\pm0.02$ &   15.4 &                                        blends \\
    Q0055$-$269 &   2.744091 &  2 &              & $15.43\pm0.04$ & $13.12\pm0.04$ &   38.0 &                                               \\
    Q0055$-$269 &   2.705788 &  2 &              & $14.97\pm0.13$ & $12.54\pm0.03$ &   11.4 &                                               \\

\hline
\end{tabular}

\begin{list}{}{}
\item[$^{\mathrm{a}}$]
Class `1', `2' and `3' refer to higher-, normal and lower-\nciv\/ systems,
as shown in red, black and sky-blue filled circles in Fig.~\ref{fig9}.
See Section~\ref{sec5:2} for details.

\item[$^{\mathrm{b}}$]
Listed only when the velocity range has to be extended from the 
default $\pm150$\,\kms\/ velocity range.

\item[$^{\mathrm{c}}$]
The associated error of the integrated column densities was calculated using the standard,
independent
error propagation method when adding up the column densities. This is {\it not} 
the error obtained using the summed column densities option in {\tt VPFIT}, 
which is usually much smaller.  

\item[$^{\mathrm{d}}$]
Only \ion{Si}{iv}, \ion{O}{vi} and \ion{N}{v} are listed.
If \ion{Si}{iv} is not detected and \ion{O}{vi} and \ion{N}{v} 
are blended or not detected, the entry is left blank.
When \ion{Si}{iv} is blended and \ion{N}{v} and \ion{O}{vi} are also blended
or not detected, the entry is noted as `blends'. The entry `out of range' indicates that
all of \ion{Si}{iv}, \ion{O}{vi} and \ion{N}{v} are out of the observed wavelength range.
A system including a saturated \ion{C}{iv} is noted as `saturated \ion{C}{iv}'.

\end{list}

\end{table*}



\begin{table*}
\caption[]{The integrated column densities of \civ\/ clumps. Only the beginning of the entire table is shown.
The full version of this table is available electronically on the MNRAS website.}
\label{tab3}
\vspace{-0.2cm}
\begin{tabular}{lcccccll}
\hline

QSO  & $z_{\mathrm{abs}}$ & $[v_{1}, v_{2}]$
 & $\log N_{\mathrm{\ion{H}{i}, \, cl}}$\,$^{\mathrm{a}}$ & $\log N_{\mathrm{\ion{C}{iv}, \, cl}}$\,$^{\mathrm{a}}$ &
  Class\,$^{\mathrm{b}}$ & Other ions\,$^{\mathrm{c}}$ \\[0.1cm]
 &    &   (\kms\/) &  &  &   &  \\
 
\hline

    Q0055$-$269 &   3.256200 & [$ -40,  +100$\,] & $15.31\pm0.02$ & $13.58\pm0.05$ &       &                     \ion{Si}{iv}, \ion{O}{vi} \\
    Q0055$-$269 &   3.248119 & [$ -15,   45$\,] & $13.95\pm0.08$ & $12.80\pm0.03$ &    uncertain &                                               \\
    Q0055$-$269 &   3.248119 & [$ +125,  +200$\,] & $15.26\pm0.02$ & $12.67\pm0.05$ &                            &                                               \\
    Q0055$-$269 &   3.190942 & [$-100,  +300$\,] & $15.67\pm0.03$ & $14.58\pm0.03$  &                            &                                  \ion{Si}{iv} \\
    Q0055$-$269 &   3.095658 & [$ -20,   +20$\,] & $15.06\pm0.06$ & $12.85\pm0.05$ &    &                                               \\
    Q0055$-$269 &   3.085889 & [$ -25,   +30$\,] & $15.36\pm0.02$ & $13.16\pm0.08$ &                               &                                  \ion{Si}{iv} \\
    Q0055$-$269 &   3.038793 & [$-210, -170$\,] & $14.23\pm0.03$ & $12.25\pm0.07$ &                             &                                               \\
    Q0055$-$269 &   3.038793 & [$-160,  -80$\,] & $14.51\pm0.03$ & $12.45\pm0.06$ &                              &                                               \\
    Q0055$-$269 &   3.038793 & [$ -25,   +30$\,] & $14.70\pm0.06$ & $12.73\pm0.03$ &                               &                                               \\
    Q0055$-$269 &   3.004992 & [$ -30,   +30$\,] & $15.21\pm0.03$ & $12.81\pm0.03$ &                               &                                        blends \\
    Q0055$-$269 &   2.950571 & [$-165,  +120$\,] & $15.71\pm0.03$ & $13.87\pm0.03$ &    &           \ion{Si}{iv} \\
    Q0055$-$269 &   2.945250 & [$ -30,   +30$\,] & $16.74\pm0.02$ & $12.91\pm0.02$ &                               &                                  \ion{Si}{iv} \\
    Q0055$-$269 &   2.913867 & [$ -40,   +40$\,] & $15.21\pm0.03$ & $12.82\pm0.03$ &                                &                                               \\
    Q0055$-$269 &   2.895563 & [$ -35,   +35$\,] & $15.46\pm0.06$ & $12.98\pm0.02$ &                                &                                        blends \\
    Q0055$-$269 &   2.744091 & [$-135,   +30$\,] & $15.43\pm0.04$ & $13.12\pm0.04$ &                               &                                               \\
    Q0055$-$269 &   2.705788 & [$ -30,   +30$\,] & $14.96\pm0.13$ & $12.54\pm0.03$ &                                &                                               \\

\hline
\noalign{\smallskip}

\end{tabular}

\begin{list}{}{}
\item[$^{\mathrm{a}}$]
Same as in Footnote `c' in Table~\ref{tab2}.

\item[$^{\mathrm{b}}$]
The blank entry indicates a well-defined clump. The `uncertain' entry 
notes a uncertain clump due to the absence of \ion{H}{i} in the velocity range of a \ion{C}{iv} 
absorption or due to the ambiguous association of \ion{H}{i} and \ion{C}{iv}. 
The `nearby \ion{H}{i}' entry means that there exists a \ion{H}{i} component within 
10\,\kms\/ from either velocity bound.
The `saturated \ion{C}{iv}' entry notes that a clump
contains a saturated \ion{C}{iv}.

\item[$^{\mathrm{c}}$]
Same as in Footnote `d' in Table~\ref{tab2}.

\end{list}

\end{table*}

\begin{figure}
\includegraphics[width=9cm]{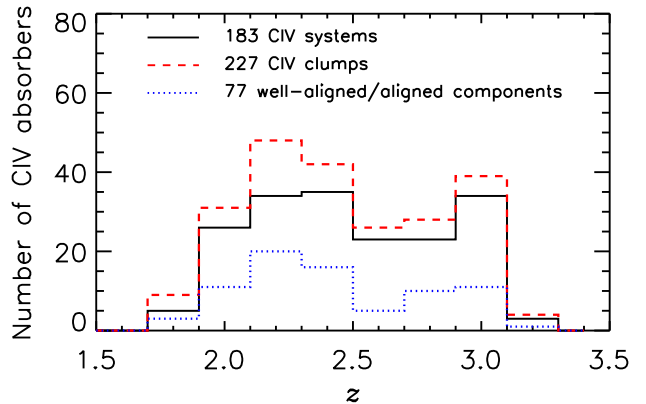}
\vspace{-0.5cm}
\caption{The redshift distribution of 183
\ion{C}{iv} systems (black solid line),
227 \ion{C}{iv} clumps (red dashed line) and 
77 well-aligned/reasonably-aligned 
\ion{H}{i}+\ion{C}{iv} components
(blue dotted line). 
Since saturated \ion{H}{i} lines at $z < 1.98$ are excluded in the sample
except toward J2233--606
due to their highly uncertain $N_{\ion{H}{i}}$, the number of the
\ion{C}{iv} systems decreases sharply.}
\label{fig5}
\end{figure}

In this working definition, our \ion{H}{i}+\ion{C}{iv}
sample consists of 183 \ion{C}{iv} systems, 
227 \ion{C}{iv} clumps,
38 well-aligned and 39 reasonably-aligned
\ion{H}{i}+\ion{C}{iv} component pairs
at $1.7 < z < 3.3$. Figure~\ref{fig5} shows their redshift distribution. 

Tables~\ref{tab2} and \ref{tab3} list the integrated 
column densities of \ion{H}{i} and \ion{C}{iv}  
of the first few \ion{C}{iv} systems and clumps, respectively,  
along with their QSO names, the system
redshift and the velocity range integrated over. The full tables are published
electronically. The 7th column of Table~\ref{tab2} is the \ion{C}{iv}-profile
weighted line width, calculated using Eq. (5) of \citet{sembach92}.
Its mean value is $<\sigma_{\ion{C}{iv}}>\,\, = 30.2$\,\kms\/. 

The 6th column of Table~\ref{tab3} indicates whether a
clump is well-defined or uncertain due to the absence of \ion{H}{i} in the
clump velocity range.
The last columns of Tables~\ref{tab2} and \ref{tab3} note any
associated \ion{Si}{iv}, \ion{O}{vi}
and \ion{N}{v}. Generally, \ion{C}{iv} systems
with higher \nhisys\/ are likely to be associated with more metal species
\citep{simcoe04, boksenberg14}. On the other hand,  
\ion{C}{iv} systems without \ion{Si}{iv} are likely to be associated with 
no other ions, and only rarely with \ion{N}{v} or \ion{O}{vi}
\citep{carswell02, schaye07}.

Table~\ref{tab4} lists the individual fitted line parameters of the two \ion{C}{iv} systems
over an integrated velocity range. All the fitted parameters for each \ion{C}{iv} system
and its velocity plot are published electronically. The format of the velocity plot 
is the same as in 
Figs.~\ref{fig2}, \ref{fig3} and \ref{fig4}, with the well-aligned
(reasonably-aligned) \ion{H}{i}+\ion{C}{iv}
components marked with thick blue (sky-blue) ticks.
We stress that we blindly include only the fitted lines within
the integrated velocity range and show only selected transitions
for simplicity and clarity, even though we have fitted almost the entire 
spectrum.
Note that we do not use the line parameters of metal ions other than \ion{C}{iv}
in this study, except to check whether a \ion{C}{iv} system is associated with other metals.
Therefore, only the line parameters of \ion{Si}{iv} are listed in Table~\ref{tab4} 
to show a \ion{Si}{iv} profile aligned with \ion{C}{iv}.
The entire fitted line lists will be published online in near future.

\section{The \nhi\/--\nciv\/ relation of \ion{C}{iv} systems and clumps}
\label{sec5}

\subsection{The integrated \nhi\/--\nciv\/ relation for \ion{C}{iv} systems}
\label{sec5:1}

\begin{table}
\caption[]{The line parameters of individual \civ\/ systems. Only the beginning of the entire table is shown.
The full version of this table is available electronically on the MNRAS website.}
\label{tab4}
\vspace{-0.2cm}
\begin{tabular}{rlr crc}
\hline

\# & Ion & $v$ & $z$ & 
$b$ & $\log N$ \\
  &     &  (km s$^{-1}$) &   &  (km s$^{-1}$)  &  \\
    
\noalign{\smallskip}

\hline
\hline
\multicolumn{2}{c}{Q0055$-$269} &  \multicolumn{2}{c}{$z =\, $3.256200} &
\multicolumn{2}{c}{$[-150, +150]$\,\kms\/} \\ 
\hline \\[-0.25cm]

   1 & \ion{H}{i}    & $   -136\pm12$ &   3.254260 & $    10.6\pm4.7$ & $ 12.04\pm0.20$ \\
   2 & \ion{H}{i}    & $    -71\pm17$ &   3.255193 & $    37.9\pm3.3$ & $ 13.73\pm0.06$ \\
   3 & \ion{Si}{iv}  & $    -25\pm16$ &   3.255850 & $    10.6\pm5.8$ & $ 11.80\pm0.18$ \\
   4 & \ion{C}{iv}   & $    -18\pm13$ &   3.255943 & $    11.9\pm2.9$ & $ 12.87\pm0.14$ \\
   5 & \ion{Si}{iv}  & $     -1\pm6$ &   3.256186 & $     9.7\pm2.1$ & $ 12.17\pm0.08$ \\
   6 & \ion{C}{iv}   & $      0\pm5$ &   3.256200 & $     9.5\pm1.3$ & $ 13.06\pm0.09$ \\
   7 & \ion{H}{i}    & $     16\pm2$ &   3.256422 & $    39.1\pm0.9$ & $ 15.31\pm0.02$ \\
   8 & \ion{C}{iv}   & $     29\pm10$ &   3.256616 & $    10.0\pm3.3$ & $ 12.57\pm0.16$ \\
   9 & \ion{Si}{iv}  & $     46\pm17$ &   3.256858 & $    22.0\pm6.5$ & $ 11.99\pm0.10$ \\
  10 & \ion{C}{iv}   & $     52\pm6$ &   3.256940 & $    13.0\pm2.3$ & $ 12.99\pm0.07$ \\
  11 & \ion{C}{iv}   & $     82\pm4$ &   3.257366 & $     9.7\pm1.3$ & $ 12.73\pm0.04$ \\
  12 & \ion{H}{i}    & $     93\pm7$ &   3.257523 & $    18.1\pm1.6$ & $ 13.51\pm0.07$ \\
  13 & \ion{H}{i}    & $    145\pm5$ &   3.258263 & $    26.1\pm2.5$ & $ 13.14\pm0.04$ \\[0.1cm]

\hline 

\multicolumn{6}{c}{Q0055-269, \hspace{0.4cm}  $z = $3.248119, \hspace{0.3cm} 
      [$-150, +250$]\,\kms\/ } \\[0.07cm]
\hline \\[-0.25cm]
   1 & \ion{H}{i}    & $   -144\pm32$ &   3.246077 & $    26.9\pm7.2$ & $ 12.70\pm0.20$ \\
   2 & \ion{H}{i}    & $    -64\pm8$ &   3.247206 & $    41.2\pm2.3$ & $ 14.43\pm0.02$ \\
   3 & \ion{H}{i}    & $    -11\pm6$ &   3.247961 & $    17.7\pm2.3$ & $ 13.95\pm0.08$ \\
   4 & \ion{C}{iv}   & $      0\pm2$ &   3.248119 & $     4.0\pm0.9$ & $ 12.50\pm0.04$ \\
   5 & \ion{C}{iv}   & $     23\pm4$ &   3.248447 & $     9.8\pm1.7$ & $ 12.49\pm0.05$ \\
   6 & \ion{H}{i}    & $     61\pm4$ &   3.248988 & $    44.7\pm2.2$ & $ 14.77\pm0.02$ \\
   7 & \ion{H}{i}    & $    161\pm2$ &   3.250407 & $    31.1\pm0.5$ & $ 15.26\pm0.02$ \\
   8 & \ion{C}{iv}   & $    166\pm10$ &   3.250474 & $    23.8\pm3.4$ & $ 12.67\pm0.05$ \\[0.1cm]

\hline
\end{tabular}
\end{table}

\begin{figure}

\includegraphics[width=8.5cm]{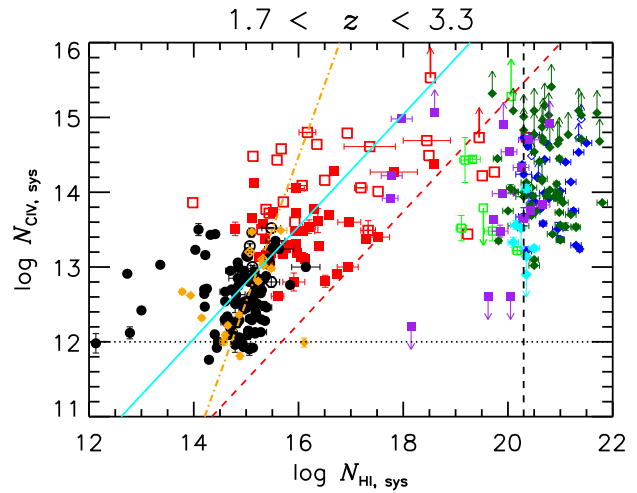}

\caption{The integrated \nhi\/--\nciv\/ relation 
including the LLSs/sub-DLASs/DLASs. Filled and open symbols
represent the \ion{C}{iv} systems integrated over
$\pm 150$\,\kms\/ and the extended
velocity ranges, respectively.
Circles and red squares indicate \ion{C}{iv} systems without
associated \ion{Si}{iv} and with \ion{Si}{iv}, respectively, while orange diamonds
represent systems whose \ion{Si}{iv} region is blended. 
For LLSs/sub-DLAs/DLAs, the data are
from \citep[filled blue diamonds]{prochaska01, prochaska03},
\citep[open light green squares]{dessauges03},
\citep[filled dark green diamonds]{fox07b},
\citep[open light green squares]{peroux07},
\citep[filled cyan diamonds]{penprase10} and 
\citep[filled purple squares]{lehner14}. Only errors larger than the symbol
size are shown for clarity. 
The red dashed line
delineates a $N_{\ion{H}{i}}$--$N_{\ion{C}{iv}}$ relation converted
from the IGM median optical depth relation between \ion{H}{i}
and \ion{C}{iv} estimated by \citet{schaye03}, while the cyan
solid line is the same relation with 20 times stronger $N_{\ion{C}{iv}}$.
See the text for details. 
The orange dot-dashed line is a robust, single power-law fit
for systems at $\log N_{\mathrm{\ion{H}{i}, \, sys}} \in [14, 16]$: 
$\log N_{\mathrm{\ion{C}{iv}, \, sys}} = (-16.52 \pm 3.77) + (1.94 \pm 0.25) \times 
\log N_{\mathrm{\ion{H}{i}, \, sys}}$.
The horizontal dotted line and the vertical dashed line note
the ``practical" \ion{C}{iv} detection limit at \lognc4\/ $=12.0$
and the border line between sub-DLAs and DLAs at
$\log N_{\ion{H}{i}} = 20.3$.
For any ambiguities, refer to the online, coloured version. 
}
\label{fig6}
\end{figure}

Figure~\ref{fig6} shows the integrated \nhi\/--\nciv\/ relation for a fixed
velocity range of $v \in [-150, +150]$\,\kms\/. 
Integrated column densities are obtained from adding up {\it blindly} all the
\ion{H}{i} and \ion{C}{iv} components over a given velocity ranges for each
system. The associated errors are calculated using the standard error
propagation for addition assuming uncorrelated errors as follows, for each {\it i}-th
component $\log N_{i} + \sigma(\log N_{i}) = y_{i} + \sigma (y_{i})$:   

\noindent $$N_{\mathrm{tot}} = \sum_{i} 10^{y_{i}},$$
\noindent $$\sigma (N_{\mathrm{tot}}) = \sqrt{\sum_{i} \sigma (y_{i}) \times 10^{y_{i}} / 0.434},$$

\noindent which leads to

\begin{equation}
\log N_{\mathrm{tot}}  + \sigma (\log N_{\mathrm{tot}}) =
        \log (N_{\mathrm{tot}}) + 0.434 \times \frac{\sigma (N_{\mathrm{tot}})}
         {N_{\mathrm{tot}}}.
\end{equation}

This is usually dominated by a few components with a large error.
Therefore, when a system contains such components,
the error is often over-estimated compared to the one obtained by using
the summed column densities option in {\tt VPFIT}.
Since the {\tt VPFIT} error is only a fit error
without a continuum fitting uncertainty in the way usually adopted in the studies
of the Milky Way interstellar medium, e.g. \citet{sembach91} and since 
our fit error is already small,
mostly with less than 0.05\,dex, we decided to use this standard error
for addition. Keep in mind that the column density defined this way 
is similar to the volume-averaged column density.
We also note that the integrated column density is in general 
measured more reliably than $N_{\mathrm{\ion{H}{i}}}$ and $b$ of individual
components (see Sections~\ref{sec3:1} and \ref{sec3:2}).

Of the 183 \ion{C}{iv}
systems, 64 are \ion{Si}{iv}-enriched, 103 are \ion{Si}{iv}-free
$\log N_{\mathrm{\ion{Si}{iv}}} \le 11.5$) and 16 have
blended or uncertain \ion{Si}{iv} systems. 
As expected, those with extended velocity range marked with open symbols 
tend to be associated with systems showing
higher-\ncivsys\/ and \ion{Si}{iv}, as
stronger \ion{C}{iv} in general consists of multiple components spread 
over a larger velocity range and is associated with other ions. 
Otherwise, there is no strong segregation
on the \nhisys\/--\ncivsys\/ plane. 
Therefore, no distinction is made in the further analysis 
between the \ion{C}{iv} systems integrated over
the $\pm150$ \kms\/ and extended velocity ranges.

The data points of LLSs/sub-DLAs/DLAs at $1.7 < z < 3.3$
are complied from literature.
Any duplicated
absorber from the literature was discarded in favour of our own measurements.
In studies on sub-DLAs and DLAs, \ion{C}{iv} is not an ion of a main interest,
therefore, not many measurements are available in the literature.
Both \nhi\/ and \nciv\/ taken from the literature do not meet our definition
of a \ion{C}{iv} system. However, both values can be considered to be
added up over the same extended velocity range, as including 
any typical forest \ion{H}{i} absorbers near sub-DLAs and DLAs in a given
velocity range has a negligible effect on their $N_{\mathrm{\ion{H}{i}}}$.

The red dashed line delineates a $N_{\ion{H}{i}}$ and $N_{\ion{C}{iv}}$ relation,
$\log N_{\ion{C}{iv}} \sim 0.75 \times \log N_{\ion{H}{i}} + 0.24$. This 
was converted from the relation between the median \ion{H}{i} and \ion{C}{iv} optical depths
($\tau_{\mathrm{\ion{H}{i}, \, med}}$ and $\tau_{\mathrm{\ion{C}{iv}, \, med}}$) 
at $z \sim 3$:
$\log \tau_{\mathrm{\ion{C}{iv}, \, med}} \sim 0.75 \times 
\tau_{\mathrm{\ion{H}{i}, med}} - 3.0$ at 
$\log \tau_{\mathrm{\ion{H}{i}, \, med}} \ge 0.1$ \citep{schaye03}.
Converting an optical depth to a column density
is not trivial and requires a $b$ parameter. We assumed that the
$\tau_{\mathrm{\ion{H}{i}, \, med}}$--$\log \tau_{\mathrm{\ion{C}{iv}, \, med}}$ 
relation holds on 
an optical depth at a line centre of \ion{H}{i} and \ion{C}{iv}, and assigned 
a single $b$ value for \ion{H}{i} and \ion{C}{iv} as 
$b_{\mathrm{\ion{H}{i}}} = 25.8$\,\kms\/ and
$b_{\mathrm{\ion{C}{iv}}} = 10.8$\,\kms\/, respectively.  
These $b$ values are the median $b$ value of \ion{H}{i} and \ion{C}{iv} in 
the analysed redshift range listed in Table~\ref{tab1}.

The optical depth
analysis uses all the \ion{H}{i} absorption regardless of its association to \ion{C}{iv}.
On the other hand, the integrated $N_{\ion{H}{i}}$--$N_{\ion{C}{iv}}$ relation 
uses only for \ion{H}{i} absorbers associated with \ion{C}{iv}. 
This difference causes that
the optical-depth-converted $N_{\ion{H}{i}}$--$N_{\ion{C}{iv}}$ relation is below
most data points in Fig.~\ref{fig6}. The cyan line presents
the same $\tau$-converted $N_{\ion{H}{i}}$--$N_{\ion{C}{iv}}$ relation 
if \nciv\/ is 20 times stronger. This proportionality constant
of 20 is not based on any fit, but is chosen to match
a majority of the data points at $\log N_{\mathrm{\ion{H}{i}, \, sys}} \in [14, 17]$.
This implies that the median $\tau_{\mathrm{\ion{C}{iv}}}$--$\tau_{\mathrm{\ion{H}{i}}}$ 
relation samples the IGM gas having about 20 times lower \ion{C}{iv} than
the individually detected \ion{C}{iv}-enriched gas. 

There are three noticeable features in the integrated \nhi\/--\nciv\/ relation:

1. At $N_{\mathrm{\ion{C}{iv}, \, sys}} \ge 12.8$,
\nhisys\/ and \ncivsys\/ display a scatter plot, independent of \nhisys\/,
even with a lack of 
\ion{C}{iv} systems at $N_{\mathrm{\ion{H}{i}, \, sys}}  \sim 18$
\citep{simcoe04}. 
Systems showing only a scatter on the \nhisys\/--\ncivsys\/
plane include most \ion{Si}{iv}-enriched \ion{C}{iv} systems shown 
in red filled/open squares including sub-DLAs and DLAs. 

2. A majority of \ion{Si}{iv}-free
\ion{C}{iv} systems at $\log N_{\mathrm{\ion{C}{iv}, \, sys}} \le 13$ shown 
in filled and open circles follow a well-defined power-law relation between 
\nhisys\/ and \ncivsys\/,
mostly concentrating at $\log N_{\mathrm{\ion{H}{i}, \, sys}} \sim 15$. 
The orange dot-dashed line is a robust, single power-law fit
to \ion{Si}{iv}-free systems at $\log N_{\mathrm{\ion{H}{i}, \, sys}} \in [14.0, 16.0]$: 
$\log N_{\mathrm{\ion{C}{iv}, \, sys}} = (-16.52 \pm 3.77) + (1.94 \pm 0.25) \times 
\log N_{\mathrm{\ion{H}{i}, \, sys}}$. 
This power-law is much steeper than from the optical depth analysis. 

For all the \ion{C}{iv} systems at $\log N_{\mathrm{\ion{H}{i}, \, sys}} \in [12, 22]$, 
\ncivsys\/ shows a steep increase at
$\log N_{\mathrm{\ion{H}{i}, \, sys}} \in [14, 16]$ from the \ion{C}{iv}
detection limit, spanning  $\sim$\,2\,dex at \nhisys\/ $\sim 15$.
Then, the relation becomes more or less
flattened at $\log N_{\mathrm{\ion{H}{i}, \, sys}} \ge 16$,
with a large scatter of $\sim$\,2.5\,dex. 
\ion{C}{iv} systems in the \nhisys\/--\ncivsys\/ plane at 
(\nhisys\/, \ncivsys\/) $\sim (17-22, \, \le \!13)$
do exist, but are rare. 
It is clear that a single power law only describes the \nhisys\/--\ncivsys\/ relation
over a short \nhisys\/ range, mainly for \ion{Si}{iv}-free \ion{C}{iv} systems.

3. There are a few outliers in the left and right sides of the orange dot-dashed 
line, especially apparent at 
$\log N_{\mathrm{\ion{H}{i}, \, sys}} \le 14$. 
Despite a large scatter expected in \ncivsys\/ for any given \nhisys\/,
hardly any  \ion{C}{iv} systems would be expected at 
$\log N_{\mathrm{\ion{H}{i}, \, sys}} \le 14$,
if naively extrapolated 
from the \nhisys\/--\ncivsys\/ relation displayed by most \ion{C}{iv} systems. 
The systems at $\log N_{\mathrm{\ion{H}{i}, \, sys}} \le 14$
belong to a class of absorbers named as high-metallicity absorbers
extensively studied in \citet{schaye07}. Their \nciv\/ is much 
higher than typical \ion{C}{iv} absorbers with a similar \nhi\/, leading 
to a higher metallicity than the typical \ion{H}{i}
forest. \citet{schaye07} argue that high-metallicity absorbers are
a transient object transporting recently metal-enriched
gas from galaxies into the IGM. 

There also exist 
systems at the right side of the orange dot-dashed line. These
systems have lower \ncivsys\/ than other typical systems having
the same \nhisys\/. There is even a system 
at $(\log N_{\mathrm{\ion{H}{i}, \, sys}}, \log N_{\mathrm{\ion{C}{iv}, \, sys}})
\! \sim \! (16, 12)$, for which a higher $\log N_{\mathrm{\ion{C}{iv}, \, sys}}$
is expected. This system is the $z = 2.326893$ absorber toward Q0109--3518. 
The saturated \ion{H}{i} Ly$\alpha$ consists of two \ion{H}{i} components
separated by $\sim\!45$\,\kms\/. A weak \ion{C}{iv} is associated with a stronger
\ion{H}{i} with \ion{C}{iii}, \ion{Si}{iii} and possibly \ion{Si}{iv}, while a weaker \ion{H}{i} component
at $v \sim +45$\,\kms\/ is associated with  \ion{C}{iii} and \ion{Si}{iii},
but not with \ion{C}{iv}. Even though two \ion{H}{i} components are very
close in the velocity space, their physical condition seems fairly different
from each other and from typical forest absorbers,
either due to the different ionising field, gas density or metallicity. 
 
\subsection{A fit to the \nhisys\/--\ncivsys\/ relation}
\label{sec5:2}
 
\begin{figure*}
\includegraphics[width=18cm]{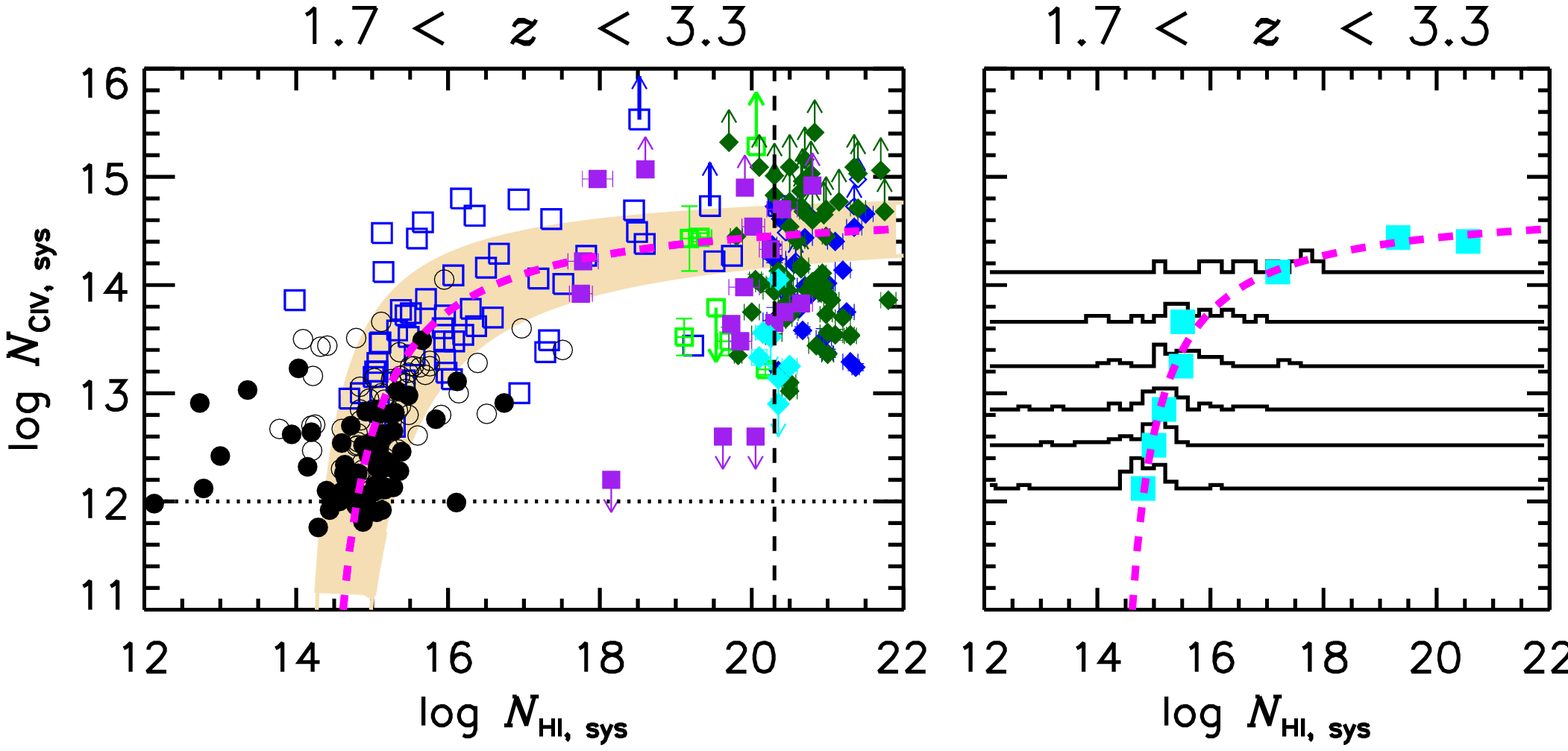}
\vspace{-0.5cm}
\caption{Left panel: The \nhisys\/--\ncivsys\/ relation for the $\pm150$\,\kms\/ range.
Filled circles, open circles and blue open squares represent a \ion{C}{iv}
system having a single-component \ion{C}{iv}, a 2-or-3-component \ion{C}{iv} 
and a multi-component \ion{C}{iv}. 
All the other symbols are the same as in
Fig.~\ref{fig6}. The rectangular hyperbola
functional fit to the filled cyan squares in the right panel
is shown as the magenta dashed curve with its 1\,$\sigma$ range.
Right panel: The histogram is the normalised 
number of \ion{C}{iv} systems as a function of \nhisys\/, with a binsize of 
$\Delta \log N_{\mathrm{\ion{C}{iv}, \, sys}} = 0.4$
at $\log N_{\mathrm{\ion{C}{iv}, \, sys}} \in [11.9, 14.3]$. 
The bottom 6 filled cyan squares overlaid
on the histogram indicate the median \nhisys\/ and \ncivsys\/
in each \nhisys\/ bin. The top two squares
represent the median \nhisys\/ and \ncivsys\/ 
at $\log N_{\mathrm{\ion{H}{i}, \, sys}} \in [18, 20]$ and [20, 22],
respectively.}
\label{fig7}
\end{figure*}

The left panel of Fig.~\ref{fig7} shows a more detailed version of Fig.~\ref{fig6}. 
Filled circles, open circles and blue open squares represent \ion{C}{iv}
systems having a single-component \ion{C}{iv}, a 2-or-3-component \ion{C}{iv} 
and a multi-component \ion{C}{iv}.
As expected, most single-component \ion{C}{iv} systems have a lower \ncivsys\/
and do not have associated \ion{Si}{iv}. 
Multi-component systems have a higher \ncivsys\/ and are associated with \ion{Si}{iv}. 
While a single power-law fit adequately 
describes at the limited \nhisys\/ range, a different
fitting function is required for the entire \nhisys\/ range.

Hydrodynamic simulations including galactic winds as a primary
IGM metal enrichment mechanism predict that metallicities 
increase rapidly with over-densities up to a 
critical overdensity, then 
flatten above this critical overdensity \citep{aguirre01a, oppenheimer12,
bordoloi14}. The slope of the metallicity--overdensity relation depends 
on wind speed, wind formation epoch and star formation rate in the
parents galaxies.
Metallicities and overdensities can be translated into 
\ncivsys\/ and \nhisys\/ in terms of observations, as both observational
quantities are similar to volume-averaged quantities.
Based on these theoretical predictions,  
we adopted a simple rectangular hyperbola fitting function to describe the 
\nhisys\/--\ncivsys\/ relation:

\begin{equation}
\log N_{\mathrm{\ion{C}{iv}, \, sys}} = 
\left[\frac{C_{1}}{\log N_{\mathrm{\ion{H}{i}, \, sys}} + C_{2}} \right] + C_{3}.
\label{eq2}
\end{equation}

To minimise the effect of a large scatter in \ncivsys\/ at a given \nhisys\/,
we take a following approach:

1. We pre-selected the \civ\/ systems having 
$\log N_{\mathrm{\ion{H}{i}, \, sys}} \le 18.0$ and  
$\log N_{\mathrm{\ion{C}{iv}, \, sys}} \le 14.5$ or having 
$\log N_{\mathrm{\ion{H}{i}, \, sys}} \ge 18.0$
and $\log N_{\mathrm{\ion{C}{iv}, \, sys}} \ge 14.0$. 
The \ion{C}{iv} systems with an 
lower limit on \ncivsys\/ were excluded. 

2. The selected \ion{H}{i}+\ion{C}{iv} pairs were binned at the \ncivsys\/ binsize of 0.4
at $\log N_{\mathrm{\ion{C}{iv}, \, sys}} \in [11.9, 14.3]$, where
where \nhisys\/ roughly increases with \ncivsys\/.

3. At each \ncivsys\/ bin, the median \nhisys\/ and the median \ncivsys\/ were 
selected as independent quantities.
This means that there is no \ion{C}{iv} system 
with the chosen median \nhisys\/ and \ncivsys\/.
These median pairs are shown
as a function of \nhisys\/ in the right panel of Fig. \ref{fig7}. 

4. At $\log N_{\mathrm{\ion{C}{iv}, \, sys}} \ge 14.0$, 
the \ion{H}{i}+\ion{C}{iv} pairs were 
binned to $\log N_{\mathrm{\ion{H}{i}, \, sys}} \in [18, 20]$ and [20, 22], 
as \ncivsys\/ is independent of \nhisys\/ with a large scatter. 
For these two \nhisys\/ bins, the median \nhisys\/ and \ncivsys\/ were again 
selected. These median column density pairs are plotted as the top 2 filled 
cyan squares in the right panel
of Fig.~\ref{fig7}. 

5. The median \nhisys\/ and \ncivsys\/ pairs 
were fitted to Eq.~\ref{eq2}, with $C_{1} = -1.90 \pm 0.55$, $C_{2} = 
-14.11 \pm 0.19$
and $C_{3} = 14.76 \pm 0.17$, respectively. 
This fit is shown in Fig. \ref{fig7}, along with a $1\,\sigma$ error range.

\begin{figure*}

\includegraphics[width=17.7cm]{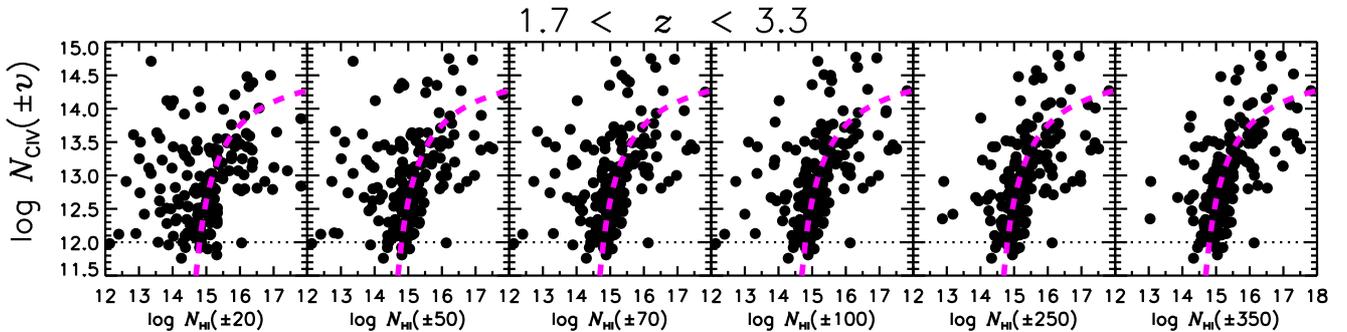}
\vspace{-0.5cm}
\caption{The \nhisys\/--\ncivsys\/ relation at 6 different integrated velocity
ranges, $\pm 20$, $\pm 50$, $\pm 70$, $\pm 100$, $\pm 250$ 
and $\pm 350$\,\kms\/,
without extending the velocity range to cover all the continuous \ion{C}{iv} 
absorptions. 
The y-axis shows the integrated $N_{\mathrm{\ion{C}{iv}}}$ 
for the same velocity range as the x-axis, with $v = \pm20, \pm50$\,\kms\/, etc.
The magenta dashed curve is the rectangular hyperbola
fit for the $\pm 150$\,\kms\/ velocity range.
}
\label{fig8}
\end{figure*}

The fitted curve follows the observed 
data points much better than a single power-law fit,
although the fit might not be the best description 
due to the lower
number of  \ion{C}{iv} systems at $\log N_{\mathrm{\ion{H}{i}, \, sys}} \in [17, 20]$.
Out of our 183 \ion{C}{iv} systems,  about 
75\,\% (137/183) lie in the shaded
region. About 16\,\% (30/183) and 
9\,\% (16/183) 
are located outside the 
1\,$\sigma$ contour at the left side (higher-\ncivsys\/) and right side
(lower-\ncivsys\/) of the fitted
 curve, respectively. We classify the \ion{C}{iv} systems into 3 groups,
at the left side of, within and
at the right side of the 1\,$\sigma$ contour as Class ``1", ``2" and
``3", respectively. 
This classification is listed in the
3rd column of Table~\ref{tab2} and is presented in the name of the 
QSO and the redshift of the \ion{C}{iv} system
on top of the velocity plot in red for Class ``1", in black for Class ``2" and
in sky-blue for Class ``3", e.g. Fig.~\ref{fig3}.

As \nhisys\/ and \ncivsys\/ are integrated quantities including many 
\ion{C}{iv}-free \ion{H}{i} components, the integrated \nhi\/--\nciv\/ relation
is likely to depend
on the integrated velocity range. Figure~\ref{fig8}
shows the \nhisys\/--\ncivsys\/ relation at  
6 different integrated velocity ranges, 
$\pm 20$, $\pm 50$, $\pm 70$, $\pm 100$, $\pm 250$ and $\pm 350$\,\kms\/.
Instead of extending the velocity range to cover all the \ion{C}{iv} absorptions
of a system, the strict velocity range was used. This results in excluding
some \ion{C}{iv} components associated with saturated \ion{H}{i} components
for stronger \ncivsys\/ systems, when the integrated velocity range is small. 
Therefore, a more scatter in the integrated \nhi\/--\nciv\/ relation is expected
for a smaller integrated velocity range.

The integrated \nhi\/--\nciv\/ relation is a scatter relation for the
$\pm 20$\,\kms\/ velocity range. 
Part of this
scatter is caused by the way the integrated column densities was calculated.
However, part of this scatter is real, implying that there is a small-scale
fluctuation in \ion{C}{iv} column densities as a function of \ion{H}{i}
column densities. 

The scatter in the integrated \nhi\/--\nciv\/ relation starts to decrease as
the integrated velocity range increases. The relation starts to converge
at the integrated velocity range larger than $\sim\!\pm100$\,\kms\/. There is
virtually no difference between the 
$\pm 250$\,\kms\/ and the $\pm 350$\,\kms\/ velocity range. This is simply due to
the fact that higher-\nhi\/ absorbers at $\log N_{\mathrm{\ion{H}{i}, \, sys}} \ge 14.5$
with which most \ion{C}{iv} systems
are associated are rare. Another such absorbers can be found at 
$\gg 1000$\,\kms\/. At the same time,
including weak-\nhi\/ absorbers in vicinity of high-\nhi\/ absorbers does
not change \nhisys\/ significantly. In short, most \ion{C}{iv} systems
follow a well-defined integrated \nhi\/--\nciv\/ relation at the velocity
range larger than $\pm 100$\,\kms\/. 
 
If the rectangular hyperbola function describing most \ion{C}{iv} systems
is extrapolated at lower-\ncivsys\/, a majority of the Ly$\alpha$ forest 
with $\log N_{\mathrm{\ion{H}{i}, \, sys}} < 14$ is expected to be 
{\it truly} \ion{C}{iv}-free, i.e. $\log \nciv\/ \ll 11.8$.

\subsection{The redshift evolution of the integrated \nhi\/--\nciv\/ relation}
\label{sec5:3}

\begin{figure}

\includegraphics[width=8.7cm]{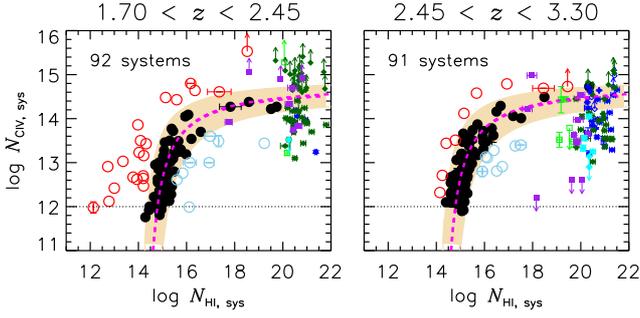}
\vspace{-0.5cm}
\caption{The integrated \nhi\/--\nciv\/ relation
at the two different redshift bins: 
$1.70 < z < 2.45$ (left panel) and
$2.45 < z < 3.30$ (right panel). The magenta curve represents a simple 
fitting function to most \ion{C}{iv} systems at $1.7 < z < 3.3$
as in Fig.~\ref{fig7}.
The shaded area indicates the 1\,$\sigma$ contour
of the fit. Red and sky-blue open circles indicate the \ion{C}{iv} 
systems which are located outside the 1\,$\sigma$ contour. 
All the other symbols are the same as in Fig.~\ref{fig6}. 
}
\label{fig9}
\end{figure}

Figure~\ref{fig9} 
shows the \nhisys\/--\ncivsys\/ relation at the two different redshift bins:
at $1.70 < z < 2.45$ 
and at $2.45 < z < 3.30$.
The high and low redshift bins were chosen simply to have 
a similar number of \ion{C}{iv} systems in the two bins.

Figure~\ref{fig9} also illustrates 
two distinct features of the integrated \nhi\/--\nciv\/ relation:

1. In the high-$z$ bin, a majority of \ion{C}{iv} systems 
are in the 1\,$\sigma$ contour defined at  
$1.7 < z < 3.3$. Even outliers
are very closely located around the 1\,$\sigma$ contour. 
In the low-$z$ bin, normal \ion{C}{iv} systems 
(Class 2, filled circles)
are also inside the 1\,$\sigma$ contour and have
 a similar \nhisys\/ spread for a given \ncivsys\/ as in the high-$z$ bin. 
This implies that most, normal \ion{C}{iv} systems
do not have any redshift evolution. 

2. Higher-\ncivsys\/ systems (Class 1, open red circles)
display the most significant
difference with redshift. In the low-$z$ bin, they spread 
into a much lower-\nhisys\/ area in the \nhisys\/--\ncivsys\/ plane. 
There are no higher-\ncivsys\/ systems 
at $\log N_{\mathrm{\ion{H}{i}, \, sys}} \le 14$ in the high-$z$ bin. 
Unfortunately, it is difficult to address the evolution of the 
lower-\ncivsys\/ systems,
since their numbers are too small.

Examination of all higher-\ncivsys\/ 
systems at low $z$ in our sample reveals that about half of them are 
isolated without any strong \ion{H}{i} absorbers
within $\pm 200$\,\kms\/, while another half are part of  
strong, saturated \ion{H}{i} absorber complexes. 
This ratio is also similar at higher $z$,  
though in this case there are only 7 higher-\ncivsys\/ systems.

\begin{figure}

\includegraphics[width=8.5cm]{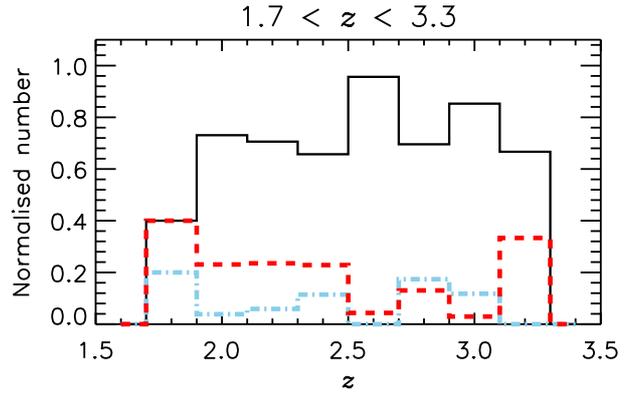}
\vspace{-1.0cm}
\caption{The normalised number of the normal, higher-\ncivsys\/
and lower-\ncivsys\/ \ion{C}{iv} systems in 
black, red dashed and sky-blue dot-dashed
histograms at 
$1.7 < z < 3.3$. The number of \ion{C}{iv}
systems in each category is normalised by the total number
of all \ion{C}{iv} systems in the same $z$ bin with the binsize of 0.2.
The first ($1.7 < z < 1.9$) and last ($3.1 < z < 3.3$) 
bins suffer from the low number of \ion{C}{iv} systems
(5 and 3 systems, respectively), while each of the other $z$ bins samples
more than 23 \ion{C}{iv} systems.
}
\label{fig10}
\end{figure}

\begin{figure}
\includegraphics[width=8.5cm]{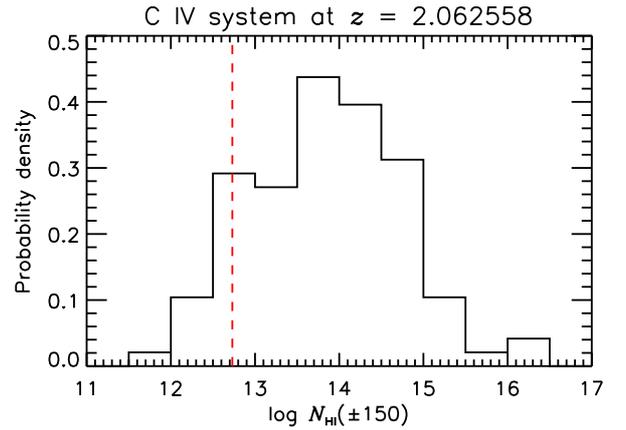}
\vspace{-0.5cm}
\caption{The probability density of 
an expected $N_{\mathrm{\ion{H}{i}}} (\pm 150)$ from
100 realisations, 
if the higher-\ncivsys\/
\ion{C}{iv} system at $z = 2.062558$ 
toward Q0122--380 is embedded in the forest at $z \sim 2.9$.
The vertical dashed line marks the real $\log 
N_{\mathrm{\ion{H}{i}}} (\pm 150) = 12.73$ of the
$z = 2.062558$ \ion{C}{iv} system.}
\label{fig11}
\end{figure}

\begin{figure*}
\hspace{-0.3cm}
\includegraphics[width=18cm]{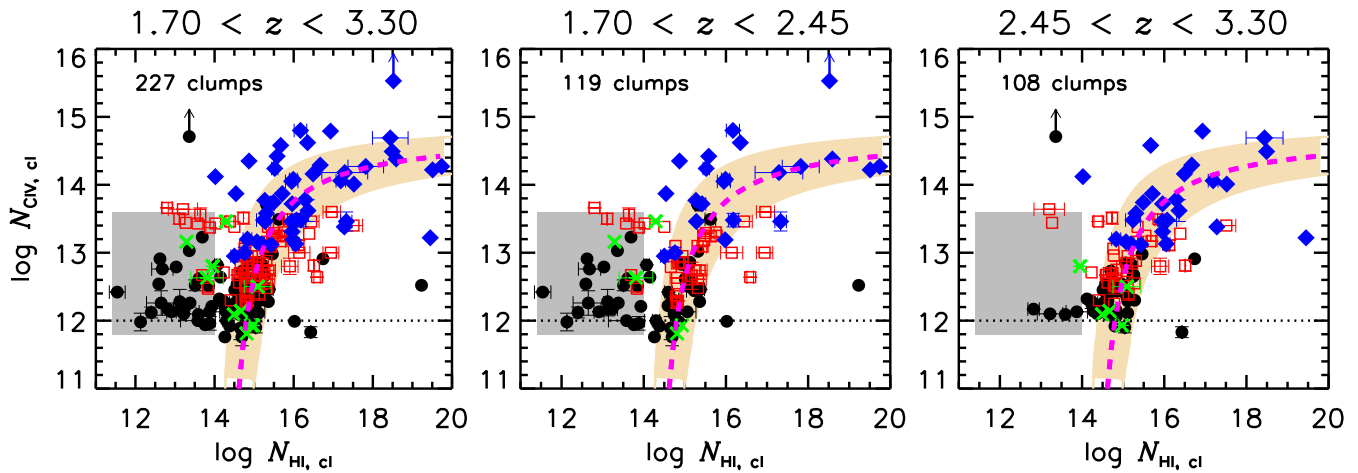}
\vspace{-0.5cm}
\caption{The integrated \nhi\/--\nciv\/ relation of \ion{C}{iv} clumps
at the 2 different redshift bins with the left panel for the full redshift range.
Filled circles and open red squares 
represent a clump with a single \ion{C}{iv}
component and 2-or-3 \ion{C}{iv} components, respectively.
Filled blue diamonds represent a clump 
with more than 4 \ion{C}{iv} components, while green crosses indicate
a uncertain clump.
All the rest of the symbols are the same as in Fig.~\ref{fig9}, including
the magenta rectangular hyperbola curve.
The grey-shaded area is where most higher-$N_{\ion{C}{iv}}$ clumps are located.
For clarity, the coloured version is available online.
}
\label{fig12}
\end{figure*}

Figure~\ref{fig10} shows the normalised number of 
3 categories of \ion{C}{iv} systems as a function of $z$, colour-coded
with the symbol colours in Fig.~\ref{fig9}. 
Although the number of our \ion{C}{iv} systems 
is not very large, there seems to be some evidence that the number of higher-\ncivsys\/
systems increases as $z$ decreases. 

However, this apparent redshift evolution of higher-\ncivsys\/ systems
could be caused by the observational bias, since the \ion{H}{i} line
number density decreases at lower redshifts \citep{kim13}.
For an integrated column density over a fixed velocity range,
stronger blending at higher $z$ would result in 
including more \ion{C}{iv}-free \ion{H}{i} components.

To illustrate this point,  
Fig.~\ref{fig11} shows the probability density of 
an expected $N_{\mathrm{\ion{H}{i}}} (\pm 150)$, 
if the higher-\ncivsys\/
\ion{C}{iv} system at $z = 2.062558$ 
toward Q0122--380 is embedded in the forest at $z \sim 2.9$.
We used part of the real spectrum at $z \sim 2.9$ of 
PKS2126--158, Q0420--388, Q0636$+$6801 and HE0940--1050.
This \ion{C}{iv} absorption feature
is initially placed at 6007\,\AA\/ ($z = 2.880$) in one of the 4 spectra.
An integrated $N_{\mathrm{\ion{H}{i}}} (\pm 150)$ was calculated for this redshift.
Then, the \ion{C}{iv} component was shifted by 4\,\AA\/ up to 6103\,\AA\/ ($z = 2.942$).
For each shift, a new integrated $N_{\mathrm{\ion{H}{i}}} (\pm 150)$ was calculated.
This process was repeated for the remaining 3 QSO spectra, with a total of
100 realisations.

Considering that the $z = 2.062558$ \ion{C}{iv} system has
the real $N_{\mathrm{\ion{H}{i}}} (\pm 150) = 12.73\pm0.04$, the
probability density peaks at $N_{\mathrm{\ion{H}{i}}} (\pm 150) \sim 14$.
In short, higher-\ncivsys\/ systems at $z  \sim 2$ could be a normal
\ion{C}{iv} systems at $z \sim 2.9$, perfectly within the 1\,$\sigma$ range
in Fig.~\ref{fig7} for its $N_{\mathrm{\ion{C}{iv}}} (\pm 150) = 12.91 \pm 0.01$.

However, this simple deduction has a one significant flaw. In the
above 100 realisations, both \ion{H}{i} and \ion{C}{iv} were treated as being
independent, which is not correct. If 
higher-\ncivsys\/ systems started to pop up mainly
due to the lower \ion{H}{i} line number density
at lower redshifts, the same logic should apply
for normal \ion{C}{iv} systems. This would shift the fitted dashed curve
in the low-$z$ bin to the left side in Fig.~\ref{fig9}, 
while no such evolution is observed.
Therefore, the increasing number of higher-\ncivsys\/ systems at lower redshifts
is likely to be real, not a consequence of less blending.

\subsection{The integrated \nhi\/--\nciv\/ relation of \ion{C}{iv} clumps}
\label{sec5:4}

\begin{figure*}
\hspace{-0.3cm}
\includegraphics[width=18cm]{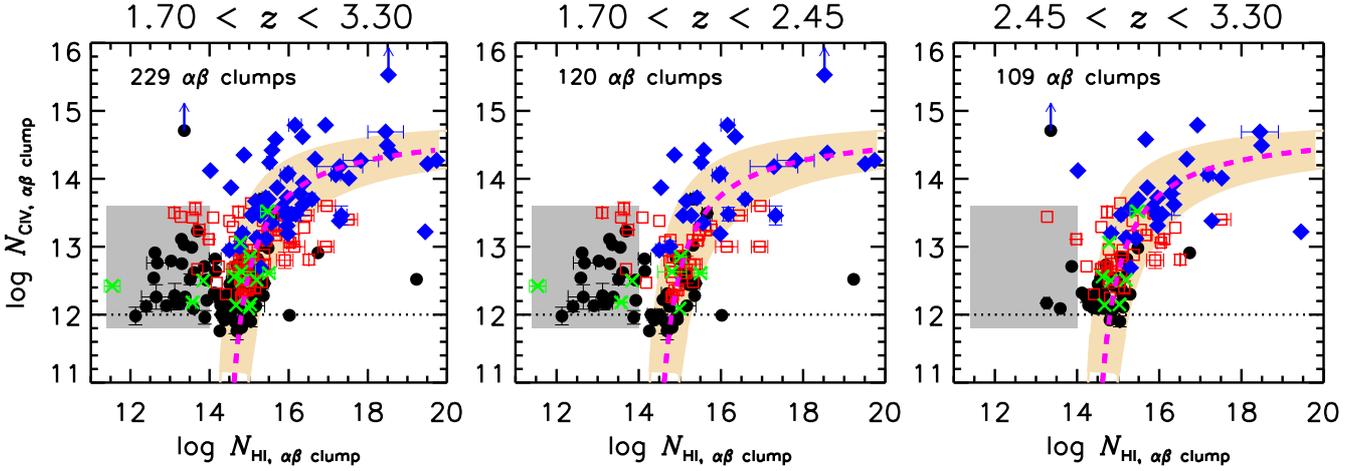}
\vspace{-0.5cm}
\caption{The integrated \nhi\/--\nciv\/ relation of \ion{C}{iv} 
Ly$\alpha\beta$ clumps
at the 2 different redshift bins with the left panel for the full redshift
range.
All the symbols are the same as in Fig.~\ref{fig12}.
The online, coloured version is available for a clear view.
}
\label{fig13}
\end{figure*}

Our definition of a \ion{C}{iv} system includes many nearby 
\ion{C}{iv}-free \ion{H}{i} absorptions. Therefore, a term ``clump" was 
introduced in Section~\ref{sec4:3},
in order to assign \ion{C}{iv} only with the clearly associated \ion{H}{i} in 
the velocity space.
First, a \ion{C}{iv} clump velocity range is defined 
for which a \ion{C}{iv} profile wing recovers
to a normalised \ion{C}{iv} flux $F_{\ion{C}{iv}} = 1$
longer than 5\,\kms\/. Then, 
only the \ion{H}{i} components in the clump velocity range are associated
with the \ion{C}{iv} clump, e.g. Figs.~\ref{fig3} and \ref{fig4}.
Depending on the \ion{H}{i} and \ion{C}{iv} profiles, 
a system can consist of a single clump or many clumps.

As defining a clump can be subjective, due to the uncertainties
in the continuum and non-unique \ion{H}{i} component
structures from the Voigt profile fitting, 
an auxiliary term ``Ly$\alpha\beta$ clump" was also
introduced. For this definition, 1) a velocity range is chosen
for a \ion{C}{iv} absorption profile wing to recover to $F_{\ion{C}{iv}} \ge 0.98$
and 2) \ion{H}{i} absorption profiles from Ly$\alpha$ and Ly$\beta$
are also clearly separable at similar relative velocities as \ion{C}{iv}. 
This definition is based largely on the profile shape.
The Ly$\alpha\beta$ clump is closest to a conventionally 
defined \ion{C}{iv} absorber/system in literature. 

Figures~\ref{fig12} and \ref{fig13} show the integrated \nhi\/--\nciv\/ relation 
for clumps and Ly$\alpha\beta$ clumps, respectively.
The \nhicl\/--\ncivcl\/ relation of 
clumps displays much more scatter than the
\ion{C}{iv} systems, which is rather similar to the the \nhisys\/--\ncivsys\/ relation
with an integrated velocity range less than $\pm50$\,\kms\/, as seen in 
Fig.~\ref{fig8}. This is expected since \nhicl\/ and \ncivcl\/ of most clumps
are measured at a smaller velocity range than $\pm150$\,\kms\/.

The scatter is mainly spread
into a gray-shaded area in the left side of the hyperbolic fit curve in 
the low-$z$ bin, where most higher-$N_{\ion{C}{iv}}$ systems are located.
In contrast,  
in the high-$z$ bin, clumps 
are distributed mostly along the hyperbolic fit curve. Similar to \ion{C}{iv} systems,
higher-\nciv\/ clumps are mostly found at lower redshifts.

Qualitatively, Ly$\alpha\beta$ clumps also show
a similar trend on the 
$N_{\mathrm{\ion{H}{i}, \alpha\beta clump}}$--$N_{\mathrm{\ion{C}{iv}, \alpha\beta clump}}$
plane. This is mainly caused by the fact that 
most \ion{H}{i} components just outside the clump velocity bound
are usually weak. In addition, clumps/Ly$\alpha\beta$ clumps in the gray-shaded area have
only one \ion{C}{iv} component and 
$\log N_{\mathrm{\ion{H}{i}, \, cl}} \le 14$, i.e. unsaturated, which enables to obtain
a reliable \ion{H}{i} component structure from a Ly$\alpha$ only. Therefore,  
an integrated velocity range for clumps and Ly$\alpha\beta$ clumps in the gray-shaded area
is the same in most cases.

\begin{figure*}
\includegraphics[width=17.9cm]{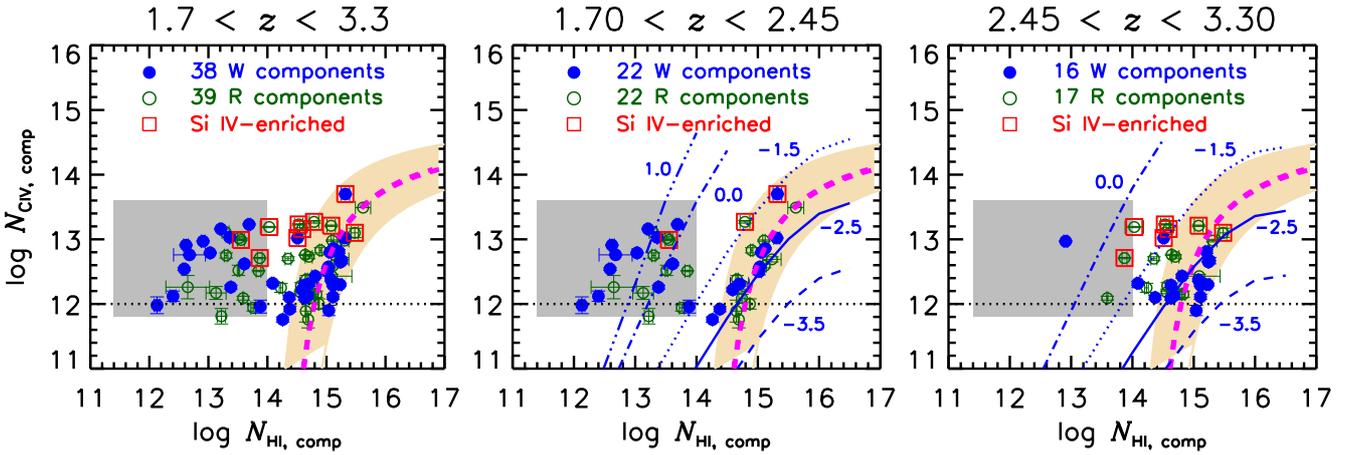}
\caption{The component \nhi\/--\nciv\/ relation at the 2 different redshift bins
with the left panel for the full redshift range.
Filled and open circles represent the 38 well-aligned (W) 
and 39 reasonably-aligned (R) 
\ion{H}{i}+\ion{C}{iv} component pairs, respectively. 
The symbols 
in larger red open squares indicate the aligned
pairs with an aligned \ion{Si}{iv} component.
The magenta curve is the same as in
Fig.~\ref{fig7}. The gray-shaded area delineates the same gray-shaded
region in Fig.~\ref{fig13},
and is the region populated mainly by higher-$N_{\ion{C}{iv}}$ systems
and clumps.
The blue dashed, solid, dotted,
dot-dashed and dot-dot-dot-dashed 
curves represent the {\tt CLOUDY} predictions with the fiducial Schaye
$n_{\mathrm{H}}$--\nhi\/ relation for [C/H] $= -3.5, -2.5, -1.5$,
0.0 and 1.0, respectively, noted with a number next to the corresponding curve.
} 
\label{fig14}
\end{figure*}

\section{\ion{C}{iv} components}
\label{sec6}

\subsection{Photoionisation modelling}
\label{sec6:1}

The integrated relations presented in Section~\ref{sec5} hold
for averaged quantities, without considering actual one-to-one 
physical association between \ion{H}{i} and \ion{C}{iv}. Therefore, 
any attempt to derive physical conditions of the absorbing gas
based on the integrated \nhi\/--\nciv\/ relation, such as the carbon abundance,
is meaningless. However, for aligned
components with the difference between the velocity
centroid of relatively clean \ion{H}{i} and \ion{C}{iv} components 
less than 5\,\kms\/, which can be assumed to be co-spatial, an
analysis based on photoionisation equilibrium becomes
possible. We note that only 12\,\% of the total of 
628 \ion{C}{iv} components are
aligned, as expected from strong evidence of
a velocity difference between \ion{H}{i} and \ion{C}{iv} 
\citep{ellison00, reimers01}. Therefore, the aligned pairs do not represent a majority 
of \ion{H}{i} and 
\ion{C}{iv} absorbing gas, but only sample the gas having a simple physical 
structure. We also note that only 13\,\% (10 
out of 77 pairs) aligned pairs have 
an aligned \ion{Si}{iv} component.

The internal thermal and chemical structures of optically-thin \ion{H}{i} absorbers with 
$\log N_{\mathrm{\ion{H}{i}, comp}} < 17.2$ such as our aligned
\ion{H}{i} components are mainly determined by photoionisation and photoheating 
from the ambient ultraviolet (UV) background radiation balanced
by adiabatic cooling in the expanding Universe, i.e. the Hubble expansion,
and by radiative cooling \citep{hui97, schaye00a, schaye01, wiersma09, dave10}. 
At $z \sim 2.4$, the cosmic mean density corresponds to the total (neutral and ionised)
hydrogen volume density $\log n_{\mathrm{H}} \sim -6.72 + 3 \times \log (1+z) \sim 
-5.13$\,[cm$^{-3}$]
\citep{wiersma09}. Any absorbers below this density need to be taken account of
the Hubble expansion, while absorbers above this density expect to collapse. 
Based on the subset of our well-aligned \ion{H}{i}+\ion{C}{iv} component pairs,
\citet{kim15} find that most \ion{H}{i} components aligned with \ion{C}{iv} and \ion{C}{iii}
at $\log N_{\mathrm{\ion{H}{i}, comp}} \le 16$ have $\log n_{\mathrm{H}} \in [-5.2, -3.3]$
at $z \sim 2.4$. Since $n_{\mathrm{H}}$ of our \ion{H}{i}+\ion{C}{iv} pairs is
close to the cosmic mean density, we assume that both Hubble expansion 
and gravitational collapse do not play a significant role, i.e. a static absorber.
This assumption leaves 
radiative cooling the dominant source of cooling of these optically-thin absorbers.
In addition, \citet{kim15} also find that most aligned, optically-thin
\ion{H}{i}+\ion{C}{iv}
absorbers have the gas temperature in K at $\log T \in [3.5, 5.5]$ peaking at $\log T
\sim 4.4\pm0.3$, implying that the photoionisation model can be considered to be
adequate for our aligned \ion{H}{i}+\ion{C}{iv} component pairs in this study.

To model the gas clouds, we used the photoionisation code {\tt CLOUDY} 
version c10.3  \citep{ferland13}.
The geometry of the gas was assumed as a uniform slab in thermal and ionisation
equilibrium, and we used the redshift-dependent
{\tt CLOUDY}-default UV background, the Haardt-Madau (HM) UVB
2005 version with contributions from both
QSOs and galaxies (Q+G). 
For the two redshift ranges at
$1.70 < z < 2.45$ and $2.45 < z < 3.30$,
we used the HM Q+G
UVB 2005 at
$z = 2.3$ and $z = 2.8$ since a large fraction of the pairs has
a redshift similar to the adopted one.
The {\tt CLOUDY}-default
solar abundance pattern was used, with the solar carbon abundance of 
(C/H)$_{\odot} = -3.61$. The carbon abundance is expressed in the usual way as
${\mathrm{[C/H]}} = \log({\mathrm{C/H}}) - \log{\mathrm{(C/H)_\odot}}$. 

For an assumed UV background, deriving physical parameters of \ion{H}{i}+\ion{C}{iv} 
gas requires an additional carbon
transition such as \ion{C}{ii} or \ion{C}{iii} to break a degeneracy
between unknown [C/H] and $\log n_{\mathrm{H}}$.
Since estimating a reliable $N_{\ion{C}{iii}}$ or $N_{\ion{C}{ii}}$ is not possible
for a majority of our \ion{H}{i}+\ion{C}{iv} pairs due to blending,  
we instead generated a set of grid models with [C/H] varying from $-3.5$ to 2.0 with 
the logarithmic step size of 0.5, 
with $\log n_{\mathrm{H}}$ from $-7.0$ to $0.0$ with the 
logarithmic step size of 0.2, and with $\log N_{\ion{H}{i}}$ as a stopping criterion
varying from 12 to 16.5 with the logarithmic step size of 0.5.
Then, to break a degeneracy, we used 
the $n_{\mathrm{H}}$--\nhi\/ relation by \citet{schaye01} as our fiducial Schaye
relation, 
$\log N_{\mathrm{\ion{H}{i}}} \sim 20.86 + 1.5 \log n_{\mathrm{H}}$,
for the low-density forest in hydrostatic equilibrium,
using his default values for the gas temperature, the UVB and the gas mass fraction.

\subsection{The \nhi\/--\nciv\/ relation of \ion{C}{iv} components}
\label{sec6:2}

Figure~\ref{fig14} shows the component \nhi\/--\nciv\/ (\nhicomp\/--\ncivcomp\/) 
relation. As with the \nhicl\/--\ncivcl\/ relation shown in Fig.~\ref{fig12}, 
the \ion{H}{i}+\ion{C}{iv} component pairs also display a scatter plot.
The pairs with an aligned \ion{Si}{iv} tend to have
higher \ncivcomp\/ and appear predominantly at higher redshifts.

The pairs occupy a
well-defined region on the \nhicomp\/--\ncivcomp\/ plane at
$(\log N_{\mathrm{\ion{H}{i}, \, comp}}, \log N_{\mathrm{\ion{C}{iv}, \, comp}})
= (12.0$--15.5, 11.7--13.5), without any pairs with strong \ion{H}{i}
and \ion{C}{iv}. This is largely due to the fact that

\begin{enumerate}
\item strong \ion{H}{i} and \ion{C}{iv} absorptions break into several weaker 
components, 

\item a majority of the \ion{C}{iv} components are not at the same velocity as
\ion{H}{i} \citep{ellison00, reimers01},
and 

\item the velocity structure of the strong \ion{H}{i} gas is less well resolved than the
strong \ion{C}{iv} gas due to the larger \ion{H}{i} thermal broadening.

\end{enumerate}

In Fig.~\ref{fig14}, we also show the 
{\tt CLOUDY} predictions, with [C/H] noted next to each curve.
At $z \sim 2.1$,
the aligned pairs are clustered as the two distinct groups. The pairs
around the hyperbolic fit 
seem to be well-modeled
with [C/H] $\sim -2.5$ and the fiducial Schaye $n_{\mathrm{H}}$--\nhi\/ relation.
Other [C/H] values do not predict the observed \nhicomp\/--\ncivcomp\/ relation
as well as [C/H] $= -2.5$, regardless of any assumed $n_{\mathrm{H}}$--\nhi\/ relations
other than the Schaye $n_{\mathrm{H}}$--\nhi\/ relation.
On the other hand, the pairs inside the gray-shaded area on the
\nhicomp\/--\ncivcomp\/ plane can be produced only with 
[C/H] $\ge 0.0$ at $\log n_{\mathrm{H}} \in [-6.0, -5.0]$ \citep{schaye07},
reinforcing that they are likely to be connected with star formations at $z \sim 2.1$.

At $z \sim 2.9$,
the aligned pairs are predominantly in
one region. The Schaye $n_{\mathrm{H}}$--\nhi\/ relation
at [C/H] $= -2.5$ is less satisfactory than at lower redshifts, implying
that the high-$z$ aligned pairs sample a wider range of physical conditions 
than those around the rectangular hyperbola curve at $z \sim 2.1$.
For the same [C/H], \ion{Si}{iv}-enriched \ion{H}{i}+\ion{C}{iv} pairs
have a higher physical density than \ion{Si}{iv}-free pairs,
as the Schaye relation assumes a higher $n_{\mathrm{H}}$
for higher $N_{\mathrm{\ion{H}{i}}}$. For a similar
$n_{\mathrm{H}}$, i.e. a similar $N_{\ion{H}{i}}$,
\ion{Si}{iv}-enriched pairs have a higher [C/H]
than \ion{Si}{iv}-free pairs.

On face value, without taking account of any possible selection bias,
our aligned \ion{H}{i}+\ion{C}{iv} pairs
seem to evolve to have lower \nhi\/ at lower $z$, if aligned pairs
are produced by the gas in a similar location, i.e. an IGM
filament gas or a galactic halo, suggested by
their low physical gas density.

\begin{figure}
\includegraphics[width=8.5cm]{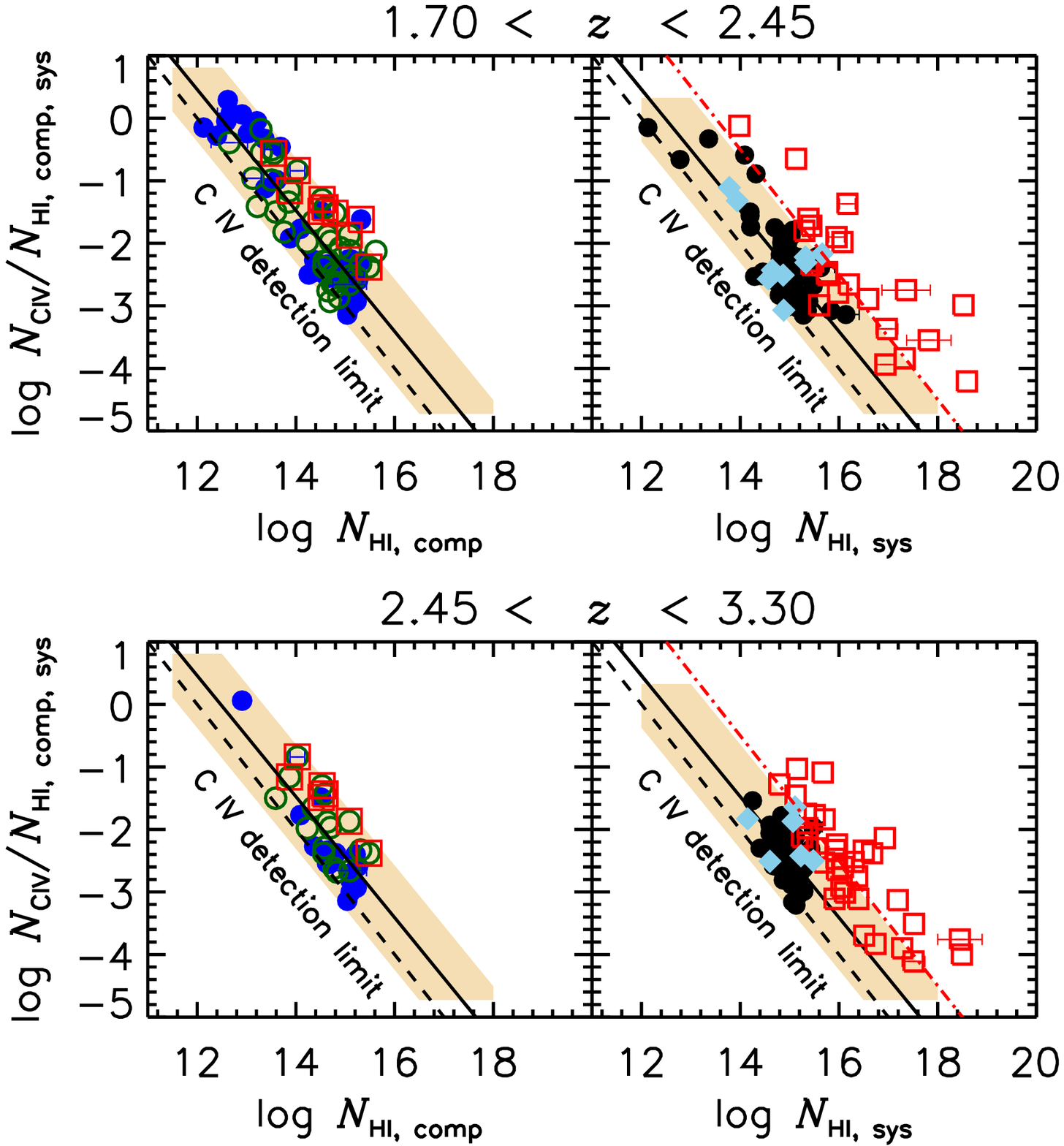}
\vspace{-0.4cm}
\caption{ 
{\it Left panels:} The \nhicomp\/--(\ncivcomp\//\nhicomp\/) relation 
at $1.70 < z < 2.45$ (upper panel) and $2.45 < z < 3.30$
(lower panel). The symbols
are the same as in Fig.~\ref{fig14}. 
Right panels: The \nhisys\/--(\ncivsys\//\nhisys\/) relation for the \ion{C}{iv}
systems. Filled circles, filled sky-blue diamonds and open
red squares indicate
\ion{Si}{iv}-free, blended/uncertain-\ion{Si}{iv} and \ion{Si}{iv}-enriched \ion{C}{iv}
systems.
In all panels, the solid line
represents a least-square-fit to the \ncivcomp\//\nhicomp\/ at the full 
redshift range:
$\log (N_{\mathrm{\ion{C}{iv}, \, comp}}/N_{\mathrm{\ion{H}{i}, \, comp}})
= (12.07\pm0.83) + (-0.97 \pm 0.06)
\times  \log N_{\mathrm{\ion{H}{i}, \, comp}}$. The light-orange shaded area 
represents the 1$\sigma$ contour. 
The black dashed line delineates a typical \ion{C}{iv} detection limit of  
$\log N_{\mathrm{\ion{C}{iv}, \, comp}} = 12.0$.
The overlaid red dot-dashed line in the right panels is the best-fit to the 
\ion{Si}{iv}-enriched \ion{C}{iv} systems:
$\log$\,(\ncivsys\//\nhisys\/) $= 13.5 - \log N_{\mathrm{\ion{H}{i}, \, sys}}$.
Only errors larger than the symbol size are displayed.
}
\label{fig15}
\end{figure}

\subsection{\ncivcomp\//\nhicomp\/ as a function of \nhicomp\/}
\label{sec6:3}

The ratio \ncivcomp\ over \nhicomp\/ provides, after ionisation corrections 
(which may be large), a measure of [C/H]. 
Not surprisingly, given that \ncivcomp\ is largely independent of \nhicomp\/ 
(see Fig.~\ref{fig14}), this ratio is very 
closely inversely proportional to \nhicomp\/ as can be seen in Fig.~\ref{fig15}. 
A formal fit over the full redshift range gives 
$\log \, (N_{\mathrm{\ion{C}{iv}, \, comp}}/N_{\mathrm{\ion{H}{i}, \, comp}})
= (12.07\pm0.83) + (-0.97 \pm 0.06)\times  \log N_{\mathrm{\ion{H}{i}, \, comp}}$, or, 
expressed slightly differently, $\log N_{\mathrm{\ion{C}{iv}, \, comp}}
= (12.07\pm0.83) + (0.03 \pm 0.06)\times  
\log N_{\mathrm{\ion{H}{i}, \, comp}}$. 

Over the column density range $\log \nhicomp\/ \in [14, 16]$ for the full redshift range, 
the median $\log\,(\ncivcomp\//\nhicomp\/)=-2.38$ for the median 
$\log \nhicomp\/=14.8$. This is close to the value $-2.46$ found by \citet{cowie95} for a 
similar \nhi\/ range at $z \sim 2.6$. For the column density range $\log \nhicomp\/ \in
[12, 14]$, the median $\log\, (\ncivcomp\//\nhicomp\/) = -0.50$
and the median $\log \nhicomp  = 13.3$.

The aligned pairs with highest \ncivcomp\//\nhicomp\/ for a given \nhicomp\/
are \ion{Si}{iv}-enriched components (symbols embedded in a larger
open red square). This suggests that the aligned 
pairs with \ion{Si}{iv} have a higher total hydrogen volume density
and/or a higher metallicity than 
\ion{Si}{iv}-free pairs as seen in Fig.~\ref{fig14}. 

Assuming that the intensity and spectral shape of the UV ionising background 
do not change significantly at $2 < z < 3.3$ \citep{bolton05, faucher08b,
boksenberg14} and that additional photons from other ionising sources are
negligible for aligned \ion{C}{iv} components at our \nhicomp\/ range,
a higher \ncivcomp\//\nhicomp\/ at lower-\nhicomp\/ implies a higher carbon 
abundance in lower-\nhicomp\/ pairs
or a higher gas volume density if a similar carbon abundance, e.g. \citet{schaye07}.

The interpretation of \ncivsys\//\nhisys\/ for \ion{C}{iv} systems 
shown in Fig.\ref{fig15} is less straightforward since they are 
averaged quantities.  
Comparing Figs.~\ref{fig8} and \ref{fig9}
with Fig.~\ref{fig14} shows that \ion{C}{iv} systems display
a larger scatter in {\ncivsys} for a given {\nhisys} than the components. 
Therefore, the \ion{C}{iv} systems display
a larger scatter in the \nhisys\/--(\ncivsys\//\nhisys\/) relation.
\ion{Si}{iv}-free \ncivsys\/ 
and blended/uncertain-\ion{Si}{iv} \ncivsys\/ 
systems follow the \nhicomp\/--(\ncivcomp\//\nhicomp\/) relation reasonably well, 
with the median $\log$\,(\ncivsys\//\nhisys\/) $= -2.40$ for the full redshift
and \nhisys\/ ranges.

On the other hand, the \ion{Si}{iv}-enriched \ion{C}{iv} systems 
(open red squares)
show an order of magnitude higher \ncivsys\//\nhisys\/ for a given \nhisys\/.
This indicates that \ion{Si}{iv}-enriched \ion{C}{iv} systems are exposed to a higher UV 
background if
the gas density and metallicity are similar to the \ion{Si}{iv}-free systems at a similar
\nhisys\/, or alternatively have a higher [C/H] and $n_{\mathrm{H}}$ for the same UVB.

\begin{figure*}
\includegraphics[width=17cm]{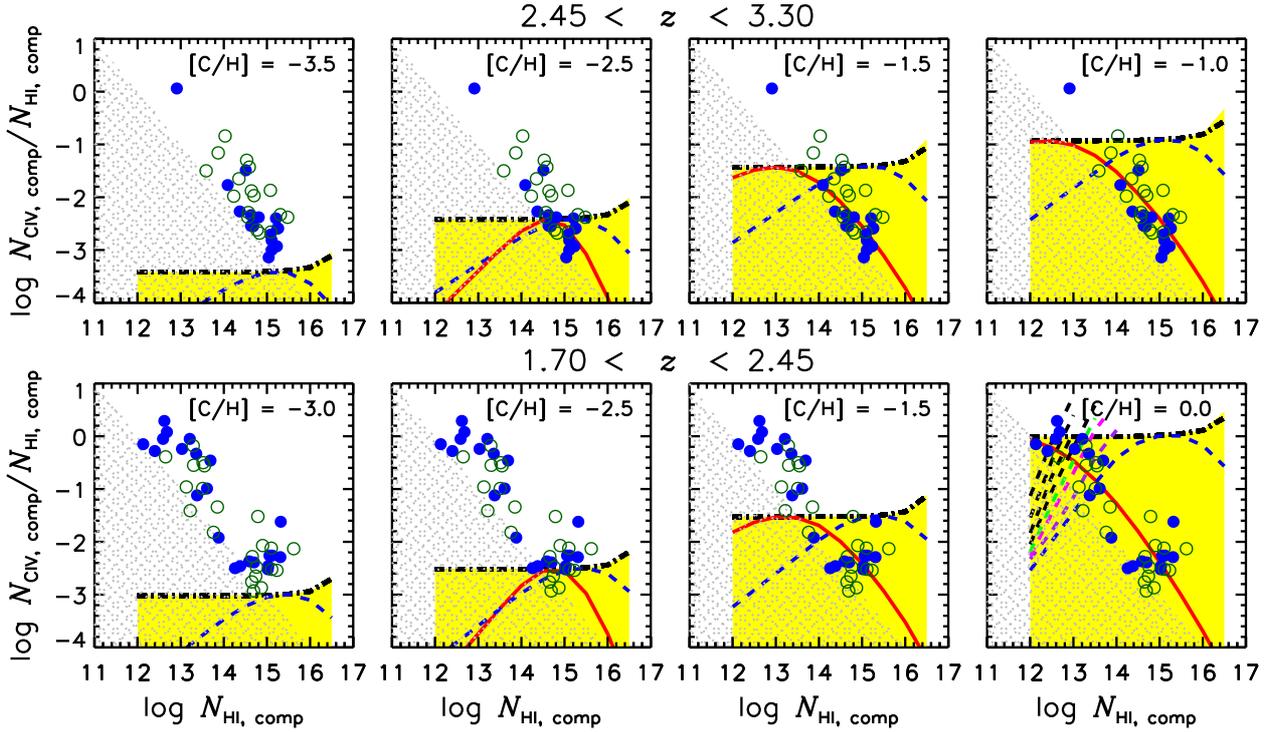}
\caption{ 
Upper (lower) panels show the \nhicomp\/--(\ncivcomp\//\nhicomp\/) 
relation at $2.45 < z < 3.30$ ($1.70 < z < 2.45$)
at four [C/H] noted in each panel. The symbols
are the observed data points as in the left panels of
Fig.~\ref{fig14}. The gray-shaded
area represents the region where the current observations cannot reach due
to detection limit. The yellow-shaded region covers the 
\nhicomp\/--(\ncivcomp\//\nhicomp\/) parameter space predicted by {\tt CLOUDY} with 
the total hydrogen volume density $\log n_{\mathrm{H}} \in [-5, 0]$. The
dot-dashed curve at the upper boundary of the yellow-shaded region
delineates $\log n_{\mathrm{H}} = -3.8$, around which the highest possible  
\ncivcomp\//\nhicomp\/ is predicted for a given [C/H]. 
Overlaid curves are the {\tt CLOUDY}-predicted
\nhicomp\/--(\ncivcomp\//\nhicomp\/) relation with a different 
$n_{\mathrm{H}}$--\nhi\/ relation. The blue dashed curve is for
the fiducial Schaye $n_{\mathrm{H}}$--\nhi\/ relation. The red solid curve
is the predicted \nhicomp\/--(\ncivcomp\//\nhicomp\/) relation
for the best-fit empirical $n_{\mathrm{H}}$--\nhi\/ relation,
$\log n_{\mathrm{H}} = A + B \times \log N_{\mathrm{\ion{H}{i}}}$.
At $2.45 < z < 3.30$, $(A, B) =
(-18.4, 1.0), (-11.6, 0.6)$ and $(-10.6, 0.55)$ for
[C/H] $= -2.5, -1.5$ and $-1.0$, respectively.
At $1.70 < z < 2.45$, $(A, B) = 
(-18.4, 1.0), (-11.6, 0.6)$ and $(-9.5, 0.5)$ for
[C/H] $= -2.5, -1.5$ and 0.0, respectively. In the lower right panel,
several dashed curves above the blue dashed curve indicates the
{\tt CLOUDY} predictions for the Schaye $n_{\mathrm{H}}$--\nhi\/ relation at
[C/H] $= 0.5, 1.0, 1.2, 1.4, 1.6, 1.8$ and 2.0, respectively.}
\label{fig16}
\end{figure*}

\subsection{{\tt CLOUDY} predictions for \ncivcomp\//\nhicomp\/ as a function of \nhicomp\/}
\label{sec6:4}

Figure~\ref{fig16} shows the observed \nhicomp\/--(\ncivcomp\//\nhicomp\/) plane
overlaid with the {\tt CLOUDY} predictions at $2.45  <  z  <  3.30$ 
and at  $1.70  <  z  <  2.45$ 
for four different carbon abundances.

For $2.45 <  z  <  3.30$:

\begin{enumerate}
\item The aligned pairs have [C/H] $\ge -3.5$. This [C/H] $\sim -3.5$  is the estimated 
median IGM metallicity from a pixel
optical depth analysis by \citet{schaye03}. That 
study uses all the \ion{H}{i} 
absorption components including those without actual \ion{C}{iv} detections
above a detection limit. At $z \sim 2.5$, the fraction of \ion{C}{iv}-enriched
absorbers at $\log N_{\mathrm{\ion{H}{i}}} \sim 15$ is about 50\,\%. This fraction
decreases rapidly as  $\log N_{\mathrm{\ion{H}{i}}}$ decreases, while the number of 
absorbers at $\log N_{\mathrm{\ion{H}{i}}} \in [13, 15]$ is about 19 times larger
than the ones at $\log N_{\mathrm{\ion{H}{i}}} \ge 15$ \citep{kim13}. Therefore,
our [C/H] limit
on the aligned pairs with detected \ion{C}{iv} absorption should be higher. 
In general, [C/H] spans at $[-3.5, -1.0]$ if $\log n_{\mathrm{H}} \sim -3.8$,
i.e. the total hydrogen volume density which produces the maximal
\ncivcomp\//\nhicomp\/ at [C/H] $\in [-3.5, -1.0]$. 
However, if we choose [C/H] $\sim -1.0$, $\log n_{\mathrm{H}}$ of most aligned
pairs ranges from $-3.2$ to $-2.3$. 

\item  At [C/H] $= -2.5$, the median \ncivcomp\//\nhicomp\/ of
the well-aligned pairs (filled blue circles)
can be reproduced approximately by the Schaye $n_{\mathrm{H}}$--\nhi\/ relation, 
with a scatter both in [C/H] and $n_{\mathrm{H}}$. We can obtain a good empirical fit for
the well-aligned pairs 
with the $n_{\mathrm{H}}$--\nhi\/ relation, 
$\log n_{\mathrm{H}} = -18.4 + \log N_{\mathrm{\ion{H}{i}}}$,
while the reasonably-aligned pairs clearly have a higher [C/H]. We note that
this empirical fit does not have any physical basis, but only depends on the observed data
and that the empirical best-fit $n_{\mathrm{H}}$--\nhi\/
relation differs for a different [C/H]. 

\item An empirical best-fit $n_{\mathrm{H}}$--\nhi\/ relation with [C/H] $= -1.0$, 
$\log n_{\mathrm{H}} = -10.6 + 0.55 \times 
\log N_{\mathrm{\ion{H}{i}}}$,
provides an overall good fit to all the aligned pairs.

\end{enumerate}

For $1.70 < z < 2.45$:

\begin{enumerate}

\item  For the data points
at (\ncivcomp\//\nhicomp\/) $\sim -2.4$, the Schaye 
$n_{\mathrm{H}}$--\nhi\/ relation
with [C/H] $= -2.5$ reproduces the observations 
reasonably well, with 
a small scatter in $\Delta$[C/H] $\sim \pm 0.5$. 

\item  On the other hand, the data points at 
$\log$\,(\ncivcomp\//\nhicomp\/) $\ge -1.5$
and $\log$\,\nhicomp\/ $\le 14$ require  
[C/H] $\ge -1.0$. 
If $\log n_{\mathrm{H}} \sim -3.8$,
their [C/H] spans from $-1.0$ to 0.0. 
With the Schaye $n_{\mathrm{H}}$--\nhi\/ relation,
[C/H] has to be $0.0 \sim 2.0$.

\item  The two distinct groups in terms of
\ncivcomp\//\nhicomp\/ and the \nhicomp\/--\ncivcomp\/ relation at $z \sim 2.1$
probably arise from different physical conditions.
 
\item  The aligned pairs grouped around at $\log$\,(\ncivcomp\//\nhicomp\/) $\sim \! -2.4$
do not seem to show a noticeable redshift-dependence on [C/H]. 

\end{enumerate}

\subsection{The velocity offset between \ion{H}{i} and \ion{C}{iv} components}
\label{sec6:5}

One of the predictions from the \nhisys\/--\ncivsys\/ relation, if we extrapolate, 
is that the 
majority of the Ly$\alpha$ forest 
at $\log N_{\mathrm{\ion{H}{i}, \, comp}}  <  14$ might be {\it truly} \ion{C}{iv}-free,
i.e. $\log N_{\mathrm{\ion{C}{iv}, \, comp}}  \ll 11.8$, especially at high redshifts.
This is in good agreement with the finding from
the stacking analysis by \citet{ellison00}. They de-redshifted the
absorption-free \ion{C}{iv} regions associated with 67 \ion{H}{i}
components at $\log N_{\mathrm{\ion{H}{i}, comp}} \in [13.5, 14.0]$
at $z \sim 3.45$ from 2 Keck/HIRES spectra. 
All the de-redshifted \ion{C}{iv} regions were then co-added
to produce a $S/N = 1250$ stacked spectrum. No absorption was
seen at down to $\log N_{\mathrm{\ion{C}{iv}, \,comp}} \sim 10.6$
(converted from the quoted 
\ion{C}{iv} detection limit of 0.15\,m\AA\/ at 4$\sigma$).

However, this result from the stacking analysis is valid only when  
the absorption centroids of \ion{H}{i} and \ion{C}{iv} occur at the same relative velocity
\citep{lu93, ellison00, pieri10}. When the velocity offset is random, 
then adopting the \ion{H}{i} redshift will result in a
smearing out any weak \ion{C}{iv} absorptions.
Indeed, \citet{ellison00} have found a velocity offset 
between \ion{H}{i} and \ion{C}{iv} 
at $\log N_{\mathrm{\ion{H}{i}}} \in [13.6, 16.0]$ 
with a dispersion of
$\sim$\,17\,\kms\/ at $z \sim 3.45$. A similar velocity offset was also
found between \ion{H}{i} and \ion{O}{vi}, and \ion{O}{vi} and \ion{C}{iv} 
at $z \sim 1.5$ \citep{reimers01}.

\begin{figure}
\hspace{-0.5cm}
\includegraphics[width=8.9cm]{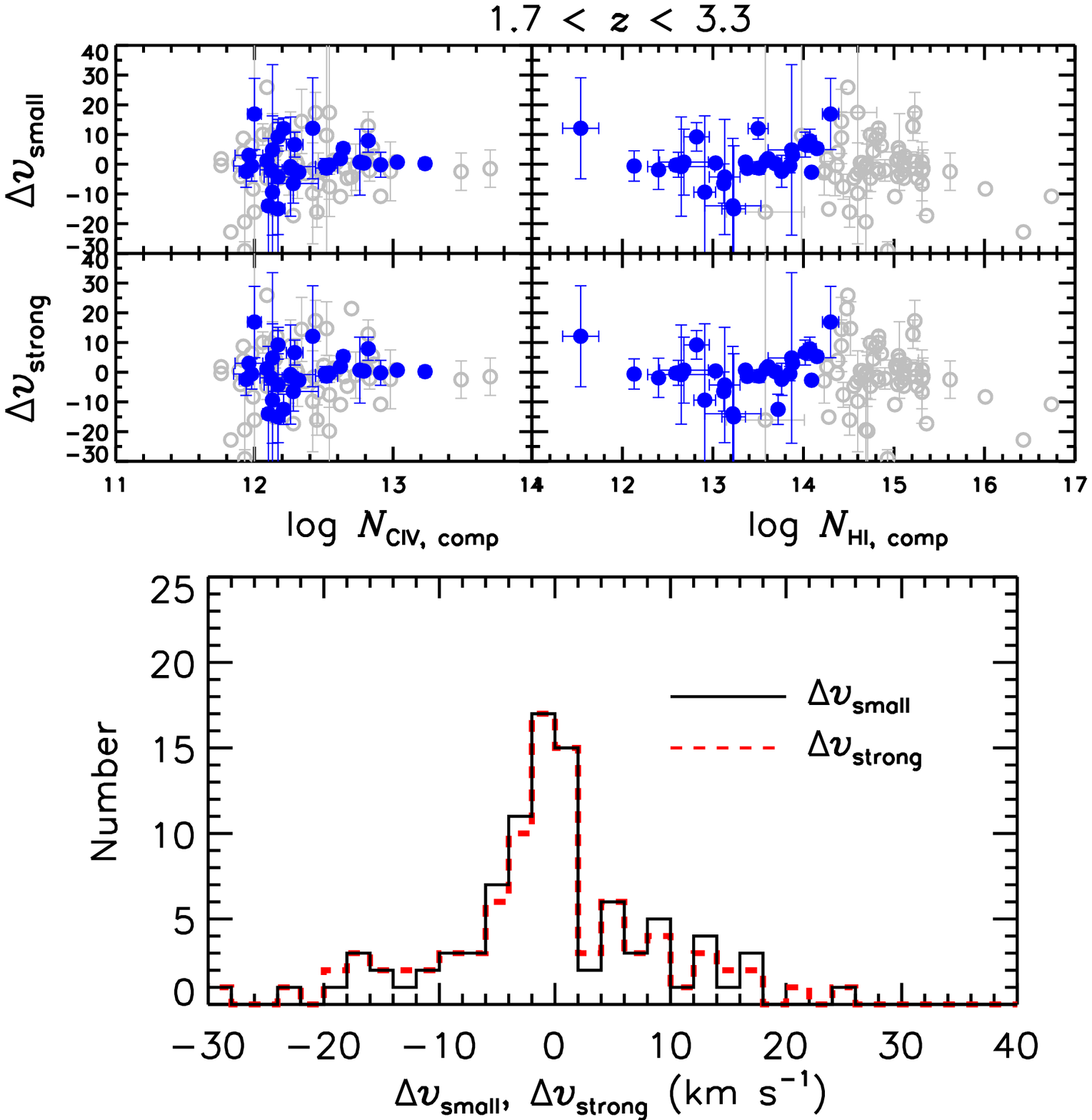}
\vspace{-0.2cm}

\caption{Upper panels: The velocity difference $\Delta v_{\mathrm{small}}$
in \kms\/ between the \ion{C}{iv}
flux minimum and the closest \ion{H}{i} component of 95 single-component
\ion{C}{iv} clumps as a function of \ncivcomp\/ (left panel) and 
\nhicomp\/ (right panel), respectively. Middle panels: 
The velocity difference $\Delta v_{\mathrm{strong}}$
between the \ion{C}{iv}
flux minimum and the strongest \ion{H}{i} component of 95 single-component
\ion{C}{iv} clumps.
Filled and open circles represent
\ion{C}{iv} components associated with unsaturated (Sample I) 
and saturated \ion{H}{i} Ly$\alpha$ components (Sample II), respectively. 
Only errors larger than the symbol size are plotted.
Lower panels: The number of all the aligned pairs as a function of
$\Delta v_{\mathrm{small}}$ (the solid histogram) and of
$\Delta v_{\mathrm{strong}}$ (the red dashed histogram).
}
\label{fig17}
\end{figure}

As our \nhi\/ was measured using higher Lyman orders, thus revealing
a more reliable \ion{H}{i} component structure, we checked
the velocity offset between \ion{H}{i} and \ion{C}{iv} centroids. 
These are shown in Fig.~\ref{fig17}. 
Among 97 single-\ion{C}{iv}-component clumps
including {\it uncertain} clumps
(the 6th column of Table~\ref{tab3}),
we selected 95 clumps 
with unsaturated \ion{C}{iv} and
$\log N_{\mathrm{\ion{H}{i}, \, cl}} \le 17$, 
since the stacking analysis 
is primarily applied for low-\nhi\/ \ion{H}{i} components with no \ion{C}{iv}
and since only the strongest \ion{C}{iv}
component would appear due to noise. 
Out of 95, 30 clumps are associated with 
unsaturated \ion{H}{i} Ly$\alpha$ (Sample I, blue filled circles) 
and 65 clumps have saturated Ly$\alpha$ 
(Sample II, gray open circles).
Note that most Sample I clumps are higher-\ncivsys\/ systems.

In both upper panels of Fig.~\ref{fig17}, the velocity offset $\Delta v_{\mathrm{small}}$
is calculated between the \ion{C}{iv} flux minimum and the closest \ion{H}{i}
component. A velocity offset is clearly present, but
without any trend with \ncivcomp\/, nor with \nhicomp\/. \ion{C}{iv} components 
with a large velocity
offset are the ones whose \ion{C}{iv} flux minimum is at the wing of
\ion{H}{i} profiles, such as the $z = 2.521424$ clump toward
Q0002--422. For 30 \ion{C}{iv} components of Sample I,
the mean and median velocity offsets 
are $0.7\pm 7.0$\,\kms\/ and 0.2\,\kms\/, respectively.
For the remaining 65 clumps of Sample II, 
the mean and median velocity offset is
$-2.5 \pm 12.8$\,\kms\/ and $-1.5$\,\kms\/. 
For the full 95 components, 
the mean and median velocity offsets are
$-1.5 \pm 11.3$\,\kms\/ and $-$0.6\,\kms\/. 
There is no clear redshift dependence on $\Delta v_{\mathrm{small}}$.

The velocity offset $\Delta v_{\mathrm{strong}}$ 
is between the \ion{C}{iv} flux minimum
and the strongest \ion{H}{i} components in the selected clumps. 
For Sample I (Sample II), the mean and median velocity offset is
$-0.1 \pm 7.0$\,\kms\/ ($-2.6 \pm 13.1$\,\kms\/)
and $-$0.2\,\kms\/ ($-1.4$\,\kms\/). 
There is no noticeable difference between
$\Delta v_{\mathrm{small}}$ and $\Delta v_{\mathrm{strong}}$, as seen
in the lower panel of Fig.~\ref{fig17}.
For the full 95 components, 
the mean and median $\Delta v_{\mathrm{strong}}$ is
$-1.8 \pm 11.6$\,\kms\/ and $-0.6$\,\kms\/.

Our standard deviation of $\Delta v_{\mathrm{small}}$
and $\Delta v_{\mathrm{strong}}$ for the full sample is about a 
factor of 1.5 smaller than the one found by \citet{ellison00}. 
About 58\% (55 out of 95 clumps) 
have $\Delta v_{\mathrm{small}}$ less than 5\,\kms\/. 
The minimum $b$ value
of \ion{C}{iv} in the single-\ion{C}{iv}-component clumps
is 4.7\,\kms\/ with a median $b$ of 11.5\,\kms\/ and a 1\,$\sigma$
of 6.0\,\kms\/. Therefore,  
the stacking analysis could decrease the absorption flux by a factor of about two.
This implies that $\log$\,(\ncivcomp\//\nhicomp\/) could decrease by about 0.3\,dex
than its true value by stacking, or increase its detection limit by the same amount.
However, considering a large observational
scatter seen in $\log$\,(\ncivcomp\//\nhicomp\/) vs $\log$\,\nhicomp\/ (see
Fig.~\ref{fig16}), a possible [C/H] range derived from {\tt CLOUDY} modelings on
the stacked data is not affected significantly. 

\section{The origin of the integrated \nhi\/--\nciv\/ relation for \ion{C}{iv} systems}
\label{sec7}

\subsection{Dependence of the \nhisys\/--\ncivsys\/ relation on other ions}
\label{sec7:1}

\begin{figure}
\includegraphics[width=8.6cm]{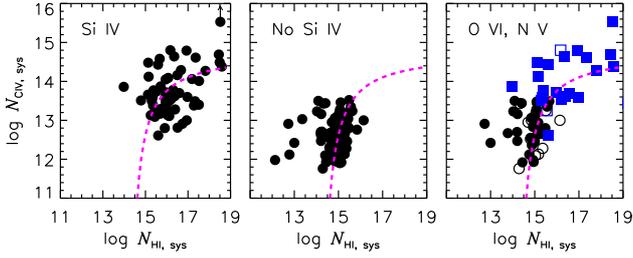}
\vspace{-0.3cm}
\caption{ 
The integrated \nhi\/--\nciv\/ relation of \ion{C}{iv} systems
associated with \ion{Si}{iv} (left panel) and no 
\ion{Si}{iv} (middle panel). Systems with blended, uncertain or
unobserved \ion{Si}{iv} regions are not included in the figures. 
In the right panel, circles
and squares represent systems without and with associated
\ion{Si}{iv}, respectively. Filled and open symbols represent
secure and uncertain detections of \ion{O}{vi} and \ion{N}{v}, 
respectively. The magenta
dashed curve is the rectangular hyperbola fit as in Fig.~\ref{fig7}.
}
\label{fig18}
\end{figure}

Figure~\ref{fig18} shows the integrated \nhi\/--\nciv\/ relation of \ion{C}{iv} systems,
separated into groups set by the existence or otherwise of \ion{Si}{iv}, \ion{O}{vi} and 
\ion{N}{v}. Any systems with blended, uncertain or not covered \ion{Si}{iv}
are excluded. The left and middle panels
show the \nhisys\/--\ncivsys\/ relation of \ion{Si}{iv}-enriched
and \ion{Si}{iv}-free \ion{C}{iv} systems, regardless of existence of
\ion{O}{vi} and \ion{N}{v}.
\ion{Si}{iv}-enriched \ion{C}{iv} systems
show a scatter around the hyperbola fit
curve, even
though they occupy in a reasonably well-defined area at
$(\log N_{\mathrm{\ion{H}{i}, \, sys}}, \log N_{\mathrm{\ion{C}{iv}, \, sys}}) \! = \! (14.0$--19.0,
12.8--15.0). However,
84\% of \ion{Si}{iv}-free \ion{C}{iv} systems 
(87 out of 103 systems) lie close to the 
curve at the lower-\ncivsys\/ end, with a few higher-\ncivsys\/ systems.

Figure~\ref{fig18} also shows
that the \nhisys\/--\ncivsys\/ relation for
systems with \ion{O}{vi} and/or \ion{N}{v}.
Due to the difficulty in detecting the \ion{O}{vi} $\lambda\lambda$\,1031, 1037 doublet
in the high-order forest region and the \ion{N}{v} $\lambda\lambda$\,1238, 1242 doublet
blended in the Ly$\alpha$ forest region,
the \civ\/ systems
shown in the right panel are not necessarily include all the systems
containing \ion{O}{vi} and/or \ion{N}{v}. 
Despite this incompleteness problem in detecting \ion{O}{vi} 
and \ion{N}{v}, most \ion{C}{vi} systems with \ion{O}{vi} and/or 
\ion{N}{v} follow a similar trend displayed for \ion{Si}{iv}-enriched and
\ion{Si}{iv}-free systems. This suggests that
\ion{O}{vi} and \ion{N}{v} sample \ion{C}{iv} systems over a wide
range of physical conditions.

\subsection{[C/H] as an origin of the steep part of the integrated \nhi\/--\nciv\/ relation}
\label{sec7:2}

If we assume that [C/H], $n_{\mathrm{H}}$ and the UV background 
in a single \ion{C}{iv} system does not vary significantly, 
the middle panel of Fig.~\ref{fig14} seems to suggest that
the steep part of the integrated \nhi\/--\nciv\/ relation could be caused by the {\it right} 
combination of [C/H] and $n_{\mathrm{H}}$. There is no
solid observational evidence on a constant [C/H], $n_{\mathrm{H}}$
and the UV background within a \ion{C}{iv} system. However,
as the steep \nhisys\/--\ncivsys\/ relation holds primarily for the \ion{Si}{iv}-free
and single-\ion{C}{iv}-component \ion{C}{iv} systems, 
they can be thought to sample a low-density region where the gas is
optically thin and no radiative transfer effects complicate the internal
structure of the absorption gas \citep{schaye00a, schaye01, shen13}.

\begin{figure}
\includegraphics[width=8.6cm]{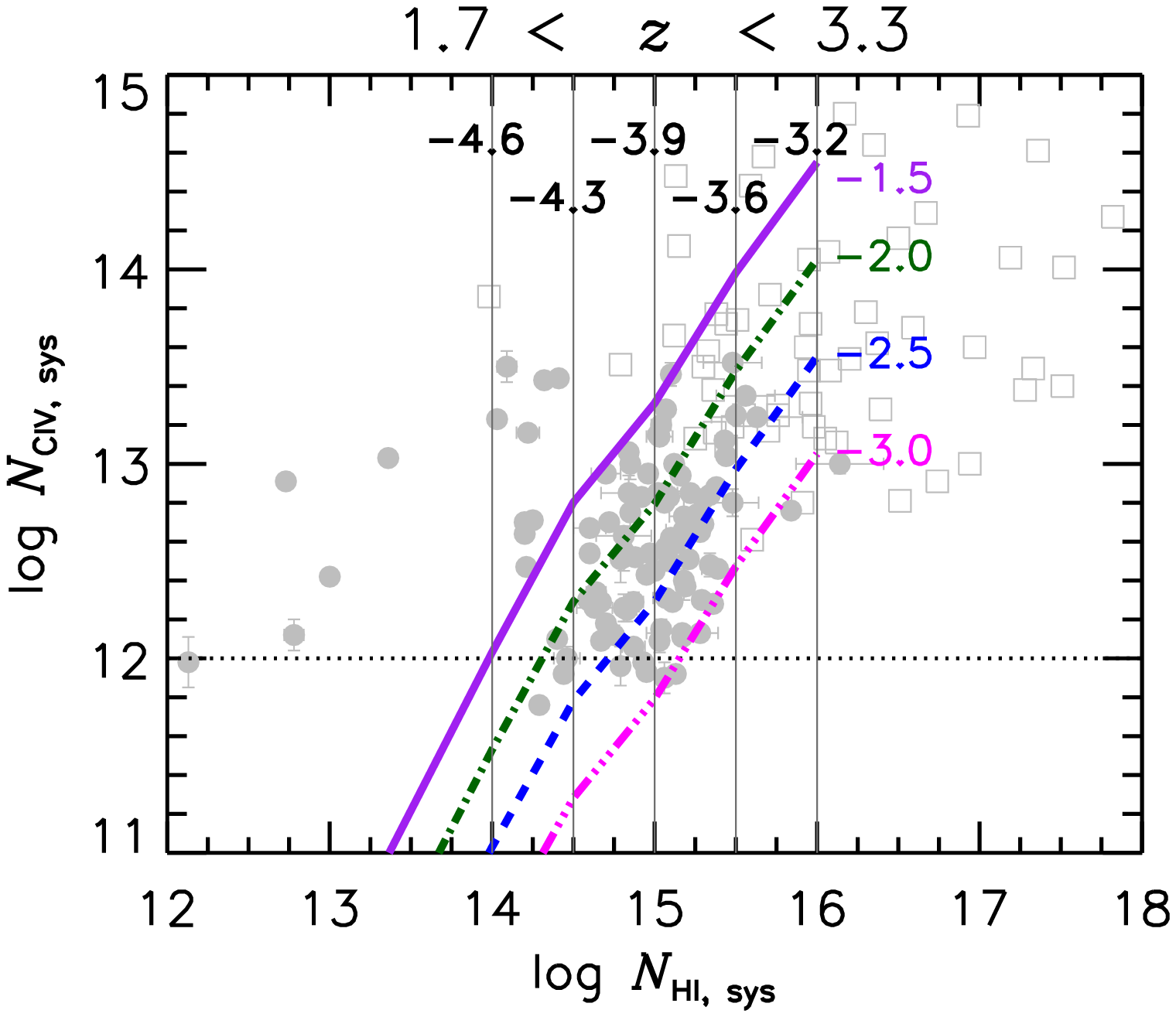}
\vspace{-0.5cm}
\caption{ 
The integrated \nhi\/--\nciv\/ relation of \ion{C}{iv} systems overlaid
with the {\tt CLOUDY} predictions for a fixed
[C/H] noted next to each curve for the fiducial
Schaye $n_{\mathrm{H}}$--\nhicomp\/
relation as a gray vertical line with its logarithmic 
$n_{\mathrm{H}}$ value
on top. The \ion{C}{iv} systems associated with \ion{Si}{iv} and without  
\ion{Si}{iv} are presented as open squares and filled circles. Systems with 
blended, uncertain or
unobserved \ion{Si}{iv} regions are not included. 
The dotted horizontal line presents the \ion{C}{iv} detection limit.}
\label{fig19}
\end{figure}

Figure~\ref{fig19} presents the steep \nhisys\/--\ncivsys\/ relation overlaid
with the {\tt CLOUDY} predictions, when the fiducial
Schaye $n_{\mathrm{H}}$--\nhicomp\/
relation was assumed. The data 
are well reproduced by
absorbers with [C/H] $\in [-3.0, -1.5]$ and $\log n_{\mathrm{H}} \in [-4.3, -3.5]$.
Note that the {\tt CLOUDY}-prediction for [C/H] $= -3.0$ even reproduces the
sharp lower edge shown at $\log N_{\mathrm{\ion{H}{i}}} \in [15.0, 16.0]$.

\subsection{Origins: Filaments and galactic halos}

If the steep \nhisys\/--\ncivsys\/ relation for \ion{Si}{iv}-free \ion{C}{iv} systems
is produced by the absorbing gas
with [C/H] $\in [-3.0, -1.5]$ and $n_{\mathrm{H}} \in [-4.3, -3.5]$, the
two best candidates for the location
of the gas are the IGM filaments close to the
star-forming galaxies or the outer 
regions of intervening halos, cf. Fig.~\ref{fig1}.
Unfortunately, calculating a column density profile of \ion{H}{i} and \ion{C}{iv}
as a function of impact parameter 
is not trivial. Even if the baryon
density profile is assumed to follow the halo dark matter profile,
conversion from the total hydrogen
to \ion{H}{i} requires an ionisation correction due to the UV background
radiation and an interaction between outflows and inflows
\citep{klar08, duffy12, thom12}. 

\begin{figure*}
\includegraphics[width=17.5cm]{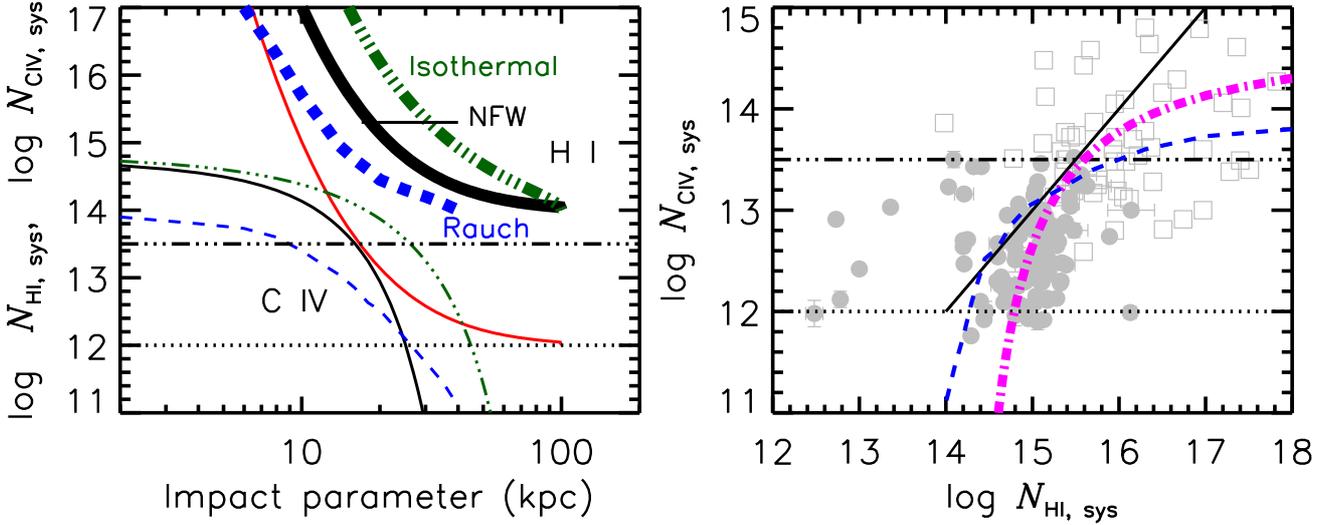}
\caption{ 
The left panel shows various profiles of \nhisys\/ (thick curves) and \ncivsys\/
(thin curves) as a function
of impact parameter $\rho$ in kpc. The thick solid and dot-dot-dot-dashed curves
are an assumed \nhisys\/ profile for the NFW
and isothermal density profiles, respectively, with the 
parameters matching $\log N_{\mathrm{\ion{H}{i}, \, sys}} \sim 14$ at 100\,kpc,
a size of an assumed intervening halo. 
The thin solid red curve
is a \ncivsys\/ profile 
with $N_{\mathrm{\ion{C}{iv}, \, sys}}/N_{\mathrm{\ion{H}{i}, \, sys}} = 0.01$
for the shown $N_{\mathrm{\ion{H}{i}, \, NFW}}$ profile.
Thick and thin blue dashed curves are
\nhisys\/ and \ncivsys\/ profiles taken from Fig.~8 by \citet{rauch97} for a filament centred
at a simulated
protogalactic clump of 
$1.2 \times 10^{9}$\,M$_{\odot}$. Their values are shown only upto 40\,kpc. 
The black solid
and dark green dot-dot-dot-dashed thin curves are the expected
\ncivsys\/ profile to match the observed \nhisys\/--\ncivsys\/ relation 
(magenta dot-dashed curve) in the
right panel for the NFW and isothermal models, respectively.
The data points in the right panel are the
same as in Fig.~\ref{fig19}. In both panels, 
the horizontal dotted line represents
our \ion{C}{iv} detection limit, while the horizontal dot-dot-dot-dashed line is the \ion{C}{iv}
detection limit of $\log N_{\mathrm{\ion{C}{iv}, \, gg}} = 13.5$ from
the galaxy-galaxy pair study at $z \sim 2.2$ by \citet{steidel10}.
}
\label{fig20}
\end{figure*}

For simplicity, we took a \citet{navarro97} (NFW) spherical dark matter density 
profile as the model for the \ion{H}{i} density profile, 
$n_{\mathrm{\ion{H}{i}, \, {\tiny{NFW}}}} (r) = \frac{A}{r/x (1+r/x)^{2}}$,
where $r$ is a radial distance and the values
of $A$ and $x$ vary from halo to halo. 
We note that the NFW profile successfully describes objects on large scales such as
clusters and filaments, while it 
has difficulties 
on galaxy scales \citep{moore99, primack09, governato12}.
We also used a power-law gas density profile with $n_{\mathrm{\ion{H}{i}, \, iso}} (r) = 
B/r^{2}$ for an isothermal sphere. 

The radial \ion{H}{i} column density profile is then obtained by integrating
$n_{\mathrm{\ion{H}{i}}} (r)$ along a line of sight through a halo at a
given impact parameter $\rho$. For our NFW model,
\begin{equation}
N_{\mathrm{\ion{H}{i}, \, NFW}} (\rho) = 2 \int^{R}_{\rho} \frac{n_{\mathrm{\ion{H}{i}, \, 
   NFW}}(r) \, r}{\sqrt{r^{2} -\rho^{2}}} \, dr,
\end{equation}
where $R$ is a radius of a spherical halo in kpc, for  
$\rho$ and $r$ in kpc, $n_{\mathrm{\ion{H}{i}, \, NFW}}$ in cm$^{-3}$
and $N_{\mathrm{\ion{H}{i}, \, NFW}} (\rho)$ in cm$^{-2}$.

For an isothermal sphere,
\begin{equation}
N_{\mathrm{\ion{H}{i}, \, iso}} (\rho) = B \left[\frac{\pi}{2\rho} - 
  \frac{\arcsin(\rho/R)}{\rho}\right].
\end{equation}

Figure~\ref{fig20} shows 
these assumed \nhisys\/ profiles, including the simulated \nhisys\/ profile by
\citet{rauch97} for an IGM filament centred at a protogalactic clump 
with the halo mass of $1.2 \times 10^{9}$\,M$_{\odot}$. 
For the NFW model,
we assumed a halo size of 100\,kpc and took a value 
of $x = 16.1$\,kpc as for the Milky Way \citep{nesti13}. Then, we varied the value of $A$
roughly to match $\log N_{\mathrm{\ion{H}{i}, \, NFW}} \sim 14$ at 
100\,kpc based on Fig.~\ref{fig7}. This leads to a dimensionless constant $A$ of $\log A = 15.0$.
We used a similar procedure to make an isothermal model
to have $\log N_{\mathrm{\ion{H}{i}, \, iso}} \sim 14$ at 100\,kpc. 

Figure~\ref{fig20} also illustrates an expected \ncivsys\/ profile to reproduce the
observed \nhisys\/--\ncivsys\/ relation for a given \nhisys\/ profile.
The Rauch \ncivsys\/ profile 
gives an overall similar shape as our \nhisys\/--\ncivsys\/ relation, but follows
the upper envelope of our data, implying that their simulation 
systematically overproduces
\ncivsys\/ for a given \nhisys\/. When the ratio of \nhisys\/ and \ncivsys\/
is constant, a linear \nhisys\/--\ncivsys\/ relation is expected (the thin solid line in the right panel).

We note that our derived \ncivsys\/ profile
depends on the assumed \nhisys\/ profile which is
not likely to be correct and that our observed \nhisys\/--\ncivsys\/ relation
is from an ensemble of many gas clouds along the line of sight passing 
galaxies and IGM filaments of different masses and sizes (see Fig.~\ref{fig1}).
As a more massive galaxy tends to have a larger halo and a higher star-formation rate
\citep{brooks11}, our toy model of a 100\,kpc halo size is too simplistic.
Fortunately, the observed \nhisys\/--\ncivsys\/ relation 
is not a function of impact parameter, i.e. both integrated column densities
are measured at the {\it same} impact parameter. 
Moreover, simulations have found that density profiles of halos
and the surrounding IGM filaments are self-similar when scaled with the
virial radius which is dependent on the galaxy mass \citep{pallottini14}.
Therefore, our assumed column density profiles
can be extended by multiplying the impact parameter axis by any number
and the comparison of \nhisys\/ and \ncivsys\/ profiles at the same impact
parameter does not depend on any assumed halo size.  

Figure~\ref{fig20} implies that the \ncivsys\/ profile should be
different from \nhisys\/ to reproduce the observed \nhisys\/--\ncivsys\/ relation, 
regardless of the assumed \nhisys\/ profile. \ncivsys\/ should have
a rapid decrease compared to \nhisys\/ at closer to the halo size or the IGM filament size. 
As a result, the extent of \ion{C}{iv} should be much smaller than \ion{H}{i}
above our detection limit of \ion{H}{i} and \ion{C}{iv}.

A similar trend has been found at $z \sim 2.2$ from the galaxy-galaxy pair study
by \citet{steidel10}. Their composite background galaxy spectrum shows a sharp decrease in
the \ion{C}{iv} rest-frame equivalent width (REW) at $\sim$80--90\,kpc above the \ion{C}{iv} 
detection limit, the \ion{C}{iv} REW of $\ge 0.15$\,\AA\/ 
(roughly $\log N_{\mathrm{\ion{C}{iv}}} \ge 13.5$).
On the other hand, \ion{H}{i} extends
to a larger distance to $\sim 250$\,kpc, but shows a similar rapid falloff in 
the \ion{H}{i} REW at $\ge 0.31$\,\AA\/  
(roughly $\log N_{\mathrm{\ion{H}{i}}} \ge 13.25$) around 250\,kpc. The rapid falloff in 
\ncivsys\/ seems to continue to our smaller \ncivsys\/ range, implying that a 
line-of-sight \ion{C}{iv} extent is not very different 
between $\log N_{\mathrm{\ion{C}{iv}, \, limit}} = 13.5$ and 
$\log N_{\mathrm{\ion{C}{iv}, \, limit}} = 12.0$.

We note that the \ncivsys\/ profiles displayed in Fig.~\ref{fig20} are
different from those observed by \citet{steidel10}. Their composite spectrum shows
\ion{C}{iv} absorption down to $\log N_{\mathrm{\ion{C}{iv}}} \sim 13.5$ at $\sim$90\,kpc,
while the same \ncivsys\/ is expected at $\rho \sim 14$\,kpc for our toy-model NFW profile.
This discrepancy is mainly due to the 
direct comparison of column density profiles as a function of
impact parameter without normalised by the virial radius. 
The Steidel sample mainly consists of
galaxies bright enough to obtain a spectrum at $z \sim 2.2$, thus a higher column density
at the same impact parameter than our 100\,kpc toy-model halo.
 
The existence of
a well-characterised \nhisys\/--\ncivsys\/ relation implies 1) a line-of-sight extent of the
\ion{C}{iv} gas is smaller than \ion{H}{i}, 2)
$N_{\mathrm{\ion{C}{iv}, \, sys}}$ decreases more rapidly than 
$N_{\mathrm{\ion{H}{i}, \, sys}}$ at the larger impact parameter,  
3) their ratio can be well-characterised by a simple parameter, such as a potential
of intervening halos or the IGM filament and 4) the integrated \ion{H}{i} column 
density can be used as a proxy of normalised impact parameter.

\subsection{Implications of the lack of evolution for the steep \nhisys\/--\ncivsys\/ relation}

If we assume that a metal-enriched gas expands according to the Hubble flow
as soon as it is placed in the surrounding IGM filament, 
ignoring any interaction with the infalling IGM, and that
the gas temperature and the UV
background do not change significantly at $z \sim 2.5$, then 
$N_{\mathrm{\ion{H}{i}}}$ at the
redshift $z_{2}$ becomes at the redshift $z_{1}$
\begin{eqnarray}
\nhi\/(z_{1}) & = & \nhi\/(z_{2}) \frac{n_{\mathrm{z_{1}}}}{n_{\mathrm{z_{2}}}} 
  \, \frac{L_{z_{1}}}{L_{z_{2}}} \nonumber\\
& = & \nhi\/(z_{2}) \, \frac{n_{\mathrm{0}} \, (1+z_{1})^{3}}{n_{\mathrm{0}} \, (1+z_{2})^{3}} 
    \, \frac{L_{\mathrm{0}} \, (1+z_{1})^{-1}}{L_{\mathrm{0}} \, (1+z_{2})^{-1}},
\end{eqnarray}
where $n_{\mathrm{0}}$ is the local gas density and $L_{0}$ is the local
line-of-sight size. 
\nhisys\/ (\ncivsys\/) of the gas with $\log \nhisys\/ = 15.2$ ($\log \ncivsys\/ = 13.6$) 
at $<z> \,\,=2.80$ 
evolves to the gas with $\log \nhisys\/ = 15.04$ 
($\log \ncivsys\/ = 13.44$) by $<z> \,\, =2.18$. This small
difference makes virtually no evolution in the \nhisys\/--\ncivsys\/ relation, given
a large spread in \ncivsys\/. 

If the majority of \ion{Si}{iv}-free \ion{C}{iv} systems occur in the outskirts of an intervening
halo due to outflows, the \nhisys\/--\ncivsys\/ relation at a given epoch
should include a signature of the outflows from
previous starburst episodes happened before that epoch.
Let us assume that the outflow velocity does not change. If
the outflow launch velocity is assumed to be 
(100\,\kms\/, 300\,\kms\/, 600\,\kms\/), the distance it travels
during two epochs $z_{2} = 2.80$ and $z_{1}=2.18$ is
(72\,kpc, 215\,kpc, 430\,kpc), respectively, for our assumed cosmology. 
Only outflows with a velocity $\ge 300$\,\kms\/ 
originating at $z > 2.8$ would have
reached the virial radius of parent galaxies with 10$^{11.7}$\,M$_{\odot}$ by $z = 2.18$. 
They would have been mixed with the surrounding IGM at $z < 2.18$
\citep{aguirre01b}.

Since more previous star-formation events are accumulated at lower redshifts 
and since more time is spent for metals to spread in the lower-\nhi\/ IGM, 
the \nhisys\/--\ncivsys\/ relation would have shown a
larger scatter at lower \nhi\/ at lower redshifts. 
The fact that no such scatter is observed implies
that 1) the terminal outflow velocity in the halo might be much lower than  
the escape velocity in general, i.e. the majority of metals stay inside a virial radius
\citep{oppenheimer08}, 
2) each intervening halo is not likely to have several previous star-formation episodes
at $z \sim 2.5$, and/or 3) the majority of outflow activities happen at $z \gg 3$, when 
a galaxy mass is lower, thus the escape velocity is also lower. Without later outflows,
the \nhisys\/--\ncivsys\/ relation at $z \sim 2.5$ is only an asymptotic behaviour of outflows
at $z > 3$, implying that 
most metals escaped to the surrounding IGM might
have already been diluted below the detection limit
\citep{aguirre01b}.

If most of \ion{Si}{iv}-free \ion{C}{iv} systems are at around the virial radius 
\citep{aguirre01b, oppenheimer08, cen11, voort11, shen13},
the \ion{C}{iv}-bearing halo gas is expected to have a similar behaviour to the \ion{C}{iv}-bearing
IGM filament gas, having its physical volume density close to the cosmic mean density.
Therefore, we do not expect any significant redshift evolution at $z \in [1.7, 3.3]$.

\section{Conclusions}
\label{sec8}

We have presented the relations between \nhi\/ and \nciv\/ of the 
183 intervening
\ion{C}{iv} absorbers at 
$1.7 < z < 3.3$, based on the 23 high-resolution ($\sim$\,6.7\,\kms\/)
spectra obtained with UVES at the VLT and HIRES at Keck, with
the detection limit of \nhi\/ and
\nciv\/ of $\log N_{\mathrm{\ion{H}{i}}} \sim 12.5$ and
$\log N_{\mathrm{\ion{C}{iv}}} \sim 12.0$. 
As \ion{C}{iv} is usually associated with saturated \ion{H}{i}, we used all the available
high-order Lyman lines to obtain a reliable component structure and a 
robust \nhi\/ from the Voigt profile fitting analysis. 

We define three terms to describe
our \ion{H}{i}+\ion{C}{iv} sample, systems, clumps and components.

{\bf Systems:}  \ion{C}{iv} system refers to all the \ion{H}{i}
and \ion{C}{iv} components within a fixed velocity range centred at the
\ion{C}{iv} flux minimum, with the default of $\pm 150$\,\kms\/. 
When a \ion{C}{iv} absorption extends over this velocity range or
when a separate \ion{C}{iv} absorption near $\pm 150$\,\kms\/ 
is seen beyond it,
the velocity interval is extended to include
additional \ion{C}{iv} in that direction by steps of 100\,\kms\/. 
Column densities
integrated over this velocity range are averaged 
quantities.

{\bf Clumps:} When the absorption wings of visibly separable 
\ion{C}{iv} profiles recover to a normalised flux of 1 and 
a closest \ion{C}{iv} absorption wing is more than
5\,\kms\/ away,
this distinct absorption feature is termed as a {\it clump}. All
the \ion{H}{i} components within the clump velocity range are assigned to that
clump. When no \ion{H}{i} exists, the clump velocity range is extended to
include nearby \ion{H}{i} components, 
depending on the profile shape of \ion{H}{i} and \ion{C}{iv}.
A clump can consist of a single
component or multiple components. 

{\bf Components:} The \ion{H}{i} and \ion{C}{iv} component pairs are grouped
{\it well-aligned} if their velocity centroid differs by $\le 5$\,\kms\/ and both
are relatively clean. If nearby \ion{H}{i} components make 
the line parameter of the aligned \ion{H}{i} less reliable or if  a \ion{C}{iv} is 
located in a low-S/N region,
the pairs are labelled as
{\it reasonably-aligned}. Photoionisation modelling can be applied for aligned
component pairs since they can be thought to be co-spatial.

From our 183 intervening \ion{C}{iv} systems, we find:

\begin{enumerate}

\item For about $\sim$\,75\,\% 
of the \ion{C}{iv} systems (137/183), 
the integrated \ion{H}{i} and \ion{C}{iv} 
column densities,
\nhisys\/ and \ncivsys\/, show a steep increase in \ncivsys\/ with \nhisys\/
at $\log$\,\nhisys\/ $\in [14, 16]$, then becomes independent of \nhisys\/ at
$\log$\,\nhisys\/ $\ge 16$, with a large scatter in
$\Delta \log N_{\mathrm{\ion{C}{iv}, \, sys}} = 2.5$\,dex for
a given \nhisys\/. 

\item This \nhisys\/--\ncivsys\/ relation is best  
approximated as a rectangular hyperbola
function at $\log N_{\mathrm{\ion{H}{i}, sys}} \in [14, 22]$ and
at $\log N_{\mathrm{\ion{C}{iv}, \, sys}} \ge 11.8$:
$$\lognc4 =
\left[\frac{(-1.90 \pm 0.55)}{\lognh + (-14.11 \pm 0.19)} \right]
   + (14.76 \pm 0.17).$$

\item The \nhisys\/--\ncivsys\/ relation does not depend on the velocity
range integrated over if it is $\ge \pm 100$\,\kms\/.
 
\item Assuming that the physical conditions in the gas do not change much
within $\pm 150$\,\kms\/, the steep \nhisys\/--\ncivsys\/ relation can be reproduced
by a gas with [C/H] $\in [-3.0, -1.5]$ and the total hydrogen volume density 
$\log n_{\mathrm{H}} \in [-4.3, -3.5]$ under the Haardt-Madau QSOs+galaxies
2005 UV background and our fiducial Schaye $n_{\mathrm{H}}$--\nhi\/ relation.
The low [C/H] and $n_{\mathrm{H}}$ suggest that the gas satisfying the 
steep \nhisys\/--\ncivsys\/ relation is likely to arise from a halo of intervening
galaxies or the surrounding IGM filaments.

\item \ion{C}{iv} systems following the steep part of the \nhisys\/--\ncivsys\/ relation at
$\log N_{\mathrm{\ion{H}{i}, sys}} \in [14, 16]$ are \ion{Si}{iv}-free, consisting
of one or two \ion{C}{iv} components. The flat part of the
relation at $\log N_{\mathrm{\ion{H}{i}, sys}} \in [16, 22]$ 
is dominated by \ion{Si}{iv}-enriched systems, implying that they are
produced by the gas with a higher physical volume density
and/or a higher metallicity, i.e. galactic discs or inner haloes.

\item The steep \nhisys\/--\ncivsys\/ relation also requires that
a line-of-sight extent of the \ion{C}{iv} gas is smaller than the \ion{H}{i} gas and that
$N_{\mathrm{\ion{C}{iv}, \, sys}}$ decreases more rapidly than 
$N_{\mathrm{\ion{H}{i}, \, sys}}$ at the larger impact parameter
above the detection limits,
regardless of the location of the \ion{H}{i}+\ion{C}{iv} gas,
i.e. intervening halos or IGM filaments. In addition, 
\nhisys\/ can be used as a proxy of normalised impact parameter by the
virial radius.

\item There is a group of \ion{C}{iv} systems (about 16\,\%, 
30/183) 
which do not follow the steep \nhisys\/--\ncivsys\/ relation at
$(\log N_{\mathrm{\ion{H}{i}, \, sys}}, \log N_{\mathrm{\ion{C}{iv}, \, sys}})
\! = \! (12.0$--14.0, 11.8--13.6). They have a higher \ncivsys\/ compared to 
\ion{C}{iv} systems having a similar \nhisys\/. This higher-\ncivsys\/ can be
obtained only by the gas with [C/H] $\in [0.0, 2.0]$ and
$\log n_{\mathrm{H}} \in [-4.3, -3.5]$ for the fiducial
Schaye $n_{\mathrm{H}}$--\nhi\/ relation,
if the gas is close to the photoionisation equilibrium, implying that
they are closely connected to star formation activities.

\item While there is no significant redshift evolution shown by the \ion{C}{iv}
systems following the \nhisys\/--\ncivsys\/ relation, the higher-\ncivsys\/ system
only shows up predominantly at lower redshifts. 
 
\item If the \nhisys\/--\ncivsys\/ relation holds at 
$\log N_{\mathrm{\ion{C}{iv}, sys}} \le 11.8$, 
we expect very few \ion{C}{iv}-enriched
\ion{H}{i} absorbers at $\log N_{\mathrm{\ion{H}{i}, \, sys}} \le 14.0$ (or \nhicomp\/),
except rare higher-\ncivsys\/ absorbers.

\end{enumerate}

As for 227 \ion{C}{iv} clumps and 
77 aligned \ion{H}{i}+\ion{C}{iv} components, we find:

\begin{enumerate}

\item At $2.45 <  z  <  3.30$, the majority of \ion{C}{iv} clumps follow the \nhisys\/--\ncivsys\/
relation. However, at $1.70 < z  < 2.45$, there is no well-defined \nhicl\/--\ncivcl\/
relation, as \ion{C}{iv} clumps start to show much larger scatters on the
\nhicl\/--\ncivcl\/ plane. For the full redshift range, there exists
no recognisable \nhicl\/--\ncivcl\/ relation.

\item For 95
single-\ion{C}{iv} component clumps, the median velocity difference 
between closest \ion{H}{i} and \ion{C}{iv} component centroids is $-0.6$\,\kms\/, 
with a 1$\sigma$ dispersion of 11.3\,\kms\/. 
The median velocity difference between
\ion{C}{iv} and strongest \ion{H}{i} component is $-0.6$\,\kms\/, with a dispersion
of 11.6\,\kms\/. The velocity smearing decreases a weak \ion{C}{iv} flux by 
a factor of about 2 in the stacking analysis, but does not change the inferred [C/H] 
significantly, given a large scatter seen in observational data. 

\item For aligned components, there is no recognisable \nhicomp\/--\ncivcomp\/
relation at both redshift ranges, with more spread in data points farther away from
the steep part of the \nhisys\/--\ncivsys\/ relation at lower redshifts.

\item There is a strong suggestion that there might be two separable \ion{C}{iv}
component groups at $1.70 \! < \! z \! < \! 2.45$. One group follows the steep part of the
\nhisys\/--\ncivsys\/ relation with $\log$\,(\ncivcomp\//\nhicomp\/) $\sim -2.4$, 
arising from the gas with [C/H] $\sim \! -2.5$ and  
$\log n_{\ion{H}{i}} \sim -4.3$. Another group has 
a wider range of \ncivcomp\//\nhicomp\/ with 
$\log$\,(\ncivcomp\//\nhicomp\/) $\in [-1, 0]$, requiring [C/H] $\ge 0.0$
for the fiducial Schaye $n_{\ion{H}{i}}$--\nhi\/ relation, 
possibly a consequence of increased star formation rates 
at $z \sim 2.1$. 

\end{enumerate}

\section*{Acknowledgments.}
We are grateful to M. Rauch, M. Viel, M. Haehnelt, J. Bolton and B. Savage
for the insightful discussions.
TSK acknowledges funding support from the European Research
Council Starting Grant ``Cosmology with the IGM" through grant
GA-257670.
RFC is also supported by the same grant for his stay at Osservatorio Astronomico
di Trieste to carry out part of this work.
TSK is also grateful to a travel support by the FP7 ERC Advanced Grant
Emergence-320596 to IoA, Cambridge, where part of this work was done.

\bibliography{literature}

\end{document}